\documentclass[a4paper,11pt]{article}
\usepackage{jcappub}
\usepackage[T1]{fontenc}
\usepackage{lineno}
\usepackage{xcolor}
\usepackage{ulem}
\usepackage{float}
\usepackage{subcaption}
\usepackage{enumitem}
\usepackage{microtype}

\newcommand{\GalR}{\textsc{GALAXY}{\fontsize{16}{16}\Large$\mathcal{R}$}\!\textsc{ATE}~}

\numberwithin{equation}{section}

\title{Gravitational Wave Bias in IllustrisTNG300 from Machine-Learned Population-Synthesis Calibrated Merger Rates}

\author[a,b,e]{Dorsa Sadat Hosseini,}
\author[f, k]{Taisiia Karasova,}
\author[g,h]{Nikshay Chugh,}
\author[e,i,j]{Alex Krolewski,}
\author[a,b,e]{Ghazal Geshnizjani}
\affiliation[a]{Department of Applied Mathematics, University of Waterloo, Waterloo, ON N2L 3G1, Canada}
\affiliation[b]{Perimeter Institute for Theoretical Physics, Waterloo, ON N2L 2Y5, Canada}
\affiliation[e]{Waterloo Centre for Astrophysics, University of Waterloo, 200 University Ave W, Waterloo, ON N2L 3G1, Canada}
\affiliation[f]{Department of Physics, Massachusetts Institute of Technology, Cambridge, MA 02139, USA}
\affiliation[g]{Department of Physics, Indian Institute of Science, Bangalore 560012, India}
\affiliation[h]{Department of Physics, The Ohio State University, Columbus OH 43210, USA}
\affiliation[i]{Department of Physics and Astronomy, University of Waterloo, Waterloo, ON N2L 3G1, Canada}
\affiliation[j]{California Institute of Technology, 1200~East California Boulevard, Pasadena, CA 91125, USA}
\affiliation[k]{Cavendish Laboratory, Department of Physics, University of Cambridge, J. J. Thomson Avenue, Cambridge, CB3 0HE, UK}

\emailAdd{d2sadath@uwaterloo.ca}
\emailAdd{karasova@mit.edu}
\emailAdd{chugh.31@osu.edu}
\emailAdd{akrolews@caltech.edu}
\emailAdd{ggeshnizjani@pitp.ca}

\abstract{ The large scale clustering of gravitational wave (GW) sources is a promising and independent probe of both cosmology and compact-binary astrophysics. Interpreting future GW clustering measurements, however, requires a detailed understanding of how compact-binary merger rates depend on host-galaxy properties and environment. We develop a simulation-based framework to model the clustering of binary black hole (BBH) mergers and to estimate the corresponding GW bias. Our method populates the IllustrisTNG300 hydrodynamical simulation with GW sources using a machine-learning emulator. The emulator is trained on the \GalR data set, which provides population-synthesis calibrated merger rate predictions by linking binary evolution to galaxy formation histories. This enables us to construct physically motivated mock GW catalogs based on galaxy-specific BBH merger rates as functions of stellar mass, star formation rate, metallicity, and redshift. To isolate the impact of host properties, we compare this framework to a simpler phenomenological host-selection model in which the merger rate depends on stellar mass alone. From both types of catalogs, we measure the GW auto-power spectrum and study how the GW and galaxy biases depend on redshift and host-galaxy properties. In both approaches, GW sources are predicted to be more strongly biased than the overall galaxy population, reflecting the higher probability in the underlying astrophysical model for BBH mergers to occur in more massive galaxies, which reside in more strongly clustered halos. However, the machine-learned merger rate model yields a stronger scale dependence of the GW bias than what the stellar mass weighting alone can explain, indicating that additional host properties also shape the clustering of GW sources. We also find that metallicity correlates positively with both the GW and galaxy biases, implying a similar enhancement for any tracer correlated with metallicity. Our results underscore the importance of physically motivated merger rate models for GW large scale structure analyses. The framework presented here provides a self-consistent link between compact-binary formation, galaxy evolution, and large scale structure, and offers a basis for interpreting GW clustering measurements from future experiments such as the Einstein Telescope and Cosmic Explorer.}

\begin{document}
\maketitle

\section{Introduction}

The direct detection of gravitational waves (GWs) from compact binary mergers including binary black hole (BBH), binary neutron star (BNS), and black hole-neutron star (BHNS) systems has opened a new observational window onto the Universe \cite{KAGRA:2013rdx, LIGOScientific:2014pky, VIRGO:2014yos, PhysRevD.88.043007, LIGOScientific:2016aoc, LIGOScientific:2018jsj, LIGOScientific:2020kqk, LIGOScientific:2021aug, KAGRA:2021duu}. These observations not only confirmed the existence of stellar mass binary black holes and neutron star mergers, but also established GWs as an independent and increasingly powerful cosmological probe. Standard (and dark) sirens provide direct measurements of the luminosity distance without relying on the cosmic distance ladder. With the growing number of detected events, these observations are also transitioning from individual detections to population-level probes of astrophysics and cosmology. As a new class of tracers, their spatial distribution can reveal information about cosmic structure formation, galaxy evolution, and the astrophysical environments in which they originate and merge. Just over the past few years, the size of the detected GW catalog has increased from the $\sim 90$ confident detections in O1-O3 \cite{Abbott_2023} to over 200 candidates in the recent GWTC-4.0 release \cite{Abac_2026, LIGOScientific:2025jau}. The rapidly expanding sample, which also includes higher-mass and higher signal-to-noise BBH events, allows for more detailed studies of how GW sources are linked to their host-galaxy environments and the surrounding large scale structure. 

Looking ahead, proposed third-generation detectors, such as the Einstein Telescope (ET) and Cosmic Explorer (CE), are expected to detect many more compact binary mergers per year, providing sub-degree localization for a large fraction of sources \cite{Iacovelli_2022,Hild_2011,Abbott_2017,Maggiore2020ETScience}. Such unprecedented merger event statistics will make it possible to measure GW clustering and transform compact binary mergers into a new probe of large scale structure.

Central to these efforts is the concept of gravitational wave bias, which quantifies how GW sources trace the underlying dark matter density field. Because compact binaries form within galaxies, GW bias is inherently linked to the connection between binary evolution, host galaxy properties, and cosmic structure formation. Several recent studies have begun to explore this multi-scale connection. In \cite{Dehghani_2025}, photometric surveys (2MPZ \cite{Bilicki2014_2MPZ}, WISExSCOS \cite{Bilicki2016_WISExSCOS}) were used to generate mock GW catalogs with stellar mass dependent host-galaxy probability functions, showing that an astrophysically motivated stellar mass-weighted selection has a strong impact on the gravitational wave bias. The specific prescription for embedding GW sources in galaxy maps or matter distributions is sometimes referred to as `\textit{seeding}' formalism. Later in the analysis of \cite{Hosseini_2026}, this phenomenological-seeding-plus-clustering pipeline was further extended to spectroscopic galaxy surveys, using SDSS DR7 \cite{Abazajian2009_SDSSDR7}, to enable a more detailed investigation of how the GW bias depends not only on stellar mass weighting selections but also on two additional host-galaxy properties: star formation rate and metallicity. In both studies, we also examined the redshift evolution and limited scale dependence of the GW bias. These works highlight that GW clustering is sensitive to the formation channels as well as the astrophysical histories of compact binary hosts, motivating models that consistently connect binary formation and merger history to galaxy evolution with large scale structure \cite{Mukherjee:2021bmw,Namikawa16, Mukherjee:2018ebj,Mukherjee:2019wcg, Mukherjee:2019oma, Calore:2020bpd, PhysRevD.103.043520, Mukherjee:2020mha, Scelfo20, Libanore21, Diaz:2021pem, Libanore22, Gagnon_2024,Marinacci_2026}.

The studies of \cite{Dehghani_2025} and \cite{Hosseini_2026}, which seed observed catalogs, are useful since galaxy surveys naturally incorporate the impact of astrophysical and cosmological processes across all scales
(from stellar evolution to large scale structure) 
on galaxy properties and their distribution.
However, this approach also faces intrinsic limitations. Observational catalogs are incomplete, magnitude-limited, and subject to selection effects, particularly at high redshift. In contrast, cosmological hydrodynamical simulations, although limited by resolution on small scales and by finite volume on large scales, are free from these observational constraints. They produce a self-consistent galaxy population within a well-defined cosmological volume, enabling controlled studies of galaxy-halo connections and large scale clustering. They also provide stellar mass, SFR, and metallicity histories for every galaxy, without observational errors.

In this paper, complementing previous work, we develop a simulation-based pipeline to study GW bias for BBH merger events as a function of redshift and the astrophysical properties of the host galaxy. Specifically, we utilize the IllustrisTNG300 simulation \cite{2015A&C....13...12N,Pillepich_2017,Springel_2017,Nelson_2017,Naiman_2018,Marinacci_2018} to produce GW source population samples and measure their spatial clustering.

Furthermore, for the seeding formalism, we take two different approaches. One of our approaches follows \cite{Dehghani_2025, Hosseini_2026}, using a phenomenologically and astrophysically motivated stellar mass weighted host selection. However, the other approach, which we present first and which our results indicate is both scientifically and computationally promising going forward, involves applying a machine-learning emulator that is trained on the population-synthesis-based seeding formalism from \GalR code \cite{Santoliquido_2022}. The population-synthesis based seeding codes combine cosmological simulations with binary population-synthesis simulations to project intrinsic merger rate predictions onto realistic cosmic histories. 
Note that hydrodynamical cosmological simulations do not yet resolve the scales of star formation or stellar binary evolution; therefore, in these seeding formalisms, the formation and evolution of simulated isolated compact binaries from their progenitor stars to their eventual mergers are embedded into the galaxy population by linking the metallicity of progenitor stars and galaxies and using galaxy star formation rates.
Our trained machine-learning emulator predicts the galaxy-specific merger rates as functions of stellar mass, star formation rate (SFR), metallicity, and redshift for different binary merger models from population-synthesis simulations, capturing key astrophysical uncertainties in compact binary formation and evolution.

It is important to emphasize that BBH formation is not expected to arise from a single astrophysical channel. In this work, the merger rate data from \GalR are based on binary population synthesis for isolated stellar binaries.
However, a growing body of work has demonstrated that dynamical formation channels may also contribute significantly to the observed BBH population. These include dense stellar environments such as globular clusters and nuclear star clusters, as well as formation and migration within active galactic nucleus (AGN) disks \cite{Barber_2025,Kritos2024RAPSTER,Ford2022AGN,Rowan_2023}. These channels can lead to different distributions of BBH masses, spins, eccentricities, and delay times, and may alter the connection between GW sources and their host environments. Therefore, the results presented in this work should be interpreted as the GW bias prediction corresponding to the isolated-binary formation channel, while extensions to include the impact of stellar clusters and AGN-assisted formation remain important directions for future work.

The structure of this paper is as follows. Section \ref{sec:illustris} summarizes the  Illustris TNG300 simulation products, the hydrodynamical simulation used in this analysis. In section \ref{sec:mergerate}, we briefly review the \GalR formalism and the data used for training our merger rate prediction and seeding emulator. In section \ref{sec:mock_gw}, we present our construction of an efficient machine-learning emulator (a gradient boosting regressor) that learns the mapping between host-galaxy properties and BBH merger rates from \GalR{} data. We then apply the trained model to all galaxies in IllustrisTNG300 to predict merger rates and generate mock GW siren catalogs, populating the full three-dimensional simulation volume with predicted merger events. Section~\ref{sec:power_bias} describes our bias-estimation pipeline, including the models used to compute the auto-power spectra of the different tracer fields (galaxies and GW sirens) and the methods employed to determine the best-fit bias parameters that characterize their clustering statistics.
Section \ref{sec:gwbias_result_msfrz} presents our best-fit GW bias estimates and their dependence on host-galaxy stellar mass, star formation rate, and metallicity, using the machine-learned seeding emulator on the IllustrisTNG300 galaxy sample. Section~\ref{sec:gwbiasresult_mass_powerlaw} also shows the GW bias measurement but this time employing a phenomenological host-selection model similar to \cite{Dehghani_2025} that depends only on stellar mass and then compares the results of these two methods. Finally, Section \ref{sec:conclusion} summarizes our findings and offers concluding remarks.


\section{The Illustris TNG300 Simulation}
\label{sec:illustris}

The IllustrisTNG project \cite{2015A&C....13...12N,Pillepich_2017,Springel_2017,Nelson_2017,Donnari_2019} is a suite of large volume, high-resolution cosmological, gravo-magnetohydrodynamical simulations of galaxy formation and dark matter evolution \cite{2010MNRAS.401..791S}. We use the \textsc{TNG300-1} simulation of IllustrisTNG, which spans a comoving box of $L_{\mathrm{box}} = 205\,\mathrm{cMpc}/h$ (volume $= 302.6^3\,\mathrm{cMpc}^3$) with a total of $2\times2500^3$ resolution elements, corresponding to a baryonic mass resolution of $m_{\mathrm{baryon}} = 7.6\times10^{6}\,M_{\odot}/h$ and a dark matter particle mass of $m_{\mathrm{DM}} = 4.0\times10^{7}\,M_{\odot}/h$. While TNG300-1 includes hydrodynamics, gravity, dark matter, and a detailed model for baryonic galaxy-formation physics, its finite resolution prevents it from resolving individual stars, compact binaries, or the detailed sub-kiloparsec structure \cite{2013MNRAS.432..176P,2017MNRAS.465.3291W}. Instead, stellar particles corresponding to unresolved stellar populations are modeled through sub-grid prescriptions. The simulation outputs are structured into group catalogs that identify dark matter halos and subhalos. In this work, we associate galaxies with subhalos according to the prescriptions outlined further below. Furthermore, we work with the subhalo data of snapshots (snapshots 99, 91, 85, 67, 50) corresponding to redshifts $z = \{0.0,\; 0.1,\; 0.2,\; 0.5,\; 1.0\}$. These specific outputs were selected because the BBH merger data set used to train our machine-learning model provides merger rate data only at these redshifts. Together, they span the redshift range where most BBH formation and merger activity occur and offer sufficient coverage to capture the redshift evolution of galaxy and merger rate properties relevant to current gravitational wave observations.

\begin{figure}[h]
    \centering
    \includegraphics[width=0.32\textwidth]{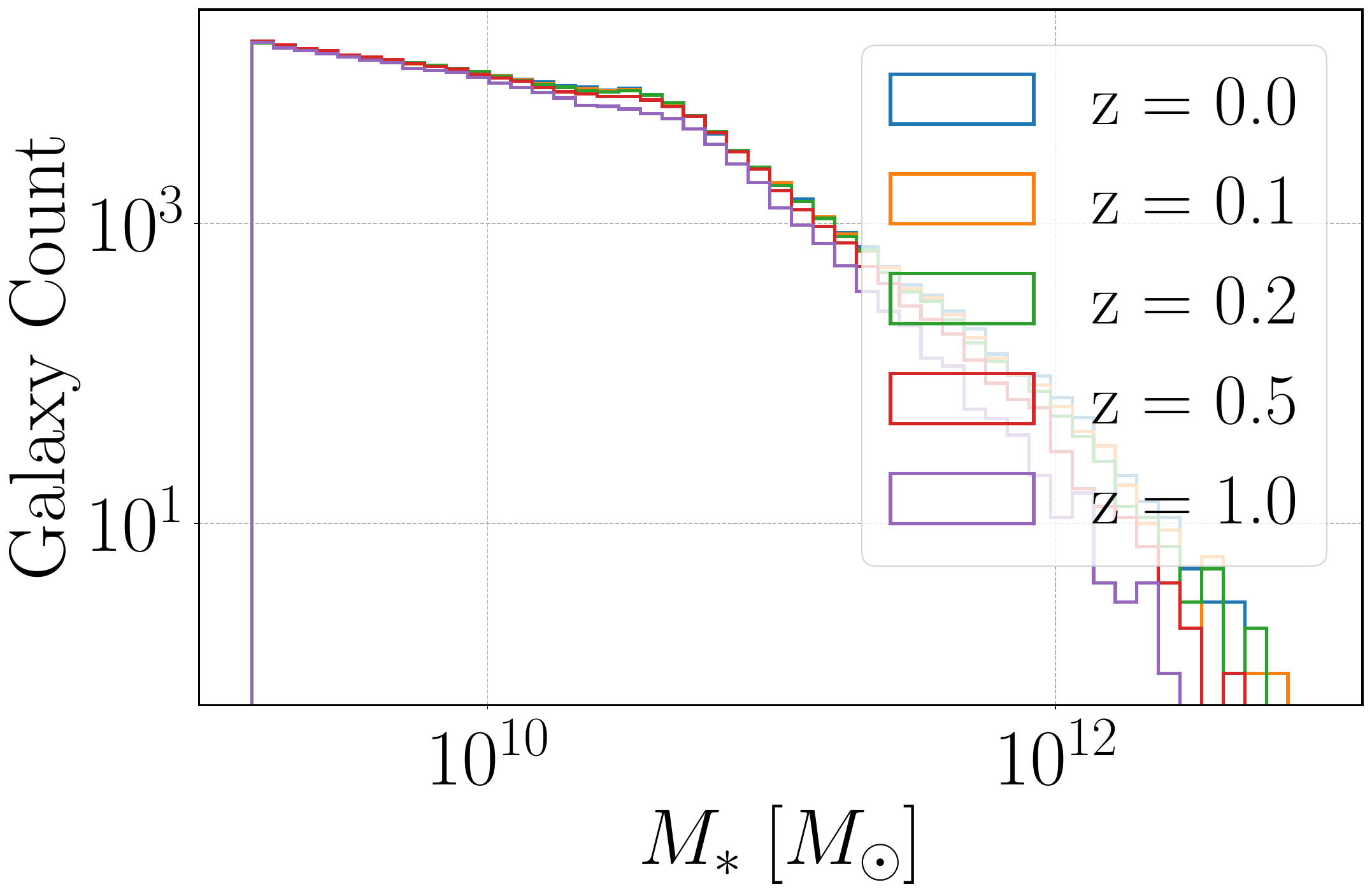}
    \includegraphics[width=0.32\textwidth]{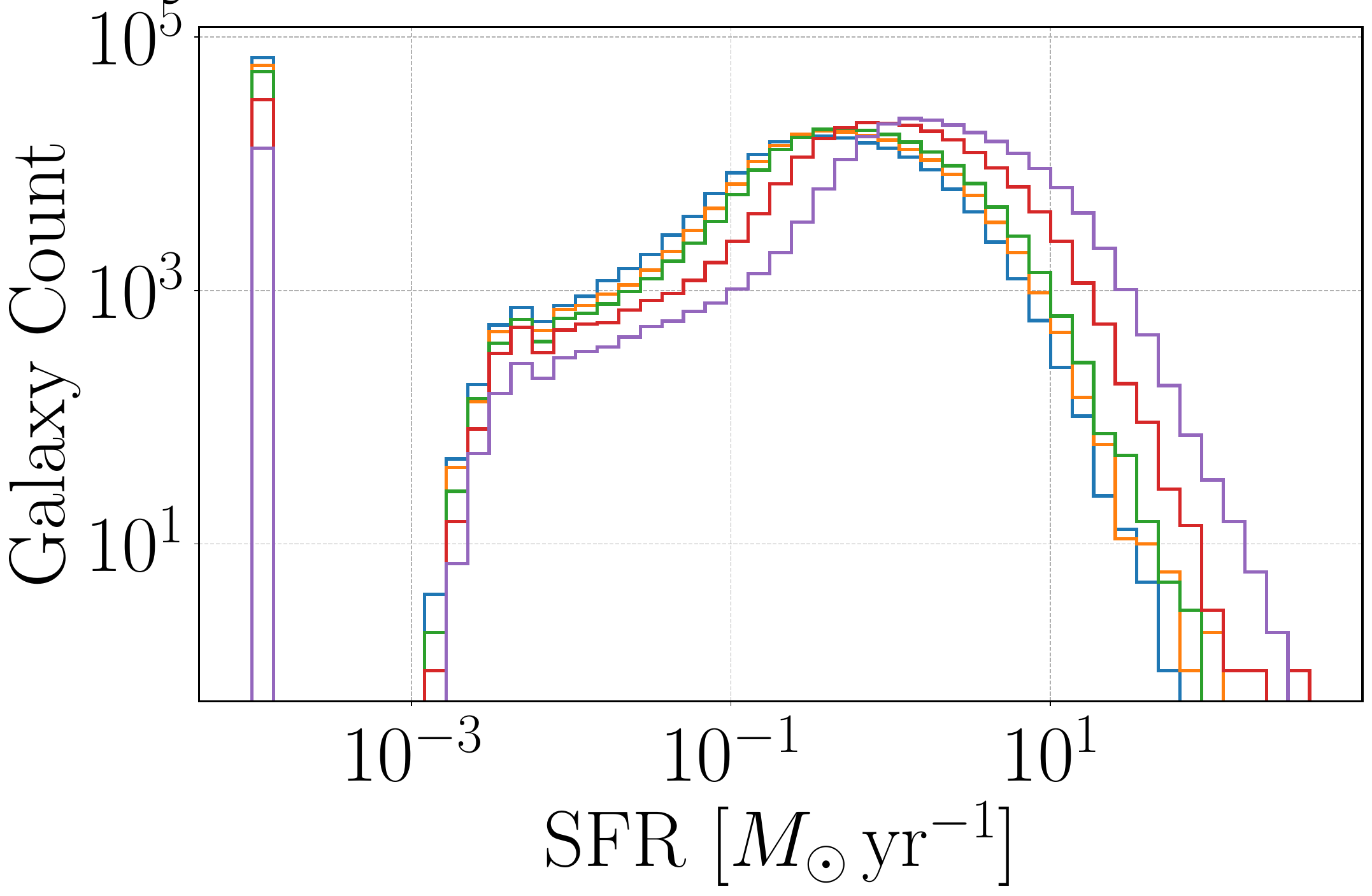}
    \includegraphics[width=0.32\textwidth]{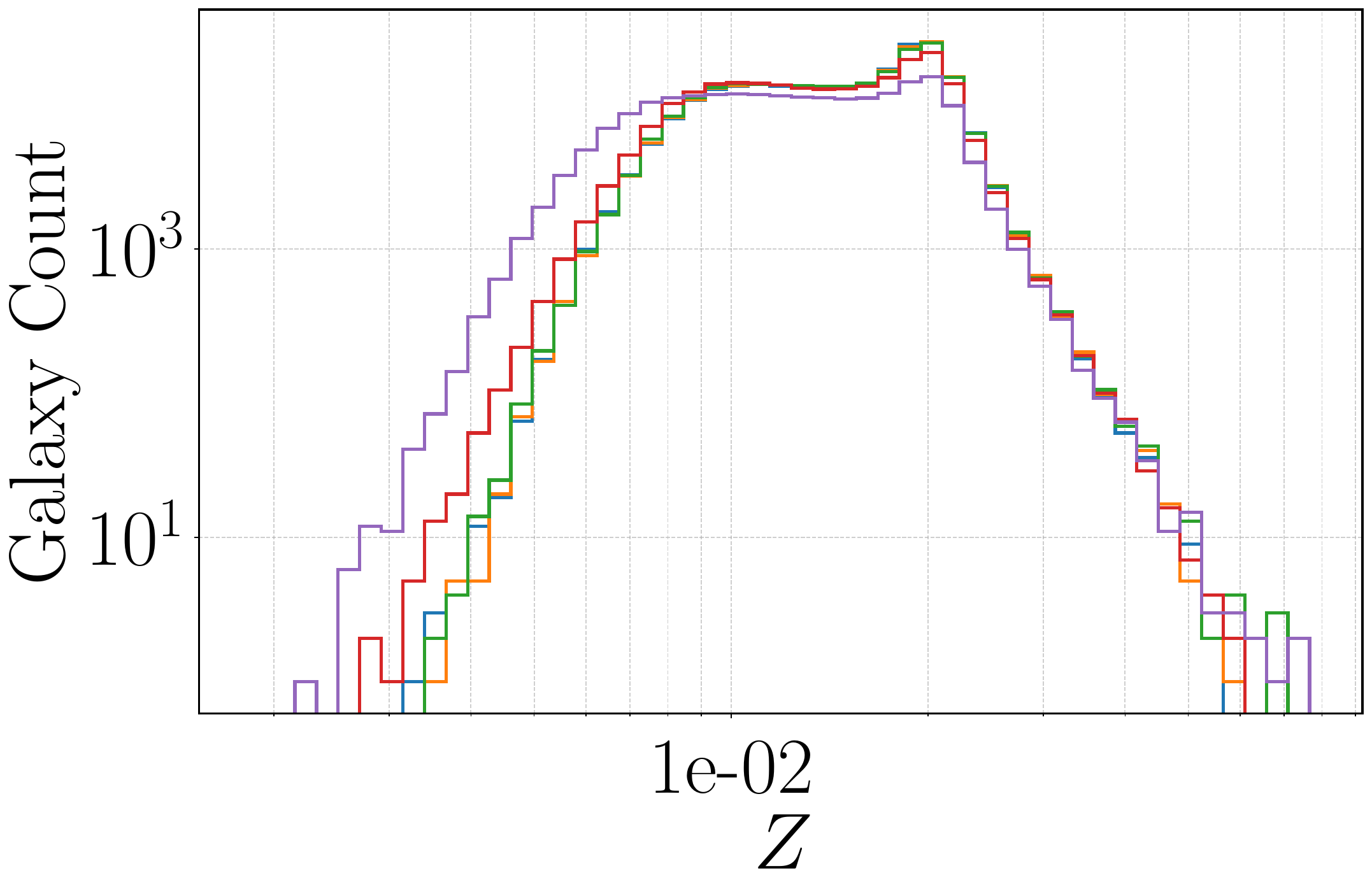}
    \caption{The stellar mass, SFR and metallicity distributions for redshifts of z=\{0.0, 0.1, 0.2, 0.5, 1.0\}.}
    \label{fig:distributions}
\end{figure}

\begin{figure}[h]
    \centering
    \includegraphics[width=0.32\textwidth]{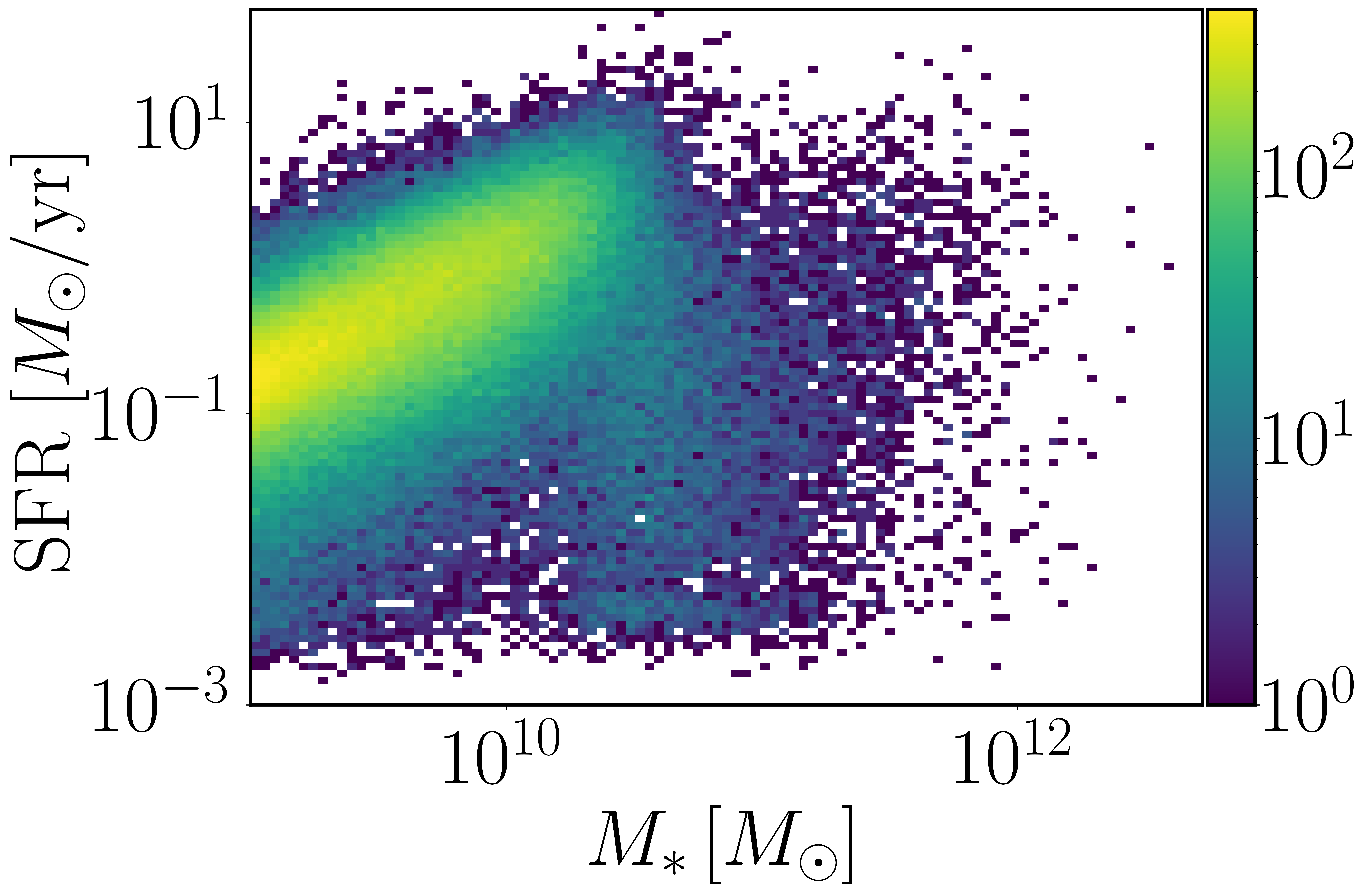}
    \includegraphics[width=0.32\textwidth]{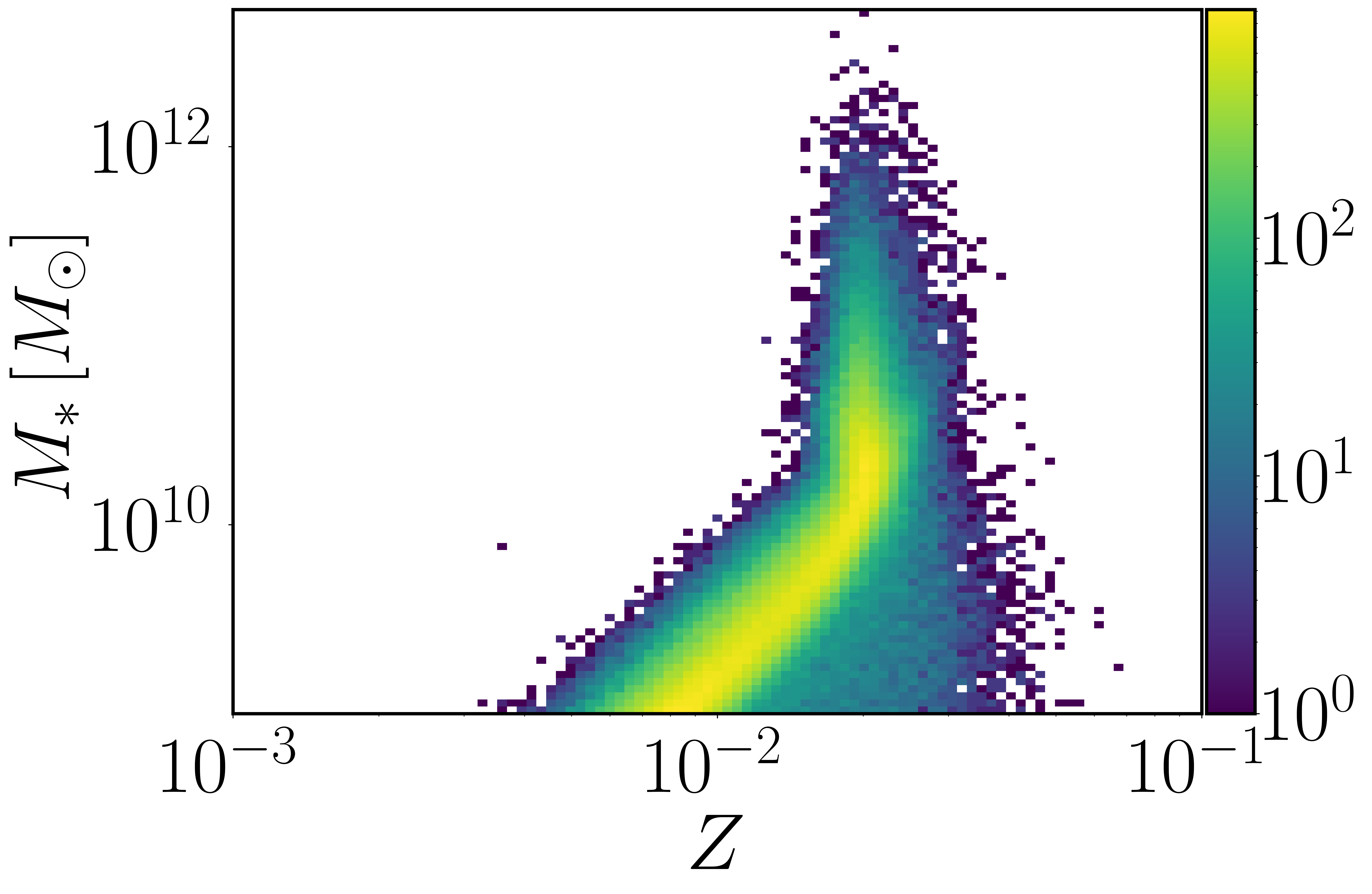}
    \includegraphics[width=0.32\textwidth]{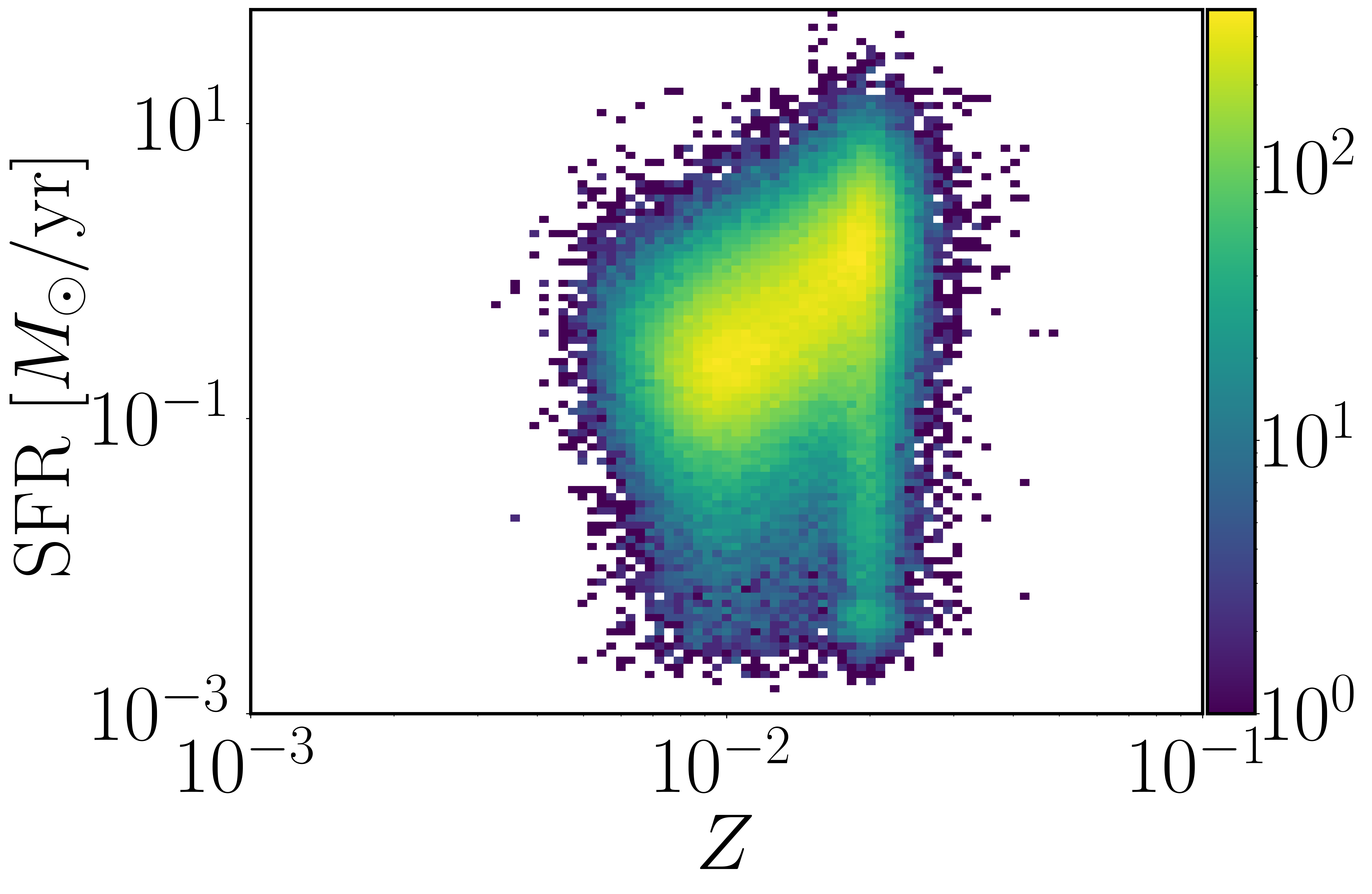}
    \caption{Distribution of galaxies in TNG-300 labeled by color with respect to stellar mass, SFR, and metallicity at $z = 0$ for star forming galaxies (SFR > 0).}
    \label{fig:ssfr_vs_stellar mass}
\end{figure}

\begin{figure}[h]
    \centering
    \includegraphics[width=0.7\textwidth]{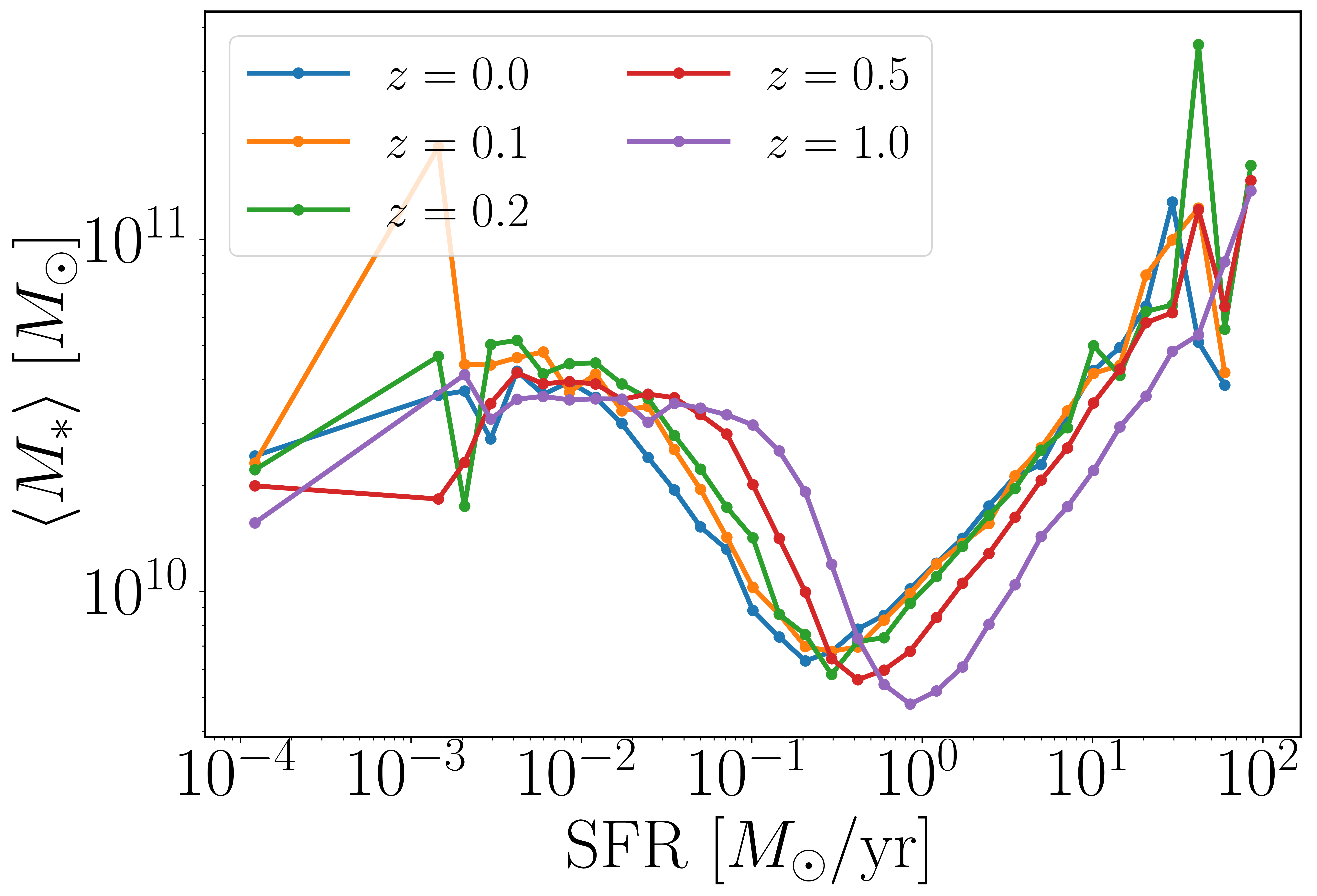}
    \caption{Mean stellar mass as a function of star formation rate (SFR) for z=\{0.0, 0.1, 0.2, 0.5, 1.0\}.}
    \label{fig:mean mass vs sfr}
\end{figure}

 We extract the following key properties from the complete catalog of the TNG300-1 subhalo to produce mock galaxy samples and later host galaxies for BBH mergers, which are required for modeling the merger rate and clustering analysis:
\begin{itemize}
\item \textbf{SubhaloMassType} This field is a six-element array that records the total bound mass of the subhalo, separated by particle type. The array entries correspond to the following components:
gas, dark matter particles, tracer particles (massless passive numerical Monte-Carlo markers used to track the Lagrangian evolution of baryonic matter), star/wind particles, and supermassive black holes. In our analysis, we extract the total stellar mass of each subhalo, which is stored in units of $10^{10}\,M_\odot/h$ and then converted to $M_\star$ in physical solar mass units by substituting $h=0.67$. This quantity includes all stellar particles gravitationally bound to the subhalo. 
To ensure adequate numerical resolution and exclude poorly resolved systems, we restrict our galaxy samples to subhalos with stellar masses $M_\ast \geq 10^{9}\,M_\odot$, corresponding to systems resolved with at least $\mathcal{O}(10^2)$ of stellar particles, as recommended by the Illustris-TNG team. 

\item \textbf{SubhaloSFR} This field provides the \textit{total star formation rate} (SFR) within a subhalo, measured in units of $M_\odot/\mathrm{yr}$. Specifically, it is the sum of the instantaneous SFRs of all gas cells that are gravitationally bound to the subhalo and are eligible for star formation according to the sub-grid model used in the TNG simulation. This quantity traces the level of ongoing star formation activity in a galaxy and is a key predictor of the likelihood that galaxies have recently formed massive stars, which may evolve into binary black holes (BBHs). In our analysis, \texttt{SubhaloSFR} is used as one of the input features in the machine learning model that estimates BBH merger rates from galaxy properties.\\

\item \textbf{SubhaloStarMetallicity} This field represents the \textit{mass-weighted average stellar metallicity} of the subhalo, calculated within twice the stellar half-mass radius ($2 R_{\rm half}$). It is defined as the average metallicity of all star particles bound to the subhalo, weighted by the mass of the star particles. Mathematically, this is expressed as:
\begin{equation}
Z = \frac{\sum_i m_i Z_i}{\sum_i m_i}
\end{equation}
where $m_i$ is the mass of the $i$-th star particle, and $Z_i$ is its metallicity.
This quantity reflects the chemical enrichment history of the galaxy and correlates with both stellar mass and star formation history. In our analysis, \texttt{SubhaloStarMetallicity} is used as another key input property for predicting BBH merger rates since metallicity is expected to influence the formation efficiency and evolution of massive binary stars that can become GW sources.

\item \textbf{SubhaloPos} This field provides the 3D comoving position of each subhalo within the periodic simulation box, specified as a triplet $(x, y, z)$ in units of comoving kiloparsecs per $h$ (ckpc$/h$). The position corresponds to the location of the most gravitationally bound particle in the subhalo i.e. the particle with the lowest gravitational potential energy, ensuring a stable and physically meaningful center. In our analysis, we use \texttt{SubhaloPos} to construct the spatial distribution of galaxies and compute clustering statistics. Specifically, we assign these positions to 3D density grids using a Cloud-In-Cell (CIC) interpolation scheme, producing density fields and enabling the calculation of power spectra for both the galaxy and gravitational wave source catalogs. To ensure unit consistency with the simulation box and Fourier analysis, we convert the subhalo positions from ckpc$/h$ to comoving megaparsecs (cMpc) considering $h=0.67$ before using CIC. 

\end{itemize}
Figure \ref{fig:distributions} shows the stellar mass, SFR, and metallicity distributions of our galaxy samples for several different redshift snapshots, where entries with $\mathrm{SFR}=0$ are assigned $\mathrm{SFR}=10^{-4}\,M_\odot\,\mathrm{yr}^{-1}$ for visualization in logarithmic SFR space. Figure \ref{fig:ssfr_vs_stellar mass} illustrates the corresponding 2D projections at $z=0$ as a function of SFR, metallicity, and stellar mass. Figure \ref{fig:mean mass vs sfr} also displays the mean stellar mass as a function of SFR for these samples \footnote{This plot does not show the scatter error bar, and the jagged features around $10^{-3}$ are expected given the small number of objects in the SFR bin, as indicated in Figure \ref{fig:distributions}.}, revealing a dip in average $M_\ast$ around $\mathrm{SFR}=1$ while higher $M_\ast$ values toward both lower and higher SFRs. 

\begin{figure}
    \centering
    \includegraphics[width=\textwidth]{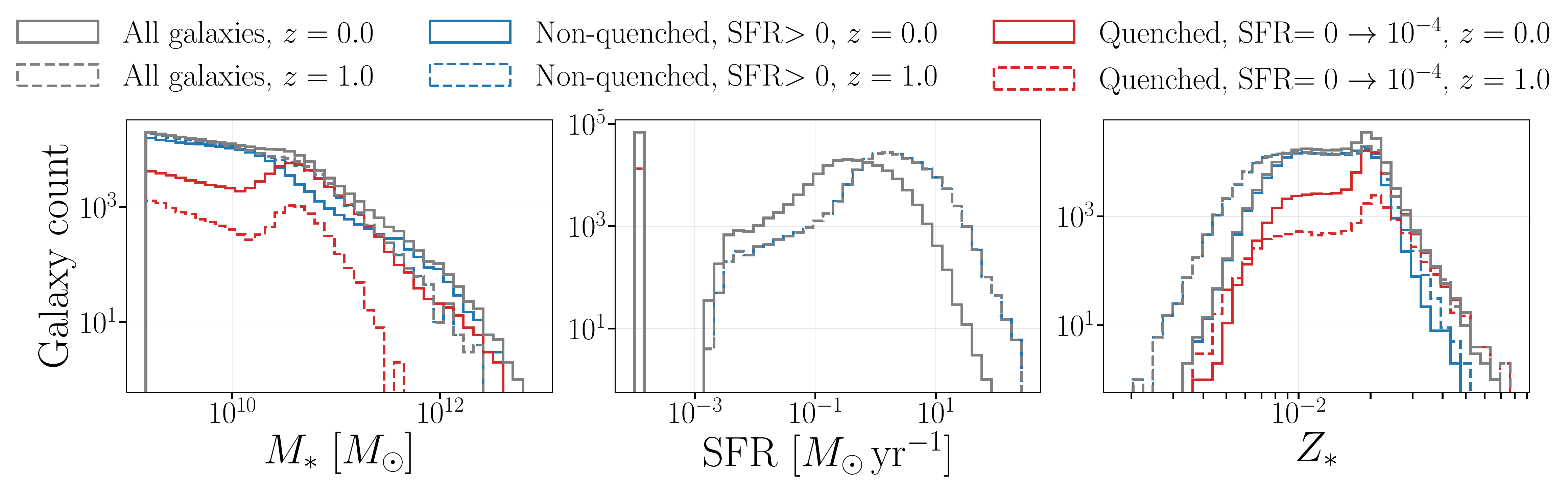}
    \caption{
    Distributions of stellar mass, star formation rate, and metallicity for galaxies at $z=0$ (solid) and $z=1$ (dashed). 
Grey, red, and blue curves show all, non-quenched ($\mathrm{SFR}>0$), and quenched ($\mathrm{SFR}=0$) galaxies, respectively. 
For logarithmic visualization, zero-SFR galaxies are shifted to $\mathrm{SFR}=10^{-4} [\,M_\odot\,\mathrm{yr}^{-1}]$.
    }
    \label{fig:galaxy_distributions_M_SFR_Z}
\end{figure}

Since we retain non-star-forming galaxies throughout our analysis of the gravitational wave (GW) bias, subhalos with $\mathrm{SFR}=0$ are consistently assigned $\mathrm{SFR}=10^{-4}\,M_\odot\,\mathrm{yr}^{-1}$ for numerical stability in logarithmic treatments while preserving their classification as quenched systems. In what follows, we refer to these subhalos in the simulation catalog as non-star-forming or quenched galaxies. We emphasize that this is an operational definition based on simulations; observational classifications of quenched galaxies are often based on an sSFR threshold or color selection. While our samples are largely composed of star-forming galaxies, leading to a strong overlap between the total and non-quenched samples, the galaxies with SFR=0 have a significant impact on the scale-dependent bias (as explored in Section~\ref{sec:gw_bias_vs_astro_properties}), so we highlight their properties and the SFR floor that we use to consistently include these galaxies in the analysis. Figure~\ref{fig:galaxy_distributions_M_SFR_Z} illustrates the one-dimensional distributions, displaying the distribution of these quenched galaxies separately at a few redshifts as well. The comparison between $z=0$ and $z=1$ also highlights the evolution of these properties, particularly the reduced quenched fraction at higher redshift.

\section{The population-synthesis calibrated seeding code, \GalR}
\label{sec:mergerate}
In this work, we adopt merger rate predictions from data provided by the \GalR \footnote{\url{https://zenodo.org/records/7376151}} code \footnote{\url{https://gitlab.com/Filippo.santoliquido/galaxy_rate_open}} \cite{Santoliquido_2022}, where binary compact objects (BCOs) samples from population-synthesis simulations (MOBSE \cite{10.1093/mnras/sty1999}) were seeded into galaxy samples by matching the metallicity of their progenitor stars to that of galaxies using the evolutionary history of galaxies from cosmological galaxy-formation simulations (EAGLE) \cite{10.1093/mnras/stu2058}.  This framework provides a self-consistent prediction for merger rates of binary compact object events (BCOs) per galaxy. 

The \textsc{MOBSE} population-synthesis code \cite{10.1093/mnras/sty1999, Mapelli_2017, Giacobbo_2017} models binary evolution starting from large ensembles of binary stars ($\sim10^7$ for each common envelope model) across a grid of metallicities, yielding a sample of BCO merger events including BBH mergers together with their corresponding delay times and metallicity-dependent formation efficiencies. The \GalR code produces synthetic galaxy populations by first sampling stellar masses from galaxy stellar mass functions \cite{10.1093/mnras/stz2057} and star-formation rates (SFR) using redshift dependent stellar mass and SFR distributions for three main groups of galaxies (main sequence, starburst, and quenched). It then estimates the metallicity from observed scaling relations, either the mass metallicity (MZR) or fundamental metallicity (FMR) relation. The progenitor of the BCO events in MOBSE sample are then populated in the galaxies as new stars based on the SFR and also by matching their metallicities to that of the formation galaxy (FG). The connection between FG where binaries originate and their descendant host galaxy (HG) where mergers occur is then established via conditional probability based on the merger trees of \textsc{EAGLE} cosmological simulation with a $(100 cMpc)^3$ volume \cite{10.1093/mnras/stw2437, 10.1093/mnras/stu2058}, which tracks the evolution of galaxies in stellar mass, star-formation rate, and metallicity across cosmic time.

The \GalR public data release provides plain-text tables containing both formation-galaxy (FG) and host-galaxy (HG) information for a range of population-synthesis and galaxy-evolution models.
Each dataset is labeled according to the adopted metallicity relation (\texttt{M57} → MZR; \texttt{M54} → FMR), the common-envelope efficiency parameter (\texttt{A1}, \texttt{A3}, \texttt{A5} corresponding to $\alpha_{\rm CE}=1,3,5$), and the binary type (\texttt{BBHs}, \texttt{BHNS}, or \texttt{BNSs}).
Files are provided for multiple snapshots ($s\in{ \{28,\dots,19\}}$), each mapping one-to-one to a redshift $z$ through the \textsc{EAGLE} snapshot-redshift relation (see Table~\ref{tableeagle}).
For each galaxy, the \GalR tables report the stellar mass $M_\star$ [$M_\odot$], star-formation rate ${\rm SFR}$ [$M_\odot{\rm yr}^{-1}$], logarithmic metallicity $\log_{10}Z$, per-galaxy merger rate $n_{\rm GW}$ [Gyr$^{-1}$], and a star-forming flag (\texttt{Type}=1 for star-forming, 0 for passive).

\begin{table}[h]
\centering
\begin{tabular}{c c c c c c c c c c c}
\hline
$z$ & $0.0$ & $0.1006$ & $0.1827$ & $0.2709$ & $0.3657$ & $0.5031$ & $0.6152$ & $0.7356$ & $0.8651$ & $1.0041$ \\
\hline
\texttt{snapshot} & 28 & 27 & 26 & 25 & 24 & 23 & 22 & 21 & 20 & 19 \\
\hline
\end{tabular}
\caption{Snapshot-redshift conversion for EAGLE.}
\label{tableeagle}
\end{table}

\begin{figure}[h]
    \centering
    \includegraphics[width=\textwidth]{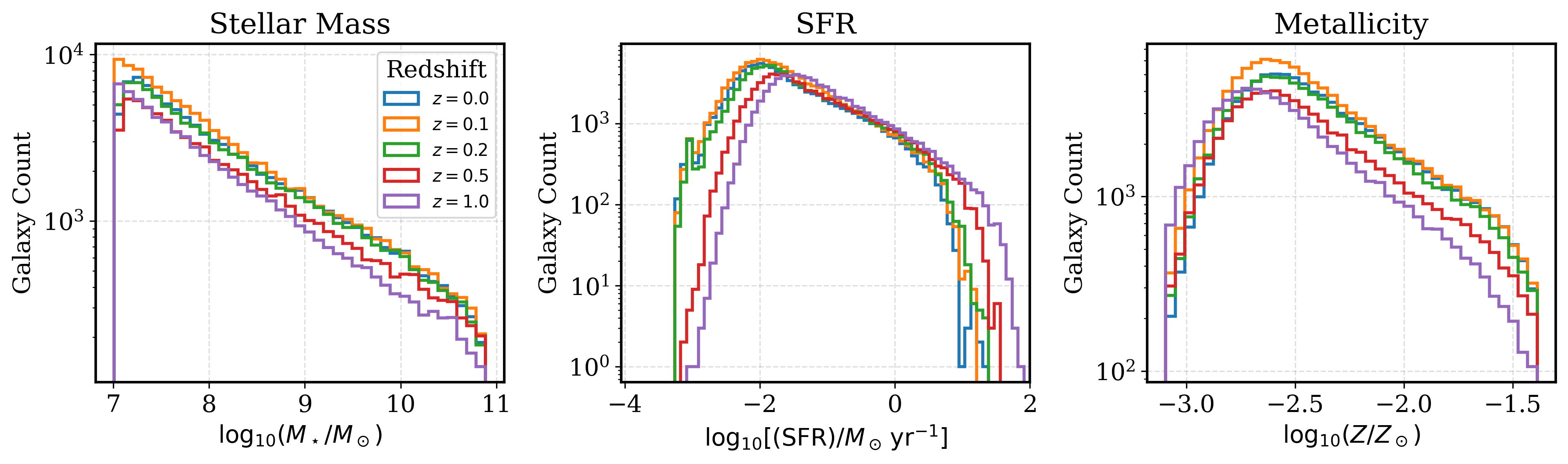}
    \caption{
        Distributions of stellar mass, star-formation rate, and metallicity of galaxy samples for \GalR data sample for FMR(M54)-$\alpha_{CE}=5$ model.}
    \label{fig:zenodo_dists}
\end{figure}

In our analysis, we will focus on the FMR(M54)-$\alpha_5$ model; however, it can be easily extended to other models as well. Figure \ref{fig:zenodo_dists} shows the stellar mass, SFR and metallicity distributions of the galaxy sample in FMR(M54)-$\alpha_5$ model used by \GalR for seeding at redshifts of $\{z = 0, 0.1, 0.2, 0.5, 1\}$.

\begin{figure}[ht]
    \centering
    \includegraphics[width=0.65\textwidth]{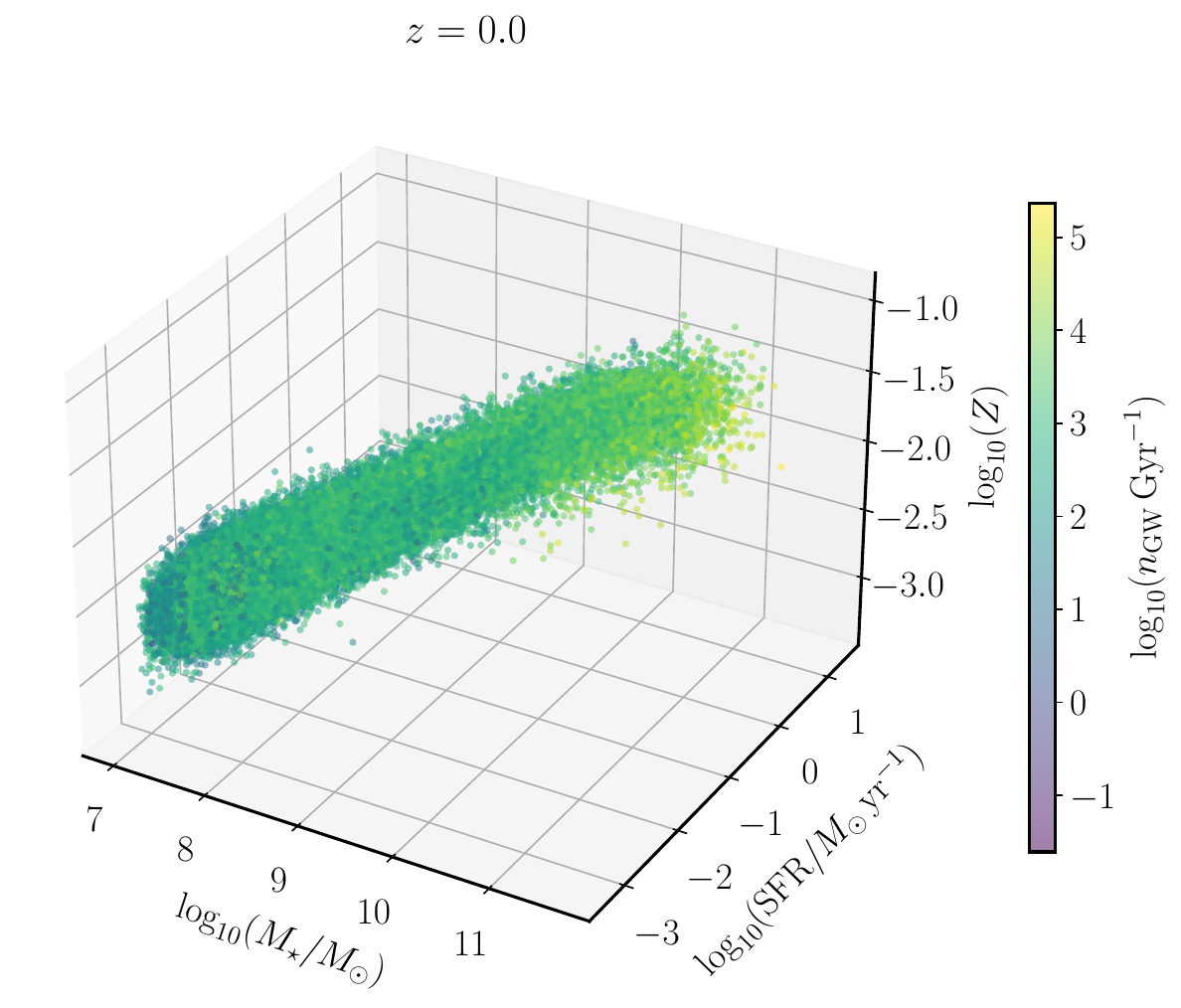}

    \caption{\textbf{Three-dimensional distribution of BBH merger rates in the FMR model.}
    The figure shows all host galaxies at $z=0.0$ for the FMR (M54) model with $\alpha=5$.
    The three axes correspond to stellar mass,
    $\log_{10}(M_\star/M_\odot)$,
    star formation rate,
    $\log_{10}(\mathrm{SFR}/M_\odot\,\mathrm{yr}^{-1})$,
    and metallicity,
    $\log_{10}(Z)$.
    Each point represents one galaxy, while the color indicates the predicted merger rate,
    $\log_{10}(n_{\mathrm{GW}}/\mathrm{Gyr}^{-1})$.
    }

    \label{fig:3d_fmr_z0}
\end{figure}

\begin{figure*}[ht]
\centering

\includegraphics[width=0.32\textwidth]{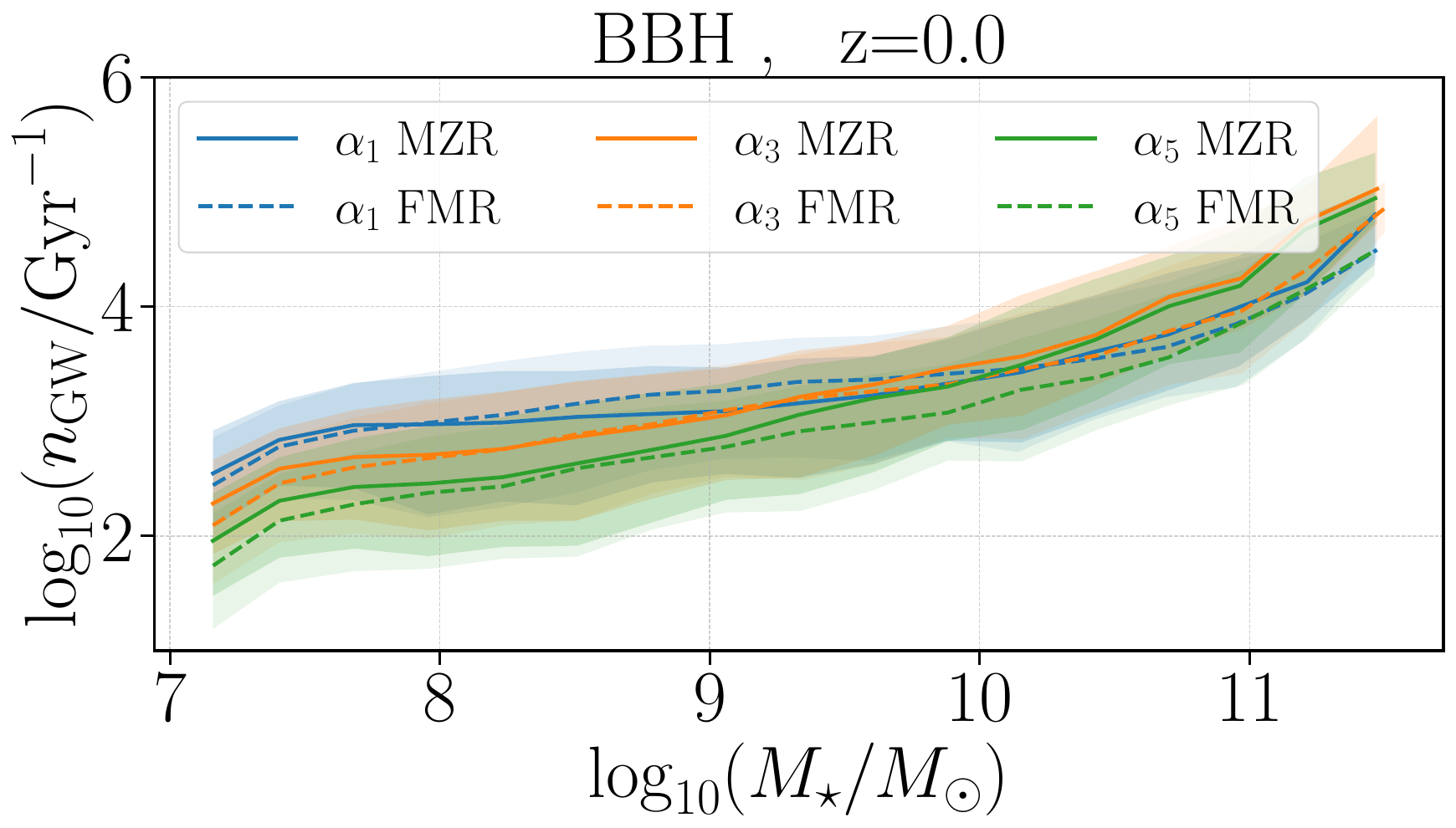}
\hfill
\includegraphics[width=0.32\textwidth]{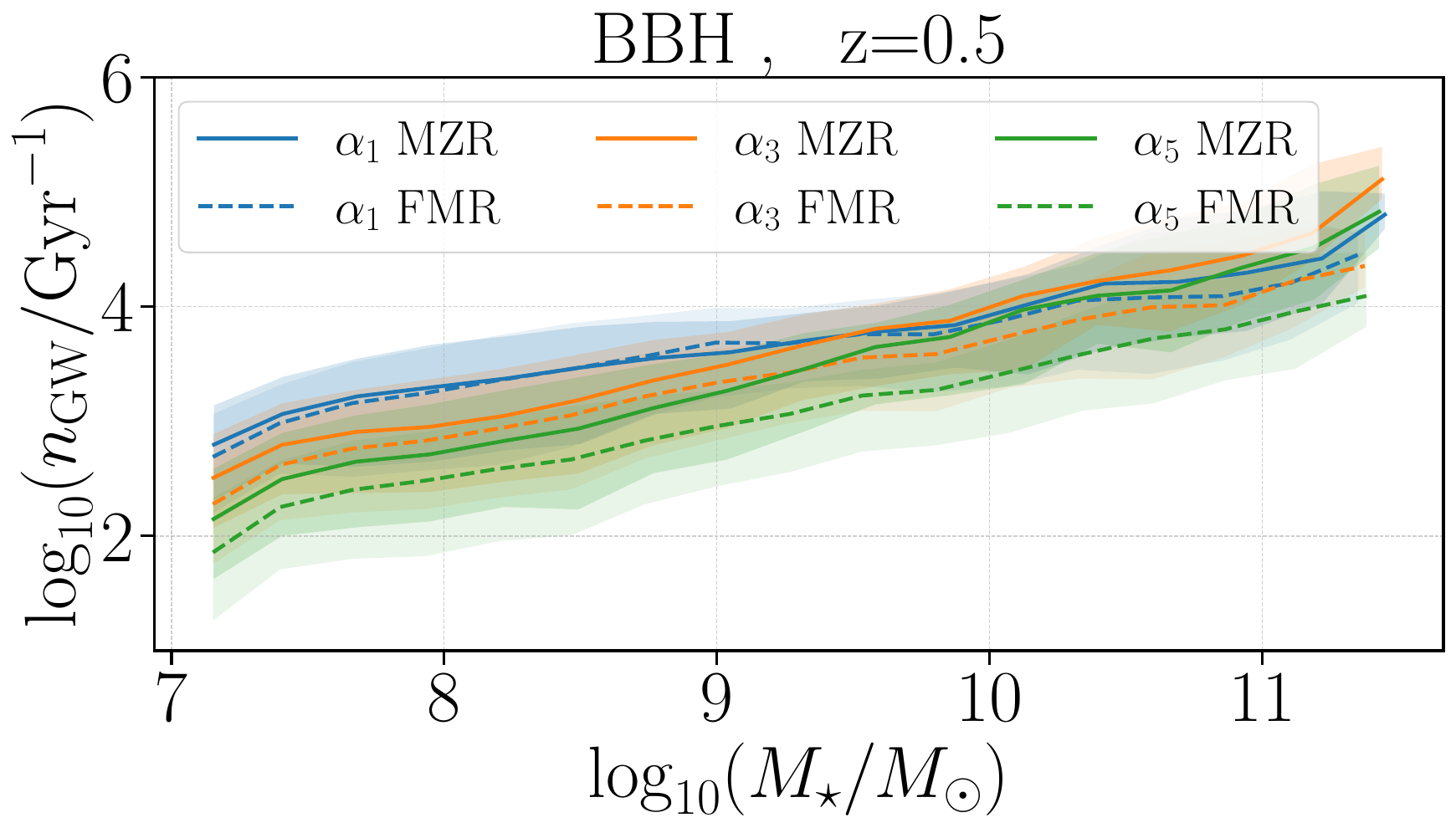}
\hfill
\includegraphics[width=0.32\textwidth]{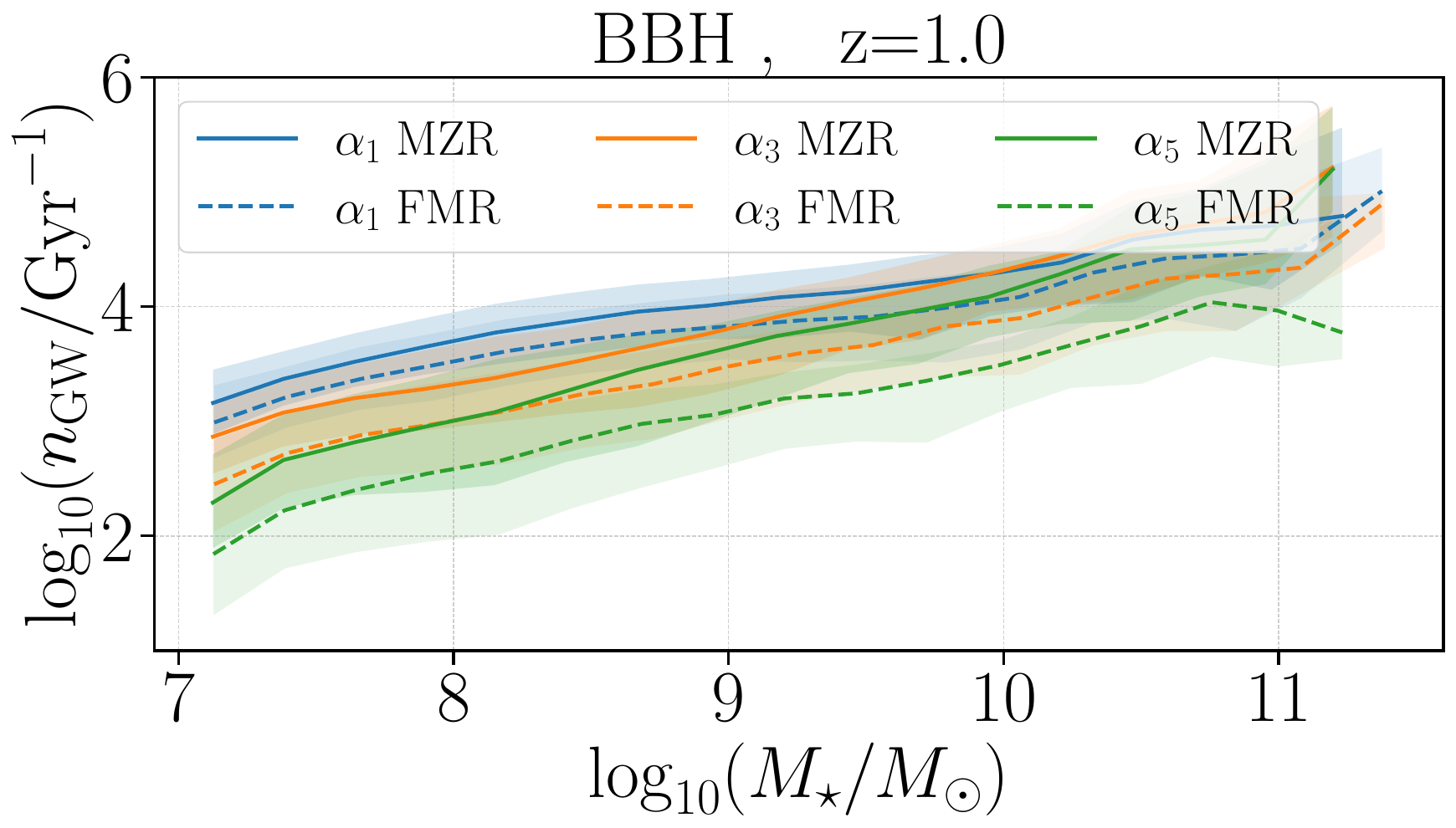}

\vspace{0.8em}

\includegraphics[width=0.32\textwidth]{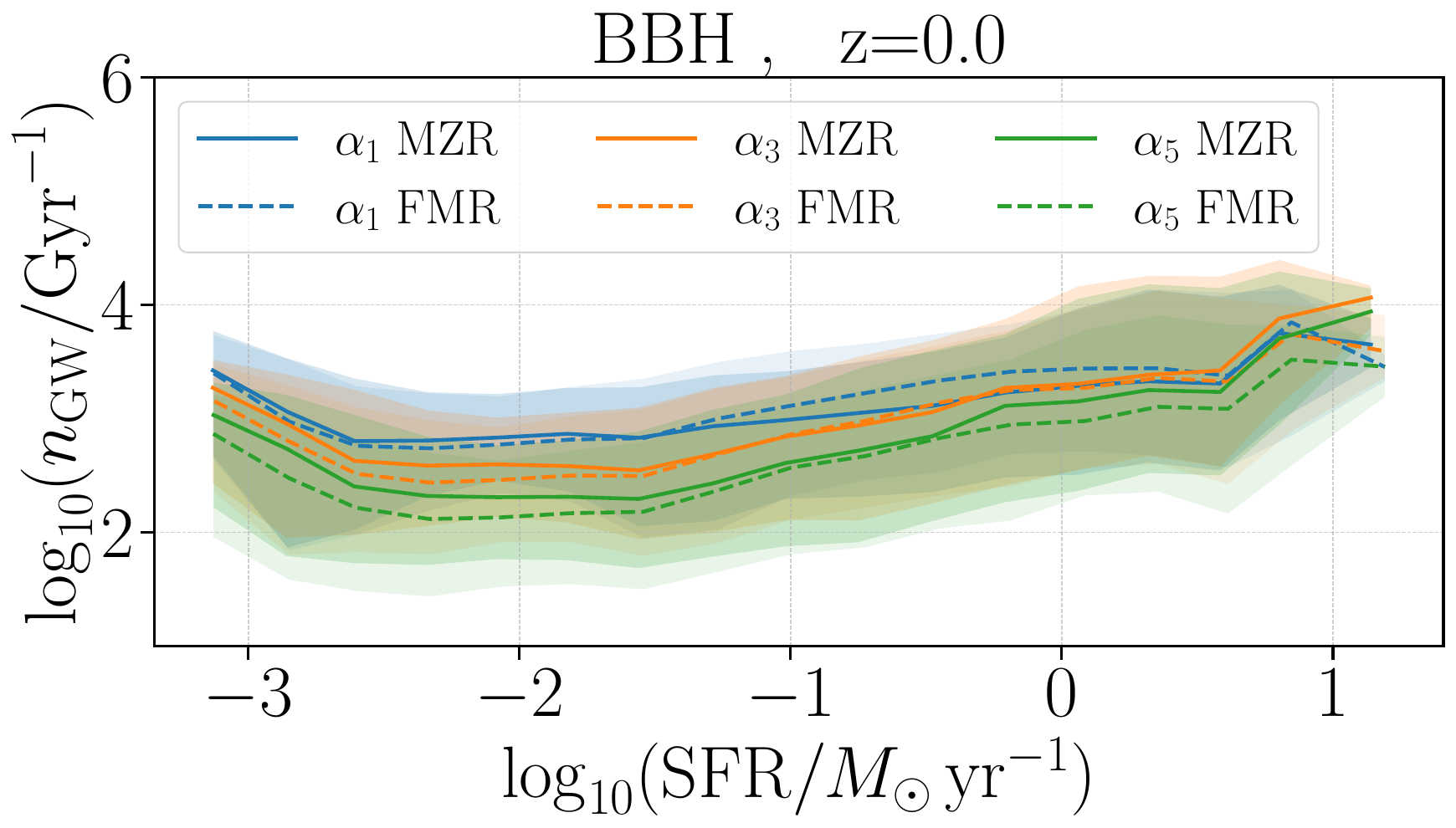}
\hfill
\includegraphics[width=0.32\textwidth]{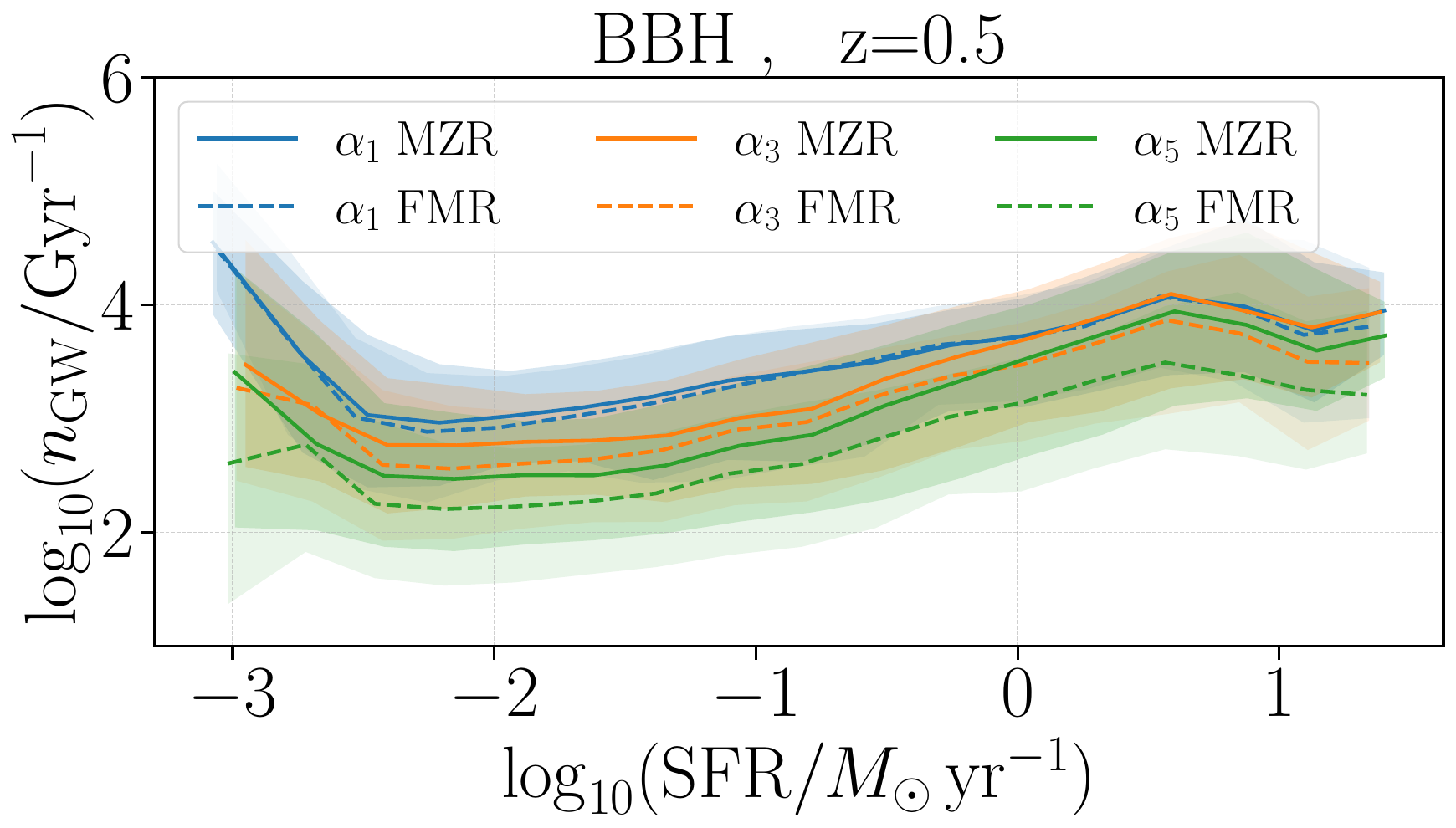}
\hfill
\includegraphics[width=0.32\textwidth]{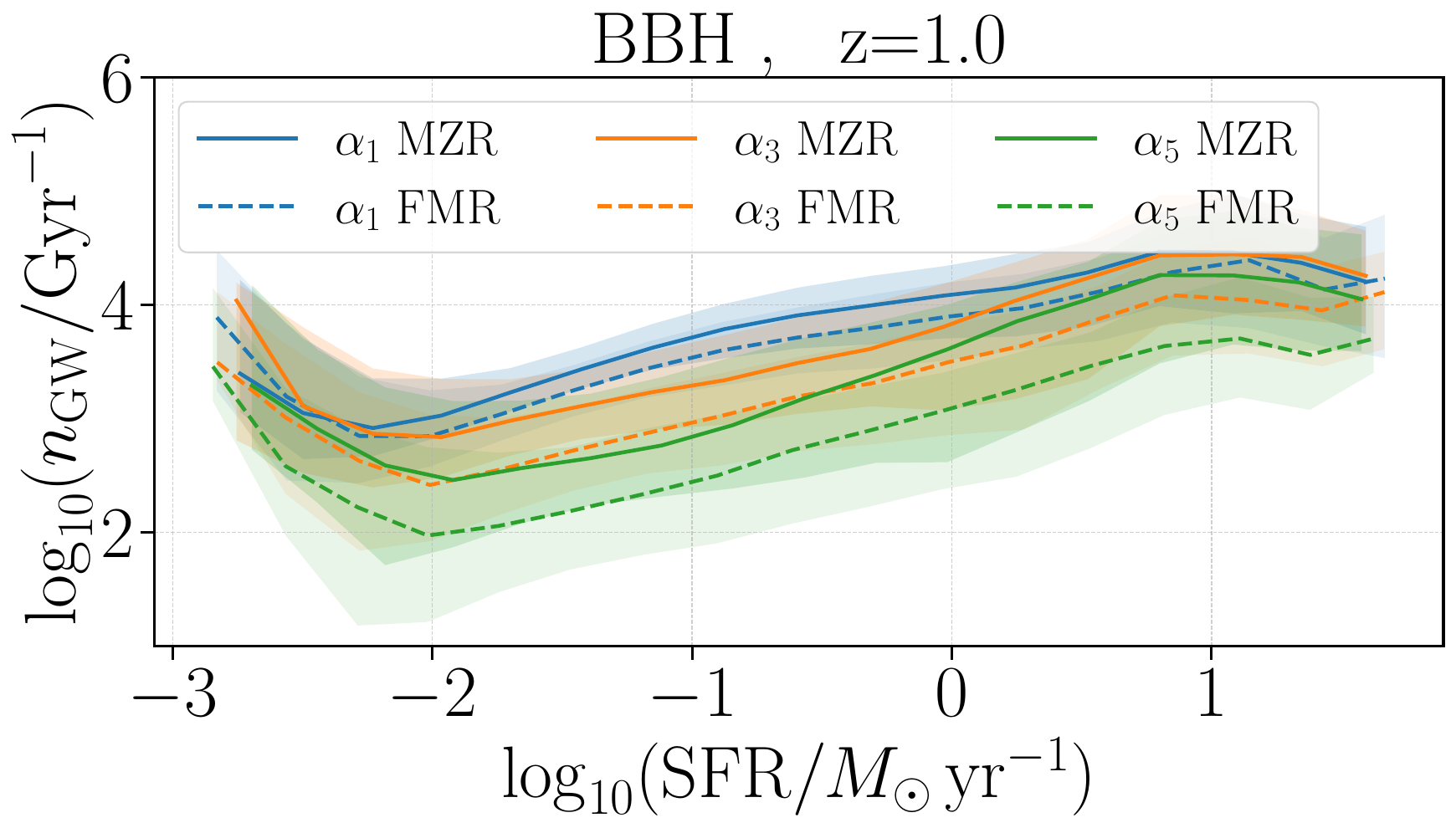}

\vspace{0.8em}

\includegraphics[width=0.32\textwidth]{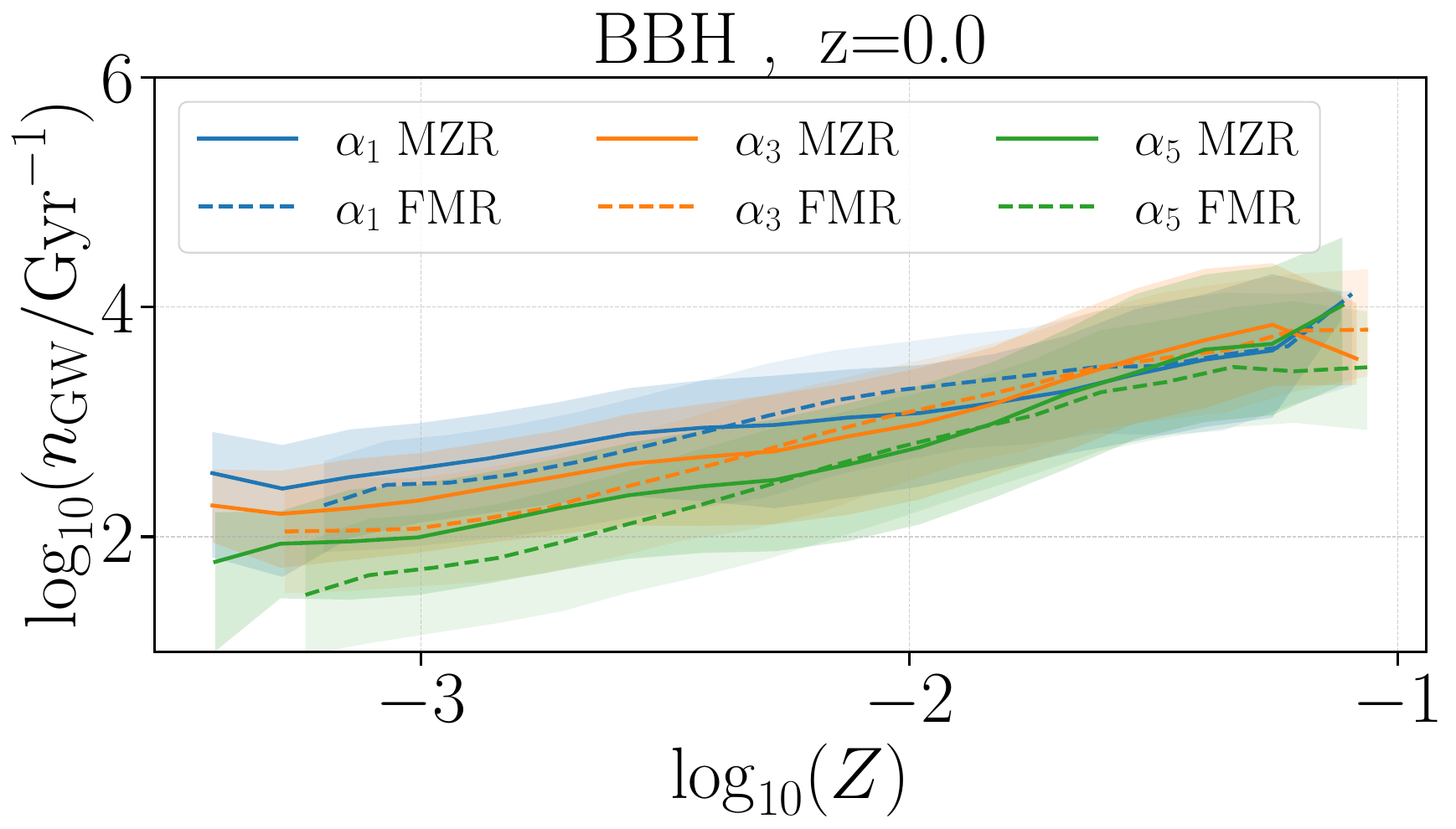}
\hfill
\includegraphics[width=0.32\textwidth]{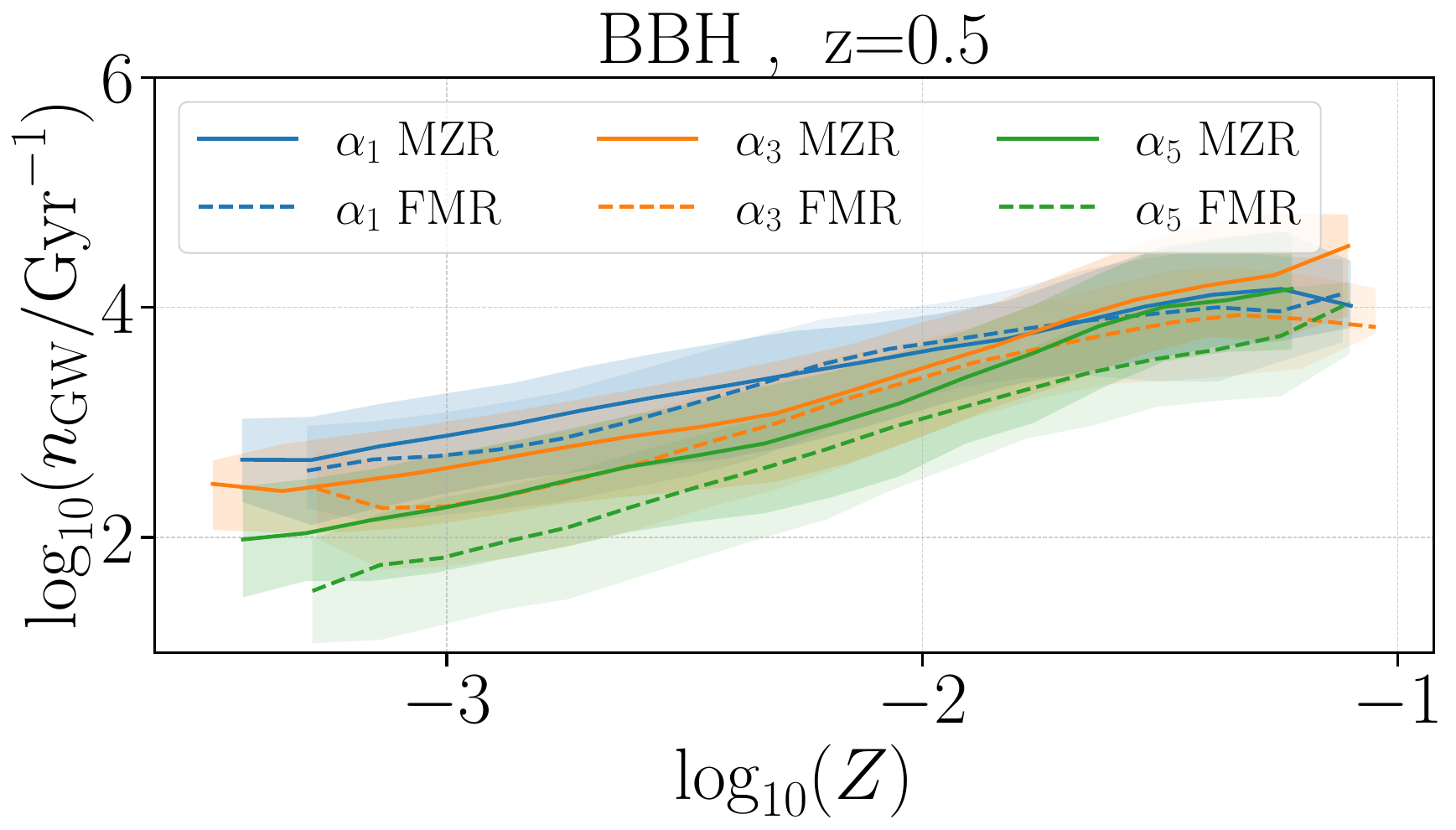}
\hfill
\includegraphics[width=0.32\textwidth]{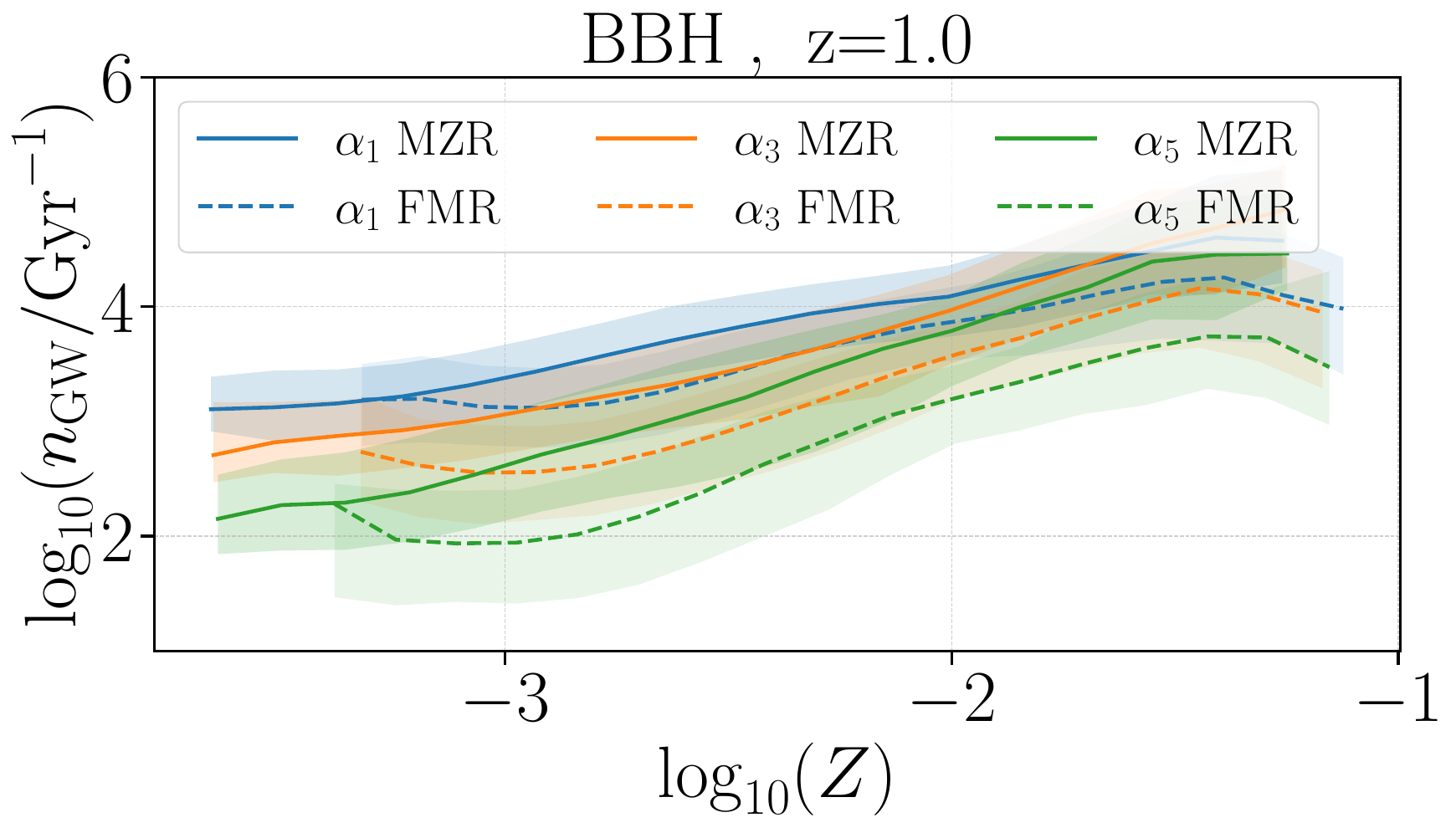}

\caption{\textbf{Dependence of the BBH merger rate on host-galaxy properties in the \GalR data.}
Rows show the median \(\log_{10} n_{\mathrm{GW}}\) as a function of host-galaxy stellar mass \(M_\star\), star formation rate SFR, and metallicity \(Z\), respectively.
Columns show redshifts \(z=0.0\), \(z=0.5\), and \(z=1.0\).
For each common-envelope efficiency parameter \(\alpha\), the MZR model (M57) is shown with solid lines and the FMR model (M54) with dashed lines in the same color.
Shaded bands indicate the 16--84\% percentile range.}
\label{fig:host-property-alpha-sweep}
\end{figure*}

As an illustrative example, Figure~\ref{fig:3d_fmr_z0} shows the merger rate-per-galaxy distribution, $n_{\rm GW}$, at $z=0$ across the stellar mass-SFR-metallicity space at the time of merger in the \GalR data files. However, this three-dimensional visualization is difficult to interpret by eye. Therefore, in Figure~\ref{fig:host-property-alpha-sweep}, we present one-dimensional projections of $n_{\rm GW}$, including other models, to intuitively understand how they correlate with the astrophysical properties of the host galaxies. For these plots, host galaxies are either grouped in $\log_{10}(M_\star)$, or $\log_{10}\mathrm{SFR}$, or $\log_{10}Z$, and for each group, the median $\log_{10} n_{\rm GW}$ is shown together with the 16--84\% percentile range. Each color corresponds to a choice of the common-envelope efficiency parameter $\alpha_{\rm CE}$; for every $\alpha_{\rm CE}$, the MZR model (M57) is plotted as a solid line, and the FMR model (M54) as a dashed line in the same color, with shaded regions indicating the intrinsic population scatter at fixed galaxy properties.

Across all redshifts and prescriptions, the merger rate exhibits the following qualitative relationships with the properties of the host galaxies at the time of merger: 

\textbf{(i) Stellar mass}: As shown in the first row of Figure \ref{fig:host-property-alpha-sweep}, the merger rate exhibits a strong monotonic increase with $\log_{10}(M_\star)$ for all models, except at $z=0$ for FMR(M54)-$\alpha_5$ where it exhibits a modest decrease at the very highest masses. Note that \GalR seeds the progenitors of BCOs into star forming galaxies at every time step using SFR. Because more massive galaxies have formed more stars over their history, they end up with more BCO progenitors. So if the delay times between formation and merger are not all very short and narrowly distributed, this explains the larger number of potential merger events in galaxies with higher stellar mass. At fixed redshift, galaxies with $M_\star \gtrsim 10^{10}\,M_\odot$ exhibit merger rates elevated by several orders of magnitude relative to lower mass galaxies.

\textbf{(ii) Star formation rate}: The second row of Figure \ref{fig:host-property-alpha-sweep} shows how the BBH merger rate varies with the host galaxy star formation rate (SFR) in the redshift range $1-0$. For this redshift range, the merger rate appears to exhibit a weaker correlation with SFR compared to stellar mass. However, at higher redshifts, there is an SFR range where we can see a consistent monotonic rise in the merger rate with SFR. This may indicate that, at early times, merger events have a higher percentage of those with short delay times, occurring in more actively star-forming galaxies, whereas, at later times, the merger population accumulates from those composed of long-delay events taking place in older galaxies.

\textbf{(iii) Metallicity}: The third row of Figure \ref{fig:host-property-alpha-sweep} shows how the BBH merger rate varies with host galaxy metallicity for different $\alpha$-models in the \GalR framework. The merger rate in most models also increases with metallicity over the range where most of the galaxies lie ($\log Z \sim -3$ to $\sim -1.5$). All $\alpha$-prescriptions exhibit a similar rising trend, followed by a gradual flattening toward higher metallicity. 

An important point to keep in mind about these figures is that the dependence of merger rates on galaxy properties is driven by both binary evolution and galaxy evolution, and these properties are not independent of one another. Consequently, the merger rates also inherit the correlations and assumed scaling relations encoded in the MZR and FMR prescriptions. For example, the increase in merger rate with stellar mass is not independent of its dependence on SFR or metallicity, as illustrated in Figure \ref{fig:3d_fmr_z0}. This motivates the development of seeding frameworks that can fully capture these multidimensional joint probabilities.

\section{Calibrated machine learning approach for generating a mock catalog of gravitational wave sources}
\label{sec:mock_gw}
In this section, we present our first approach for seeding IllustrisTNG300 and constructing the mock gravitational wave (GW) source catalog. In this approach, we first train a machine-learning (ML) emulator on the \GalR dataset, which links host-galaxy properties (stellar mass, star-formation rate, and metallicity) to their expected BBH merger rates. The trained emulator is then applied to all galaxies in the IllustrisTNG300 simulation to predict the BBH merger rate for each galaxy. This means that in this approach, the overall cosmological merger statistics, as well as the \GalR population-synthesis calibration, are encoded in the merger rate assignments. Finally, we generate the GW source catalog by probabilistically sampling merger events from these galaxies based on their predicted merger rates.
The complete end to end workflow is summarized in what follows, describing each stage of the pipeline in detail, from training the merger rate emulator to generating the final cosmologically normalized mock GW source catalogs.

\subsection{Training a Machine-Learning Emulator on the \GalR Dataset for Merger Rate prediction}
\label{subsec:ML_method}
In this work, the \GalR catalogs serve as the supervised training data to learn the empirical mapping from host-galaxy properties to BBH merger rates. We work in logarithmic space and define
\[
  \mathbf{x} \equiv \bigl(\log_{10} M_\star,\;\log_{10}{\rm SFR},\;\log_{10} Z\bigr),
  \qquad
  y \equiv \log_{10} (n_{\rm GW}\;/{\rm Gyr}^{-1})\, .
\]

The curated \GalR pairs $(\mathbf{x},y)$ serve as training data for the emulator introduced next. For each redshift snapshot, we train a separate model and later apply it to the IllustrisTNG300 galaxy catalogs at snapshots  $\{99,\, 91,\, 85,\, 67,\, 50\}$ corresponding to $z=\{0.0,\,0.1,\,0.2,\,0.5,\,1.0\}$. The emulator produces per-galaxy merger rate predictions that we store alongside the catalogs and use to build the siren samples for downstream analyses.

To select a final ML emulator, we performed a controlled comparison between two supervised machine-learning models commonly used for nonlinear regression: a fully connected neural network (NN) and a gradient boosting machine (GBM). Both models were trained on identical input features, redshift snapshots, and train--test splits, and were individually optimized via hyperparameter searches to ensure a fair comparison. Across all redshifts considered, the GBM consistently achieved lower test-set mean-squared error, higher $R^2$, and more faithful reproduction of the underlying population trends as functions of stellar mass, star formation rate, and metallicity. A comprehensive quantitative and qualitative comparison of the two models is presented in Appendix \ref{sec:AppendixA}. Guided by the results of this comparison, we adopted a Gradient Boosting Regressor (GBR) implemented in the \textsc{scikit-learn} library as the fiducial regression model in our emulator. The GBR predicts the logarithmic merger rate, $\log_{10}(n_{\rm GW}/{\rm Gyr}^{-1})$, from the host-galaxy properties $(\log_{10}M_\star,\, \log_{10}\mathrm{SFR},\, \log_{10}Z)$. To capture nonlinear interactions among these physical variables while avoiding excessive model complexity, we expanded the input space using second-order interaction-only polynomial features. All features are standardized using a \texttt{StandardScaler} fitted exclusively on the training subset to prevent information leakage.

For each redshift snapshot, we performed a 90/10 train--test split on the polynomial feature matrix. 
Hyperparameters were selected via a 5-fold cross-validation search conducted exclusively on the training subset to avoid information leakage. The explored hyperparameter grid included the number of boosting stages, learning rate, maximum tree depth, subsample fraction, and the minimum number of samples required in leaf and split nodes.

For each combination of parameters, we recorded the mean cross-validation $R^{2}$ and the mean-squared error (MSE) where $R^{2}=1$ corresponds to perfect prediction, $R^{2}<0$ indicates performance worse than predicting the sample mean. The configuration that maximizes the mean $R^{2}$ was adopted as the optimal setting for that snapshot, and MSE was used as a secondary tie-breaker. Using the best hyperparameters for a given redshift (see table \ref{tab:gbm_hyperparams_M54A5}), we then refitted a final \texttt{StandardScaler} on the full training split, transformed both the training and test data, and trained a final GBR model on the scaled training set only. Performance was evaluated on the held-out test set using  $R^{2}$ and MSE. All other GBR parameters adopted their \textsc{scikit-learn} defaults (with the loss function being squared error).
The polynomial transformer, the final training-set scaler, the best-fit GBR model, and the cross-validation summary table were saved separately for each redshift snapshot for subsequent inference.

\subsection{ Implementing the merger rate predictions for galaxies in IllustrisTNG300 simulation}
\label{subsec:merger_prediction_illustris}
After training the host to merger rate emulators on the \GalR dataset, we apply them to all galaxies in the IllustrisTNG300 simulation to estimate per-galaxy BBH merger rates at the corresponding redshift snapshots.  
For each TNG subhalo, we extract stellar mass, star-formation rate, and stellar metallicity, and convert them to the physical units and logarithmic format required by the emulator:
\[
\log M_\ast=\log\!\left(M_{\ast,\mathrm{TNG}}\right),\qquad
\log{\rm SFR}=\log({\rm SFR}+10^{-4}),\qquad
\log Z=\log Z.
\]
The small offset $10^{-4}$ is used to prevent non-star-forming galaxies in the catalog from causing numerical singularities. Each subhalo with feature vector
\[
\mathbf{x}_i=\big(\log M_{\ast,i},\,\log{\rm SFR}_i,\,\log Z_i\big)
\]
will be assigned a predicted logarithmic merger rate using the trained GBM:
\[
\widehat{y}_i=\log \widehat{n}_i,
\]
where $\widehat{n}_i$ is the expected number of BBH mergers per galaxy (per Gyr), inferred based on the learned mapping between host properties and population-synthesis rates.

This procedure effectively transfers the physical calibration of the \GalR seeding to the cosmological galaxy population in IllustrisTNG300, automatically incorporating the effects of galaxy evolution, chemical enrichment, and star-formation history into the inferred BBH merger rates based on the host properties.
The output of this step is the redshift-resolved catalogs of TNG galaxies with the three spatial coordinates and astrophysical properties ($M_{\ast,i},\,{\rm SFR}_i,\,Z_i$), along with the assigned merger rate predictions ($\widehat{n}_i$) for each specific astrophysical model. In other words, merger rate predictions are obtained and implemented on TNG on a per-galaxy basis from the three-dimensional multi variable host-galaxy model that simultaneously incorporates stellar mass, star-formation rate, and metallicity. These enriched galaxy datasets serve as the basis for constructing the Poisson-sampled GW-siren catalogs and for subsequent clustering analysis. 

\begin{figure}[h]
    \centering
        \includegraphics[width=0.4\linewidth]{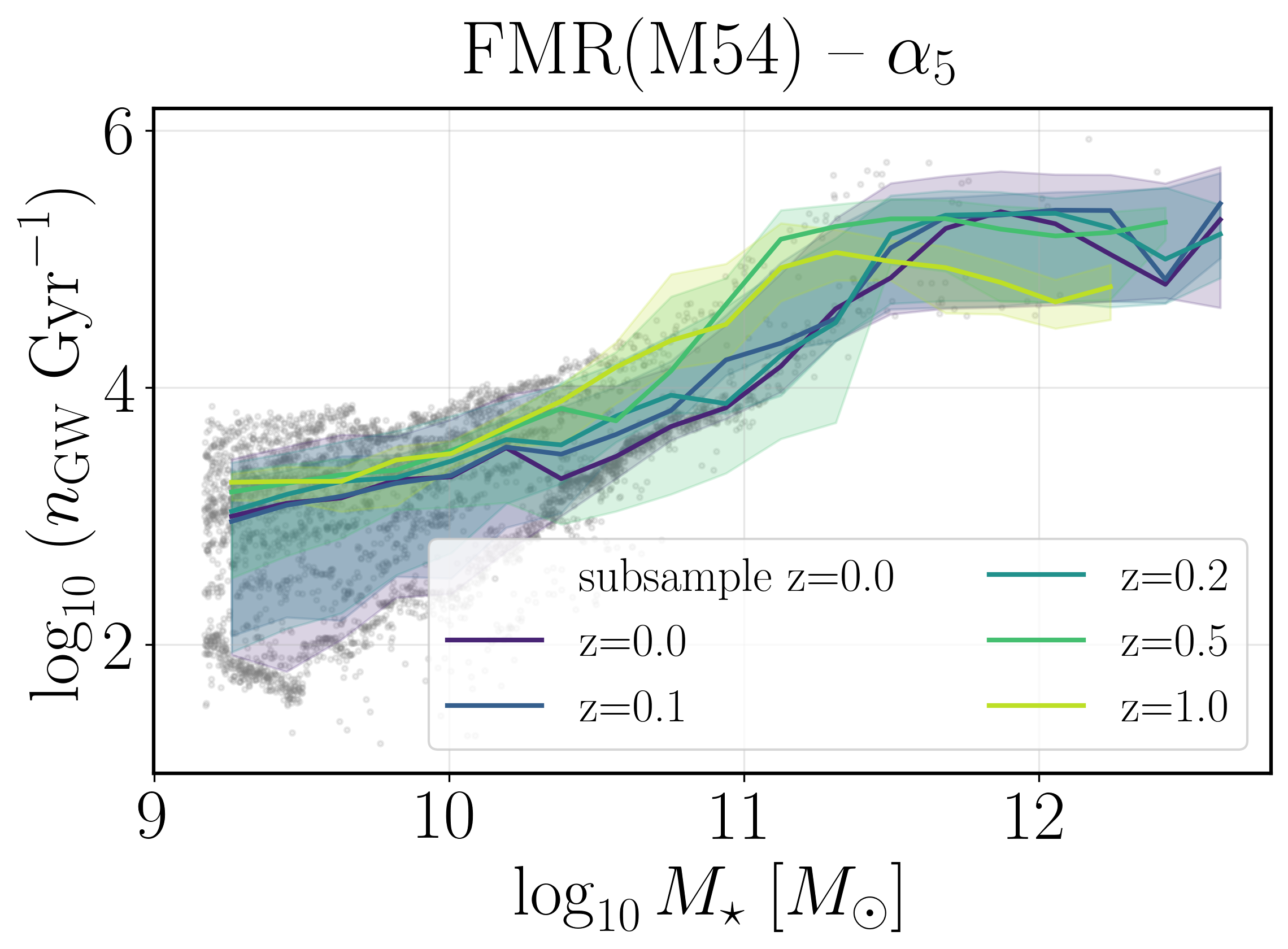} 
        \includegraphics[width=0.4\linewidth]{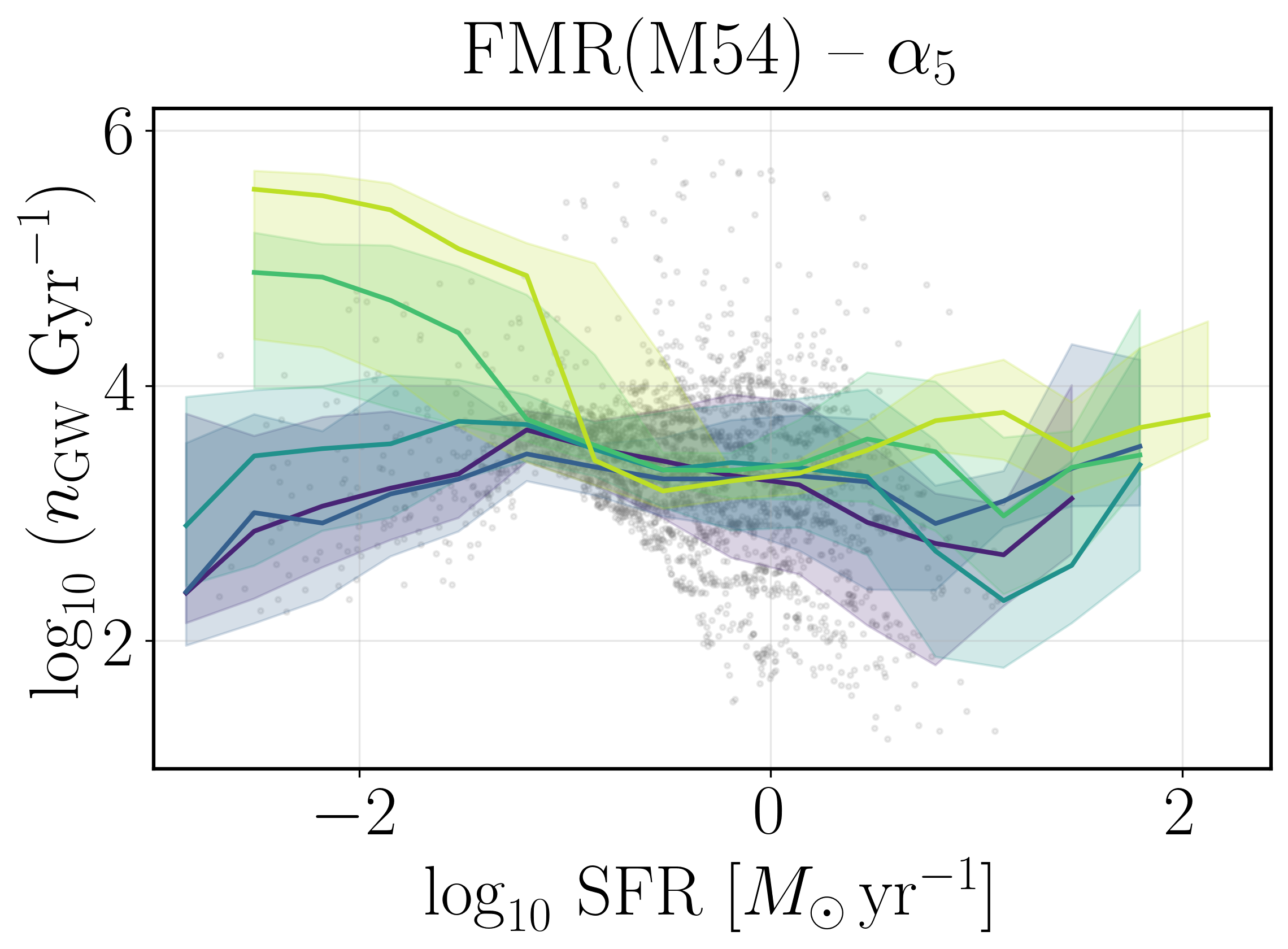}
        \includegraphics[width=0.4\linewidth]{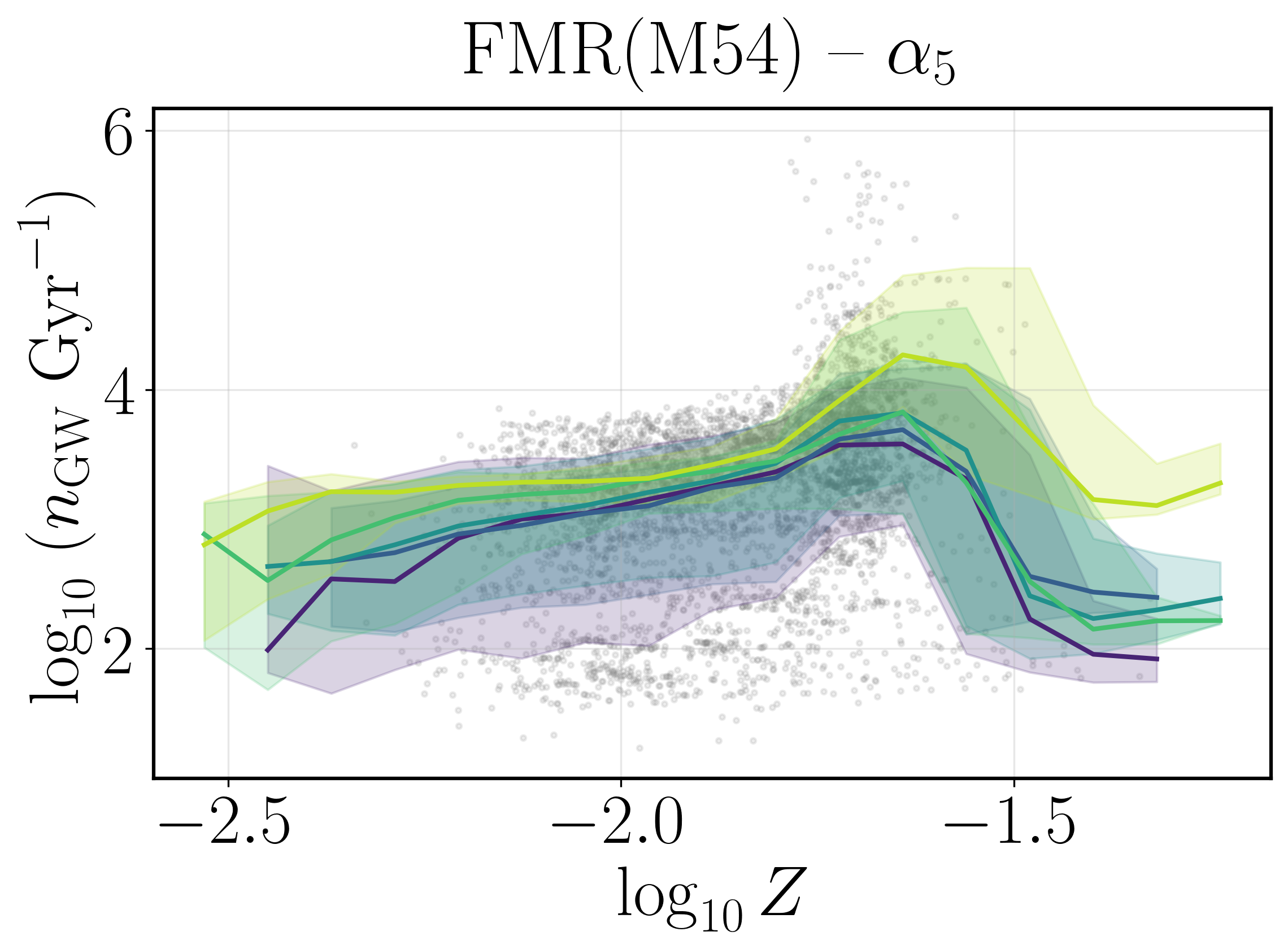}
    \caption{Distribution of the predicted BBH merger rate versus host galaxy properties in IllustrisTNG300 across redshift bins, $z\in \{0.0,\,0.1,\,0.2,\,0.5,\,1.0\}$. The blue dots are the distribution for $z=0$, solid curves show the fit to the median value of the $\log_{10}(n_{\rm GW}/{\rm Gyr}^{-1})$ within each bin and the shaded bands are $16$--$84$ percentile scatter for (a) stellar mass, (b) star-formation rate, and (c) stellar metallicity. These plots correspond specifically to the $\alpha_{CE} = 5$ model using the FMR (M54) metallicity prescription.}
    \label{fig:mergerrate_vs_galprops}
\end{figure}

Figure~\ref{fig:mergerrate_vs_galprops} shows the distribution of the implemented BBH merger rates for the FMR (M54)–$\alpha_{5}$ model, on the galaxies in the TNG galaxy sample at $z = 0$, as well as the median of the logarithm of merger rate at $z=0$ and several other redshifts, plotted against various TNG host-galaxy properties. The plots are constructed by binning galaxies along a single host-galaxy property. Solid curves show a fit to the median value of $\log_{10}(n_{\rm GW}/{\rm Gyr}^{-1})$ within each bin, with shaded regions indicating the 16th--84th percentile range. We would like to emphasize that these plots capture population-level behavior of merger rate distribution rather than conditional probabilities at fixed values of the other host-galaxy properties. We can see from these projected plots of the implemented BBH merger rate distributions in IllustrisTNG300, the prominent features observed in Figure~\ref{fig:host-property-alpha-sweep} are preserved within the scatter. In particular, there is a strong, nearly monotonic increase with stellar mass across all redshifts, while the dependences on star-formation rate and metallicity are comparatively weaker, showing milder variation over the explored redshift range. 
There are some differences, but these are expected, as the full statistical distributions of $(M_{\ast,i},\,\mathrm{SFR}_i,\,Z_i)$ for the IllustrisTNG300 galaxy sample and the \GalR sample are not identical, and the merger rate is an intrinsically three-dimensional function. Therefore, once marginalized, both the scatter and the apparent shape of the distributions naturally differ. The shaded percentile bands quantify the intrinsic scatter in the predicted merger rates, reflecting both stochasticity in the \GalR training sets and the diversity of the other two galaxy properties at fixed stellar mass/star-formation rate/metallicity in the IllustrisTNG300 population.

Predicted merger rate trends for additional population synthesis and host-galaxy models at different redshifts are presented in Appendix \ref{sec:mergerrate_other models} (Figure ~\ref{fig:three_panel_merger_rates}). As we see, the changes in the model can have an impact on merger rates.

\subsection{Mock GW Source Catalog Generation from merger rate Modeling}

Next, we populate the galaxy samples with mock GW merger events according to the predicted per-galaxy merger rates obtained from the prescription described above. In each redshift snapshot of IllustrisTNG300, GW source catalogs are constructed by Poisson sampling host galaxies according to their ML-predicted BBH merger rates. These mocks are designed to trace the spatial distribution of the underlying BBH merger population and enable a direct evaluation of the corresponding clustering bias for GW sources. In this framework, galaxies with larger predicted merger rates contribute with higher probability to the GW catalog, while the three-dimensional positions of the selected hosts are inherited from the simulation. The resulting catalogs, therefore, preserve the connection between compact-binary formation and host-galaxy properties, as well as the large scale structure of the simulated volume.

Note that in our framework, the total number of mock GW events assigned to each simulation snapshot should not be interpreted as the number of events that a specific detector would observe in a given redshift shell over a fixed observing time for a specific experiment. First, the total astrophysical merger events in a comoving volume are distinct from the number of detectable events for a specific experiment during an observing run, which depends on detector sensitivity, source properties, selection effects, and the redshift interval considered. 
Second, in \GalR the total merger rate for a snapshot is automatically fixed to the sum of the individual galaxies' merger rates predicted by \GalR from population-synthesis-calibrated models, rather than being observationally normalized to reproduce the local LVK merger rate density. Furthermore, since our goal is not to forecast measurements based on detectable events in LVK, ET, or CE, but to construct high-statistics mock catalogs that robustly sample the merger-rate-weighted host distribution and allow stable measurements of the theoretical GW bias based on population-synthesis models, we allow for large sampling times. That is, in practice, we set the number of sampled events large enough that the GW power spectrum and bias are not dominated by Poisson noise, while still providing a controlled statistical representation of the underlying rate-weighted host population. Depending on the total predicted rate in a snapshot and the redshift, this choice could correspond to an effective normalization time of order $10^2$--$10^4,{\rm yr}$, and should therefore not be regarded as a literal observing duration.

Concretely, we amplify the expected number of target mock events in a given snapshot so that it is not dominated by noise 
\begin{equation}\label{eq:N_exp}
    N_{\rm exp}(z)= A \sum_j \hat n_j.
\end{equation}
We then define a per-galaxy Poisson mean ($\mu_i$) proportional to the ML-predicted merger rate ($\hat n_i$).
\begin{equation}\label{eq:N_exp_normalization}
\mu_i\equiv A \,\hat{n}_i=
N_{\rm exp}(z) 
\frac{\hat n_i}{\sum_j \hat n_j},
\qquad
\sum_i \mu_i = N_{\rm exp}(z).
\end{equation}
The number of BBH merger events assigned to galaxy (i) is then drawn as
\[
n_i \sim \mathrm{Poisson}(\mu_i).
\]
Since this is a Poisson realization of galaxy-specific merger rates, hosts are effectively sampled with replacement, and a single galaxy may contribute more than one mock merger event. Each event inherits the comoving three-dimensional position of its host galaxy, given by \texttt{SubhaloPos}, and the resulting collection of events defines the mock GW source catalog at that redshift. As mentioned, the number of events and, consequently, the sampling time is selected such that the relative rate weighting, overdensity field, and clustering amplitude are preserved while the catalog remains well sampled. To be specific, in each snapshot, we draw approximately $1.8\times10^4$ mock GW events, corresponding to a modest fraction of order $\sim 20\%$ repeated hosts depending on the redshift. We tested varying this number for some of our models and found that the inferred GW bias on large scales is very stable, and on small scales varies at the $\lesssim10\%$ level. Thus, the mock-event number should be understood as a statistical sampling choice for estimating the theoretical GW bias, rather than as a detector-specific event count. With these weighted three-dimensional catalogs in hand, we can proceed to estimate the GW bias and its redshift and scale dependence.


\section{Gravitational Wave Bias Estimation Method}
\label{sec:power_bias}

The gravitational wave bias parameter, $b_{\textbf{GW}}$, quantifies how GW sources trace the underlying matter distribution. 
On sufficiently large scales, a constant (scale-independent) bias provides a good approximation to the relation between tracer and matter overdensities. 
In this work, however, we allow for a leading-order scale dependence, in case the explicit connection between binary black hole (BBH) mergers and host-galaxy properties introduces some mild scale-dependent effects.

We also evaluate the galaxy and GW biases by computing the auto power spectra for the three fields, the galaxy overdensity, the siren (BBH GW events) overdensity and the dark matter overdensity fields, in the IllustrisTNG300 comoving volume ($L=205\,\mathrm{Mpc}/h$) and taking the ratio of the former two to the latter. 
All fields are constructed on a cubic mesh with $N_{\mathrm{grid}}=256$ using the Cloud-in-Cell (CIC) mass-assignment scheme and the \texttt{Pylians} libraries \texttt{MAS\_library} and \texttt{Pk\_library} \cite{Pylians}.

The dark matter overdensity is obtained directly from IllustrisTNG snapshots by reading dark matter particle positions and depositing them onto the mesh via CIC, and normalizing the resulting density field as
\begin{equation}
\delta_{DM}(\mathbf{x}) = \frac{n_{DM}(\mathbf{x})}{\langle n_{DM} \rangle} - 1 \, .
\end{equation}

In principle, the GW (siren) and galaxy bias is defined with respect to the total matter overdensity field, $\delta_m$, which includes both dark matter and baryons. 
In this work, however, we use the dark-matter overdensity field, $\delta_{\rm DM}$, evaluated for the IllustrisTNG300 snapshots as our reference field when computing the matter power spectrum. 
This choice is motivated by the fact that on sufficiently large scales (small $k$), the clustering of dark matter closely tracks that of the total matter field, whereas baryonic effects primarily induce noticeable deviations on smaller, nonlinear scales. As illustrated in Figure 4 of \cite{Springel_2017}, the dark-matter and total-matter power spectra are nearly indistinguishable at low wavenumbers, while differences become progressively more apparent toward the high-$k$ end of the spectrum. 
We therefore adopt $P_{\rm DM}(k)$ as the baseline matter power spectrum in the range $0.04 < k/(h\,{\rm Mpc}^{-1}) < 1.00$, and interpret the inferred $b_{\rm GW}$ as the GW bias with respect to the underlying matter distribution on the large scales most relevant for our bias fitting.

The galaxy overdensity is constructed from Illustris subhalo catalogs and galaxy positions are painted onto the grid using CIC with unit weights, yielding a number-density field $n_g(\mathbf{x})$ and overdensity
\begin{equation}
\delta_g(\mathbf{x}) = \frac{n_g(\mathbf{x})}{\langle n_g \rangle} - 1 \, .
\end{equation}
The GW source overdensity is also constructed using CIC and from GW event catalogs generated by selecting GW events in host galaxies according to their predicted merger rates from the machine-learning model discussed in the last section. Each siren is assigned an identical per-event weight $w_{\mathrm{event}}$.
Since the analysis is performed on overdensity fields of the form $\delta = n/\bar{n} - 1$, this constant weight cancels identically, making the field construction effectively unweighted.

For each field $X \in \{DM,\, g,\, GW\}$, we compute the isotropic auto power spectrum $P_{XX}(k)$ using the \texttt{Pylians} estimator, which returns the values of the binned wavenumbers $k$, the corresponding values of monopole power spectrum, and the number of independent Fourier modes $N_{\mathrm{modes}}(k)$ contributing to each spherical $k$-shell. 
Note that we restricted the analysis to the range $0.04 < k/(h\,\mathrm{Mpc}^{-1}) < 1.00$.

Under the Gaussian approximation for Fourier modes, the standard deviation of the auto-spectrum estimator in a $k$-shell is
\begin{equation}
\sigma_X(k) = \sqrt{\frac{2}{N_{\mathrm{modes}}(k)}}\,P_{XX}(k),
\label{eq:sigma_modecount}
\end{equation}
where $N_{\mathrm{modes}}(k)$ is the number of independent Fourier modes in the bin returned by \texttt{Pylians} after binning the discrete Fourier grid. For tracer fields (galaxies and GW sources), the measured auto-spectrum $P_{tt}^{\rm raw}(k)$ ($t\in\{g, GW\}$) includes a Poisson shot-noise contribution. We subtract this contribution before fitting the bias model, so that the fit is performed on the shot-noise-subtracted spectrum $P_{tt}(k) \equiv P_{tt}^{\rm raw}(k) - P_{\rm SN}$. The Gaussian mode-counting uncertainty, however, is evaluated using the raw (pre-subtraction) spectrum, consistent with the fact that shot noise still contributes to the sample variance of the estimator even after its mean has been subtracted. In the bias fits, we weigh only by the tracer uncertainties and neglect the matter field error because the matter field is densely sampled and its statistical uncertainty is subdominant.

To estimate tracer bias, we fit the shot-noise-subtracted tracer auto-power spectrum to the following scale-dependent bias model
\begin{equation}
P_{tt}(k) = \big[b_0 + b_1\,(k-0.1)\big]^2\,P_{DM}(k),
\label{eq:P_tt}
\end{equation}
where $P_{DM}(k)$ is the dark matter power spectrum interpolated onto the tracer $k$-bins. 
Parameters $(b_0,\, b_1)$ are obtained by minimizing 
\begin{equation}
\chi^2(b_0,b_1) = \sum_{k_i}
\frac{\Big[P_{tt}(k_i) - \big(b_0+b_1 (k_i-0.1)\big)^2\,P_{DM}(k_i)\Big]^2}
{\sigma_t^2(k_i)} ,
\label{eq:chi2}
\end{equation}
where $\sigma_t(k)$ denotes the tracer uncertainty from mode counting in \eqref{eq:sigma_modecount}. 
The parameter $b_0$ represents the large scale bias in the linear regime, while $b_1$ captures leading-order scale dependence. The minimization is performed using \texttt{scipy.optimize.curve\_fit}, initialized at $(b_0,\, b_1)=(1,\, 0)$ and weighted by the Gaussian errors. 
The fitting routine returns the $2\times2$ covariance matrix $\mathbf{C}$ for $(b_0,b_1)$, from which we extract the parameters $\sigma_{b_0}=\sqrt{C_{00}}$, $\sigma_{b_1}=\sqrt{C_{11}}$, and their correlation coefficient $\rho=C_{01}/(\sigma_{b_0}\sigma_{b_1})$. 
For all figures presenting $b(k)$ or the model power spectrum (discussed in the next section), we use the covariance matrix $\mathbf{C}$ returned by the fit to compute uncertainty bands using standard linear error propagation from $(b_0,b_1)$ to the derived quantities.


\section{GW bias from \GalR training technique}
\label{sec:gwbias_result_msfrz}

In this section, we summarize our findings on GW bias for the mock GW source catalogs, which were generated from the IllustrisTNG300 simulation based on host galaxy properties and host probabilities derived from the emulator trained on the \GalR data set. 

As we will present, one of our main results is that the GW bias is systematically higher than the galaxy bias, indicating that the host galaxies of BBH merger events are more strongly clustered than the underlying galaxy distribution from which they are drawn. This enhanced clustering, which arises from the astrophysically based selection of galaxy hosts in merger rate training model, seems to be primarily due to the preferential selection of standard-siren events in more massive galaxies (top-left plot in Figure \ref{fig:mergerrate_vs_galprops}). As is well established, galaxy bias increases with stellar mass, a trend we have also confirmed for our galaxy sample, shown in Figure \ref{fig:bias_gal_binned_and_gw_vs_pivot} (see the brightly colored horizontal lines, which indicate the galaxy bias in different stellar mass bins). Moreover, when we compared the mean stellar mass of the full galaxy sample to that of the inferred GW host population in our GW mocks, we found that the latter were, on average, about an order of magnitude more massive. For instance, at $z = 0 $ the mean stellar mass increases from $\langle M_\ast^g \rangle \sim 2 \times 10^{10} \, M_\odot$ for the total galaxy population to $\langle M_\ast^{hg} \rangle \sim 2 \times 10^{11} \, M_\odot$ for GW host galaxies. We also found a $\sim 20$--$50\%$ shift toward higher mean metallicity in the GW host-galaxy sample, however this shift could be mostly due to the narrower and stronger stellar mass-metallicity relation in the underlying galaxy sample (see Figure \ref{fig:ssfr_vs_stellar mass}) and not the GW host probability selection (Figure \ref{fig:mergerrate_vs_galprops}). In contrast, we found an analogous comparison for the mean SFR to be less conclusive, particularly at low redshift, consistent with the combined impact of the trends shown in Figure \ref{fig:mergerrate_vs_galprops} and Figure \ref{fig:ssfr_vs_stellar mass}.

\subsection{GW bias dependence on redshift and scale}
\begin{figure}
  \centering
  \includegraphics[width=0.8\linewidth]{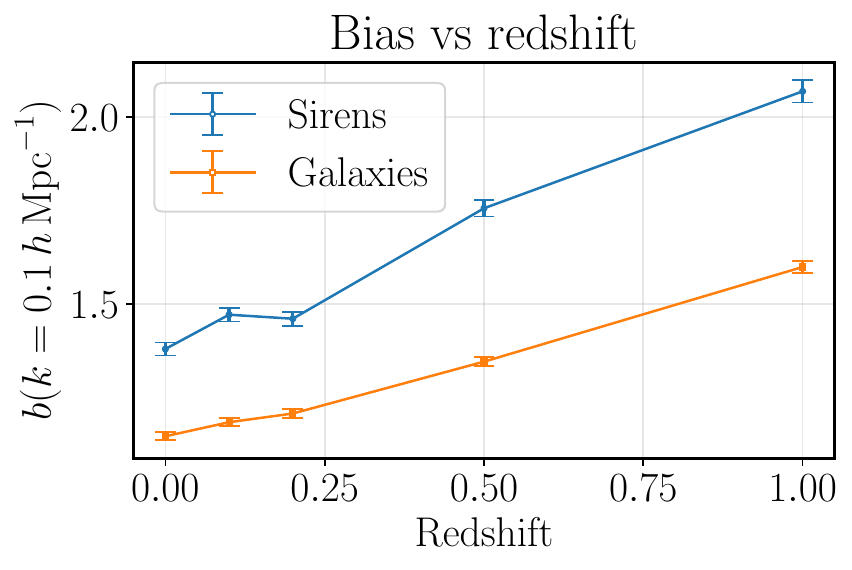}
  \caption{Bias as a function of redshift, evaluated at fixed wavenumber
  $k=0.1\,h\,\mathrm{Mpc}^{-1}$. Blue circles (orange squares) show the GW
  (galaxy) bias; lines denote linear fits. Error bars are $1\sigma$ from the fit parameter uncertainties propagated to $b(k)$. Note that this plot corresponds specifically to the $\alpha_{CE} = 5$ model using the FMR (M54) metallicity prescription.}
  \label{fig:bias-vs-redshift}
\end{figure}

Figure~\ref{fig:bias-vs-redshift} presents our results for the GW bias compared to the galaxy bias measured from the underlying IllustrisTNG subhalo population across different redshift bins. Since we allowed for linear scale dependence, the biases in these plots are evaluated at $k = 0.1\,h\,\mathrm{Mpc}^{-1}$. It can be observed that across all redshift bins, the GW bias is systematically higher than the galaxy bias, verifying that the hosts for BBH merger events are more strongly clustered than the galaxy distribution from which they originate. The error bars represent the full covariance propagation of the linear-bias parameters, $b_0$, demonstrating that the difference between the two biases is statistically significant in all redshift bins.
In addition to the amplification of bias, Figure~\ref{fig:bias-vs-redshift} shows a clear monotonic increase with redshift for both galaxies and sirens. This behavior is consistent with the expected redshift evolution of halo bias for tracer populations that are strongly correlated with stellar mass and act as indirect proxies for halo mass \footnote{At higher redshifts, halos of a given mass correspond to rarer peaks in the density field and are therefore more strongly clustered relative to the underlying matter distribution, leading to a larger effective bias, and rare peaks exhibit stronger clustering because a large scale overdensity greatly increases the chance that an otherwise unlikely fluctuation will exceed the collapse threshold.}.
The redshift evolution of the galaxy sample seen here is also consistent with previous studies of this simulation \cite{Springel_2017}\footnote{See figure 20 and table 20 in their analysis.}.  

\begin{figure}
\includegraphics[width=\linewidth]{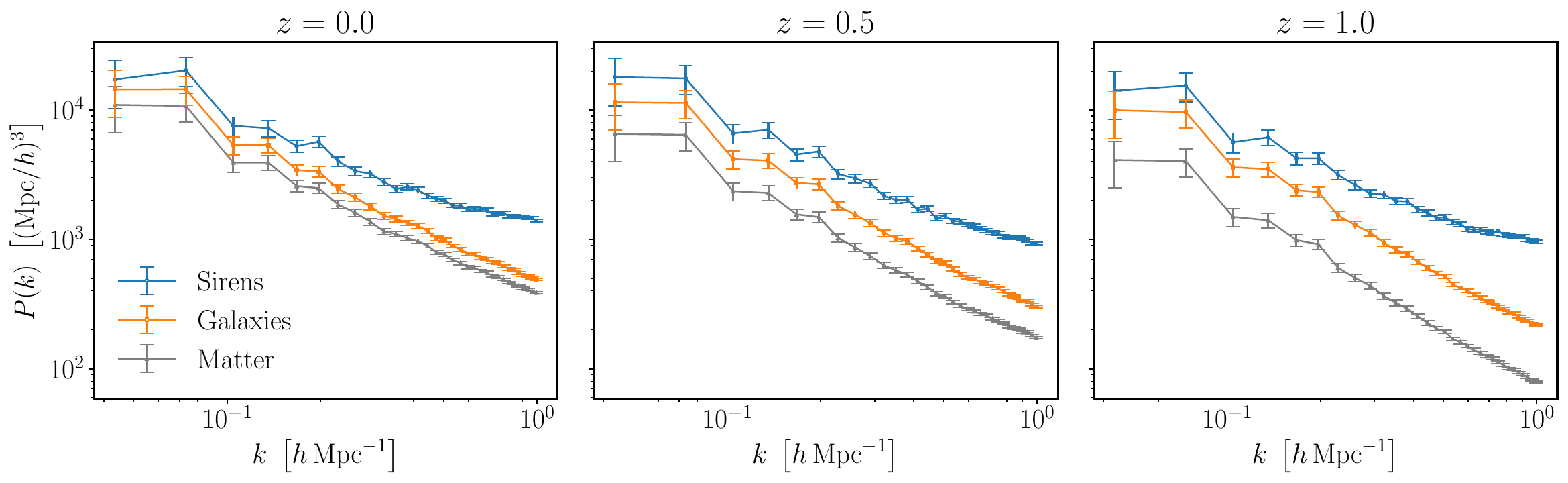}
\caption{Power spectra \(P(k)\) for dark matter, galaxies and sirens at redshifts
\(z=\{0.0,\,0.5,\,1.0\}\).
Points show measured \(P(k)\) for sirens (blue), galaxies (orange), and matter (gray), with corresponding error bars. Plots correspond to the $\alpha_{CE} = 5$ model and using the FMR (M54) metallicity prescription.}
\label{fig:pk_allz}
\end{figure}

\begin{figure}
\includegraphics[width=\linewidth]{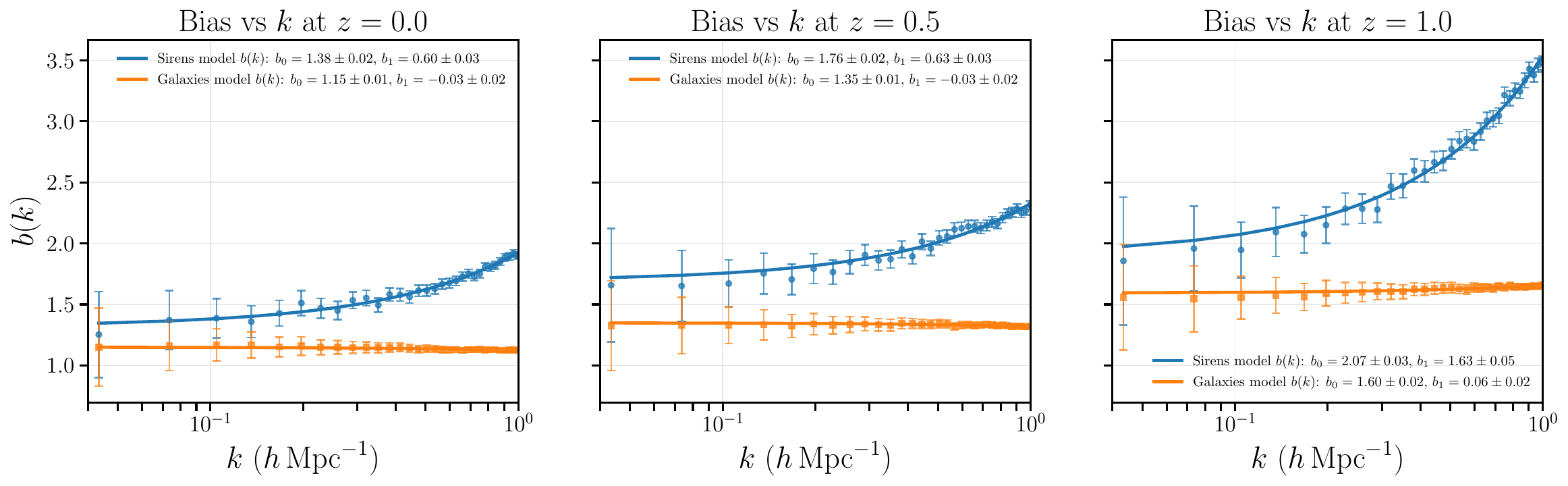} 

\caption{The scale dependence of GW bias at redshifts
\(z=\{0.0,\,0.5,\,1.0\}\). Lines denote the best-fit bias and points show measured bias for sirens (blue circles), galaxies (orange squares). Similar to Figure \ref{fig:pk_allz} the plots correspond to $\alpha_{CE} = 5$ model using the FMR (M54) metallicity prescription. }
\label{fig:scaledep_allz}
\end{figure}

Next, we investigate how different biases vary with scale. We begin by plotting the auto-power spectra of GW sources, galaxies, and dark matter. Figure~\ref{fig:pk_allz} shows the scale dependence of these power spectra for three different redshift bins $z=\{0.0,\,0.5,\,1.0\}$. For these measurements, the raw monopole $P(k)$ is obtained from the IllustrisTNG volume using CIC deposition on a grid $256^{3}$. As expected, the dark matter field exhibits a lower correlation amplitude compared to both galaxies and GW sources across all scales. The higher amplitude of the sirens' power spectrum relative to that of galaxies, as previously explained, is an expected result, given that the merger rate model places the majority of BBH mergers in galaxies with higher stellar mass, which are more clustered. 
Across all redshifts, the siren, galaxy, and matter power spectra also exhibit the expected decline toward smaller scales. Furthermore, while the overall matter and the tracers' power spectra are lower at higher redshifts due to the smaller level of linear growth at earlier times, the ratios of the siren to galaxy spectra increase, consistent with what we observed in Figure~\ref{fig:bias-vs-redshift}. The siren to galaxy power enhancement confirms that GW sources can constitute a more clustered tracer of the matter distribution compared to the total galaxy population, especially at earlier times. In addition to this overall amplification and redshift dependence, we also see a distinction in the scale-dependence of the two; that is, while the slope of galaxy power spectrum still follows the matter power spectrum on small scales up to $k\sim 1$ hMpc$^{-1}$, the slope of siren power spectrum starts to deviate from the other two around $k\sim 0.2 $hMpc$^{-1}$. This feature can be seen more explicitly in terms of GW bias. Figure \ref{fig:scaledep_allz} shows the scale dependence of galaxy and siren biases $b(k)$, for the same redshift bins around $z=\{0.0,\,0.5,\,1.0\}$. The best-fit bias parameters $b_0$ and $b_1$ introduced in \eqref{eq:P_tt} are obtained using the method described in section \ref{sec:power_bias}.
In these plots, the consistently stronger clustering of the sirens relative to the galaxy population at all scales, as well as the increase with redshift, are more clearly visible. These trends are also reflected in the fitted $b_0$ values and in the increasing ratio of $b_0$ for sirens relative to galaxies.
We can also observe explicitly that the GW bias has stronger scale-dependence compared to the galaxy bias, particularly at smaller scales. In all redshift bins $b_1$ for the GW bias is at least one order of magnitude higher than $b_1$ for the galaxy bias. Note also that both parameters $b_0$ and slope $b_1$ grow systematically with redshift. This could also be understood by noting that rarer, higher-density peaks in the matter distribution are more strongly clustered on small scales than typical regions and as the GW host galaxies are drawn preferentially from more massive galaxies that reside in these regions. This is qualitatively consistent with the \cite{Springel_2017} analysis of the TNG300 data and the apparent scale dependence of bias for different $M_*$-selected galaxy samples (Figure 20 in that paper).  

\begin{figure}[h]
\includegraphics[width=0.5\linewidth]{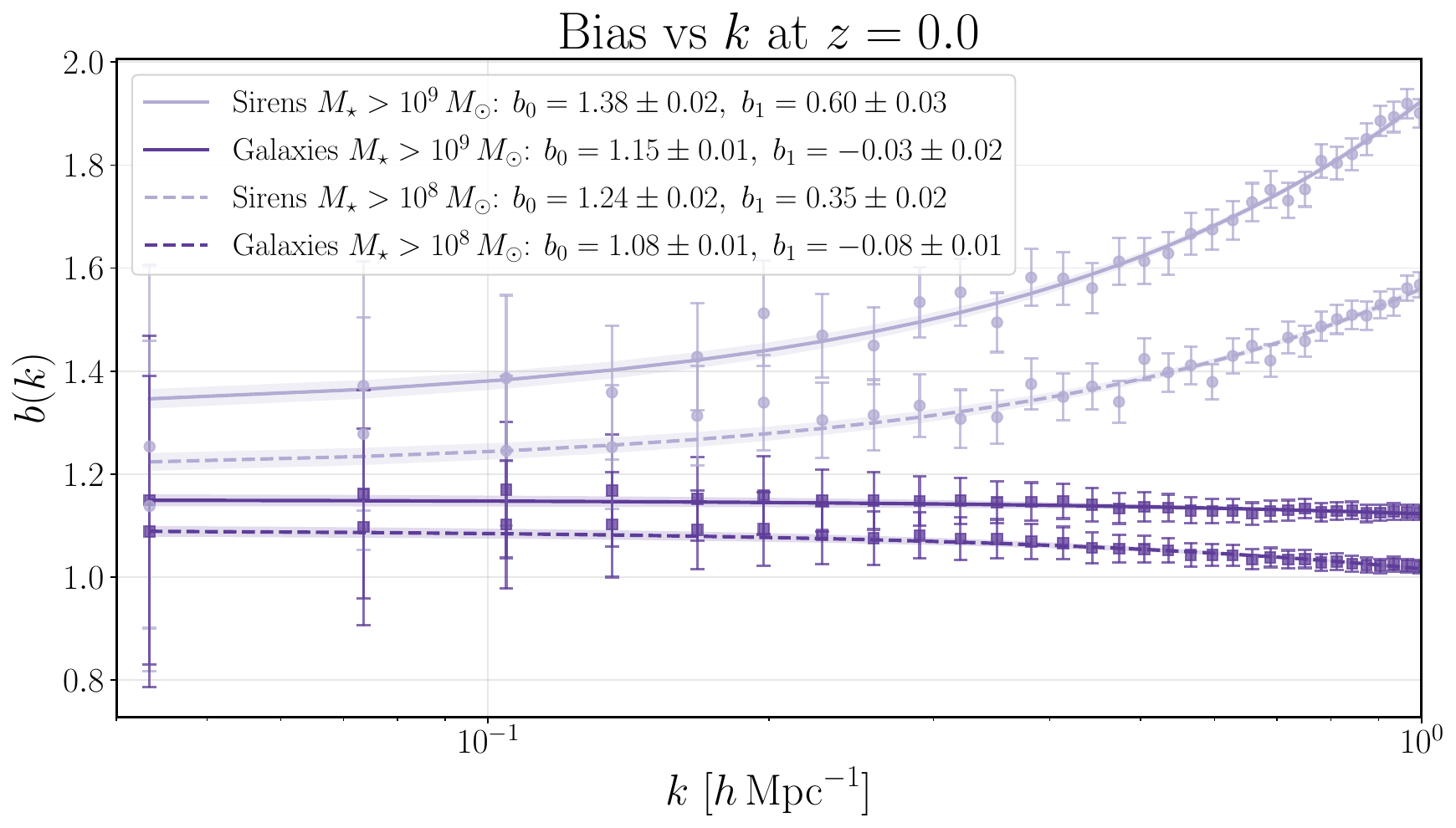}
\includegraphics[width=0.5\linewidth]{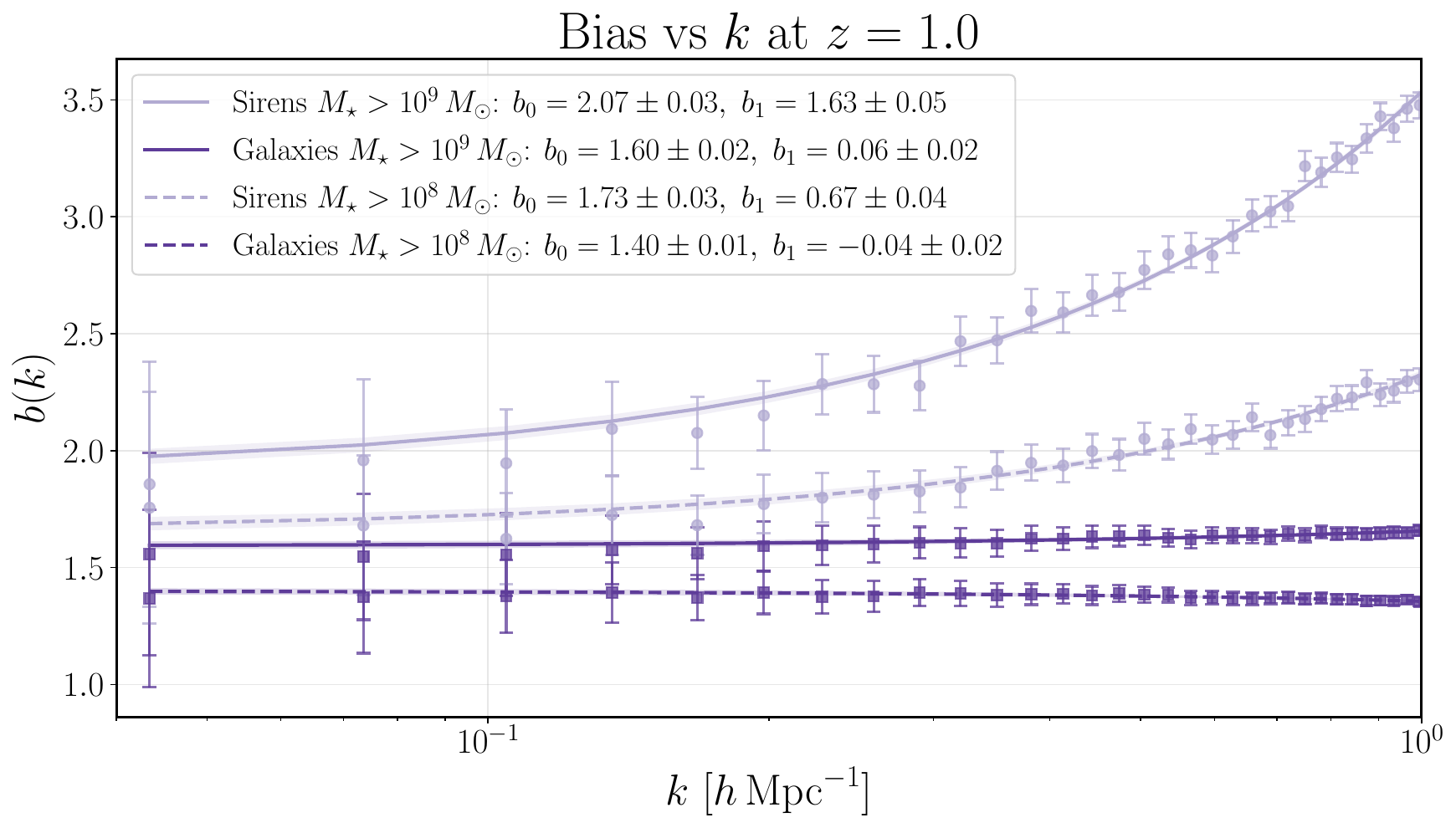}
\caption{Bias $b(k)$ for sirens and galaxies across redshifts, comparing two stellar mass selections.
Solid (dashed) curves show best-fit siren (galaxy) bias with $1\sigma$ bands over $0.04 \le k \le 1.0\,h\,\mathrm{Mpc}^{-1}$ and points show the measured bias with corresponding error bars. This figure again corresponds to the $\alpha_{\mathrm{CE}}=5$ model with the M54 (FMR) metallicity prescription.}

\label{fig:bias_k_compare_5panels}
\end{figure}

We also explored whether further lowering the stellar mass cutoff toward the TNG300 resolution limit and including lower-mass subhalos in the sample significantly affects the scale dependence of the bias. Figure \ref{fig:bias_k_compare_5panels} compares the results obtained using two different choices of the cutoff, $M_\star \gtrsim 10^{9}\,[M_\odot]$ and $M_\star \gtrsim 10^{8}\,[M_\odot]$, for two redshift bins $z=\{0.0,1.0\}$. As we see, lowering the stellar mass threshold reduces the overall bias amplitude for both tracers, reflecting the inclusion of less massive and more abundant galaxies that trace the underlying matter distribution. It does not significantly affect the scale dependence of the galaxy bias, but it weakens the scale dependence of the \textit{siren} bias by including lower-mass hosts; nevertheless, the slope parameter $b_1$ remains significantly positive. This suggests that the stellar mass selection of host galaxies can shift the value of bias up or down, but the positive scale dependence should persist. Therefore, for precise theoretical predictions and observational measurements of bias, it is important to carefully account for the selection criteria when performing cosmological inference; however, adjusting for this shift may not significantly affect the qualitative scale-dependent features.

Next, we will examine in more detail how the GW bias obtained with this approach correlates with the astrophysical properties of the host galaxies.

\subsection{GW bias versus astrophysical properties}
\label{sec:gw_bias_vs_astro_properties}
\subsubsection{GW bias vs.\ stellar mass relation}

\begin{figure}[h]
    \centering
    \includegraphics[width=0.8\linewidth]{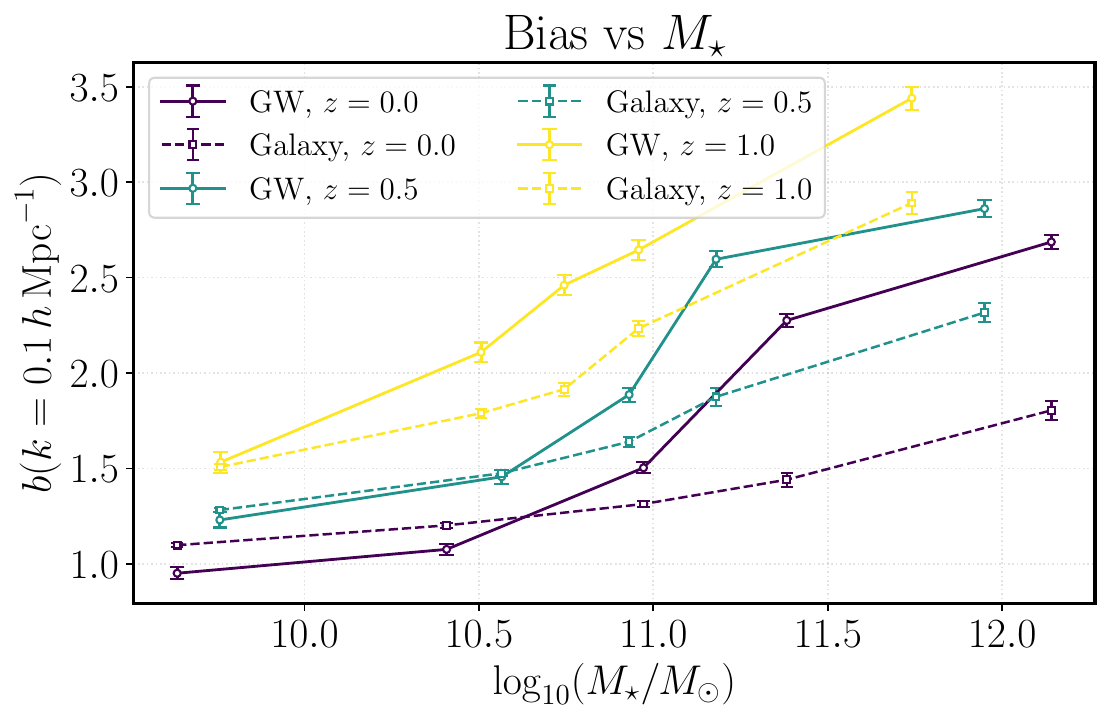}
    
    \caption{
    The best-fit bias of galaxies and sirens $b(k)$ as a function of stellar mass across redshift. The bias is obtained from a linear fit to the auto-power spectra, over the range $0.04 \le k \le 1.0\,h\,\mathrm{Mpc}^{-1}$, and evaluated at $k=0.10\,h\,\mathrm{Mpc}^{-1}$ for equal-count quantile bins in $\log_{10}(M_\star/M_\odot)$. Error bars represent the $1\sigma$ uncertainty on the fitted bias at this scale. The boundaries of resulting mass bins in $\log_{10}(M_\ast/M_\odot)$ for redshifts of $z\in\{0.0, \,0.5, \,1.0\}$ are $ \{ 9.170,\, 10.101,\,10.714,\,$$11.230,\,11.536,\,12.746\}$ 
$\{9.169,\,10.345,\,10.786,\,$$11.076,\,11.284,\,12.615\}$ and  $\{9.170,\,10.350,\,10.662,$ $10.827,\,11.089,$ $12.393\}$, respectively. This plot corresponds to the $\alpha_{\rm CE}=5$ model with the FMR (M54) metallicity prescription.
    }

    \label{fig:bgw-vs-mstar-allz}
\end{figure}

Figure~\ref{fig:bgw-vs-mstar-allz} illustrates how the evaluated GW and galaxy biases depend on stellar mass across different redshift bins. To obtain these plots for each redshift, the GW hosts and underlying galaxy samples are divided into five bins of equal count in $\log_{10}(M_\star/M_\odot)$. We measure the auto–power spectrum in each subsample, fit a linear scale dependent bias over the range $0.04 \le k \le 1.0\,h\,\mathrm{Mpc}^{-1}$ (with the lower bound set just above the Illustris box fundamental mode), and evaluate the resulting bias at a fixed wavenumber $k=0.1\,h\,\mathrm{Mpc}^{-1}$. 

At all redshifts, the inferred biases increase monotonically with stellar mass, reconfirming the underlying halo mass dependence of galaxy clustering, with stellar mass acting as a proxy for host halo mass. 

\begin{figure}[h]
    \centering
    \includegraphics[width=\linewidth]{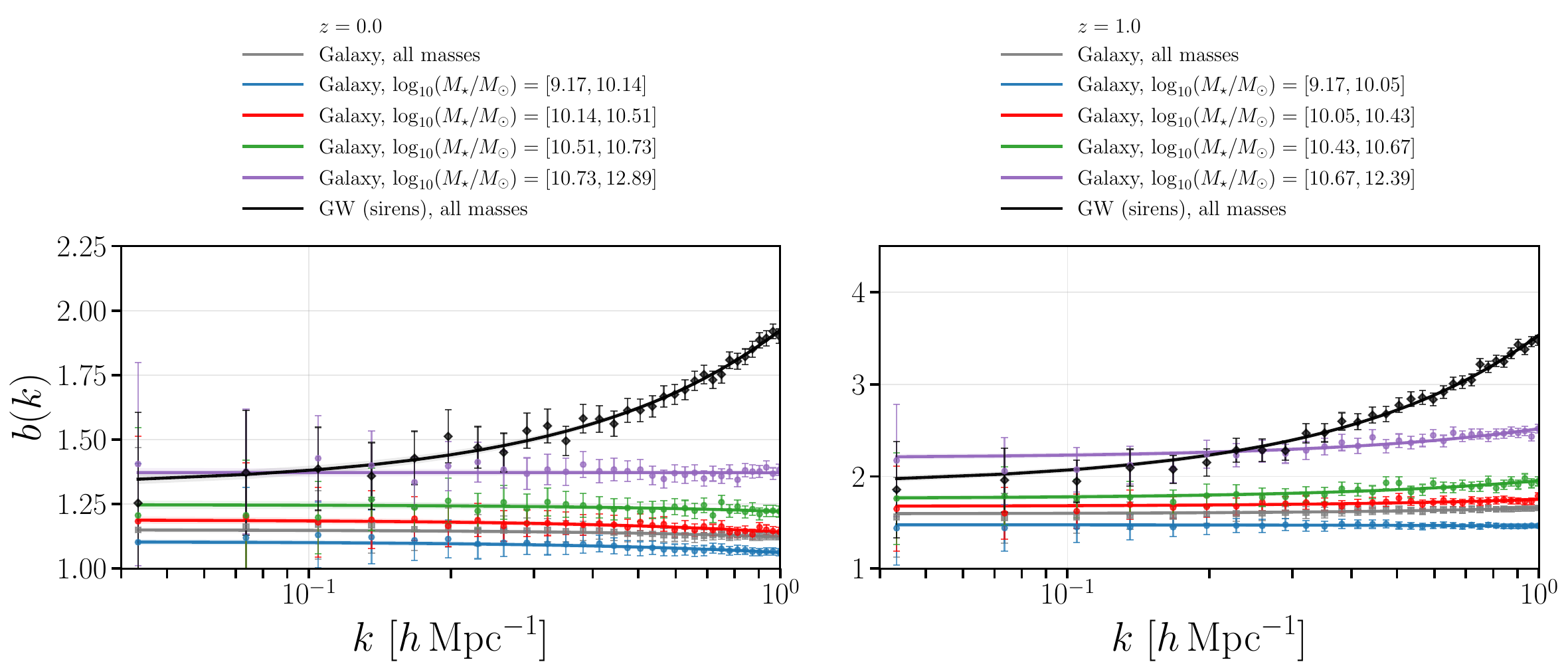}
    \caption{ Comparison between the scale-dependent bias $b(k)$ for all GW sirens and galaxies in different stellar mass bins, at $z = 0$ (left) and $z=1$ (right).
Points show the measured bias with corresponding error bars and curves show the best-fit bias model obtained over $0.04 \le k \le 1.0\,h\,\mathrm{Mpc}^{-1}$. This figure corresponds to the $\alpha_{\mathrm{CE}}=5$ model with the M54 (FMR) metallicity prescription.
    }

    \label{fig:bias_mass_bins}
\end{figure}

Interestingly, the GW and galaxy bias values are initially very similar, even at $z=1$, with the GW bias being slightly lower in the low stellar mass bins. At higher stellar masses, however, the GW bias surpasses the galaxy bias evaluated at the same mean stellar mass. This difference likely reflects the non-uniform weighting of galaxies by their predicted BBH merger rates. In other words, even within a given stellar mass bin, each galaxy contributes to the siren density field according to its expected number of BBH mergers, with an effective weight that may be greater than unity, rather than all galaxies in a specific mass range contributing equally. As shown in Figure \ref{fig:mergerrate_vs_galprops}, these weights also depend on host properties other than stellar mass, including metallicity and SFR, which further impact the clustering of the siren population.
This interpretation is further supported by Figure \ref{fig:bias_mass_bins}, which shows the full scale-dependent bias $b(k)$ for galaxies in different stellar mass bins together with the GW bias estimated for the full, merger-rate-weighted GW-host sample at $z=0$ and $z=1$. On large scales, the GW bias is closest to the galaxy bias of the upper stellar mass bins; however, on small scales, it rises much more steeply than that of these galaxy samples.

Taken together, these results suggest the siren population inherits the large scale clustering of galaxies, and because galaxy bias exhibits a strong intrinsic dependence on stellar mass, this dependence is likewise transmitted into the GW bias. Nevertheless, there is an additional enhancement on top of this effect, which becomes increasingly significant at small scales.

\subsubsection{GW bias vs.\ SFR relation}

\begin{figure}[h]
  \centering
  \includegraphics[width=0.7\linewidth]{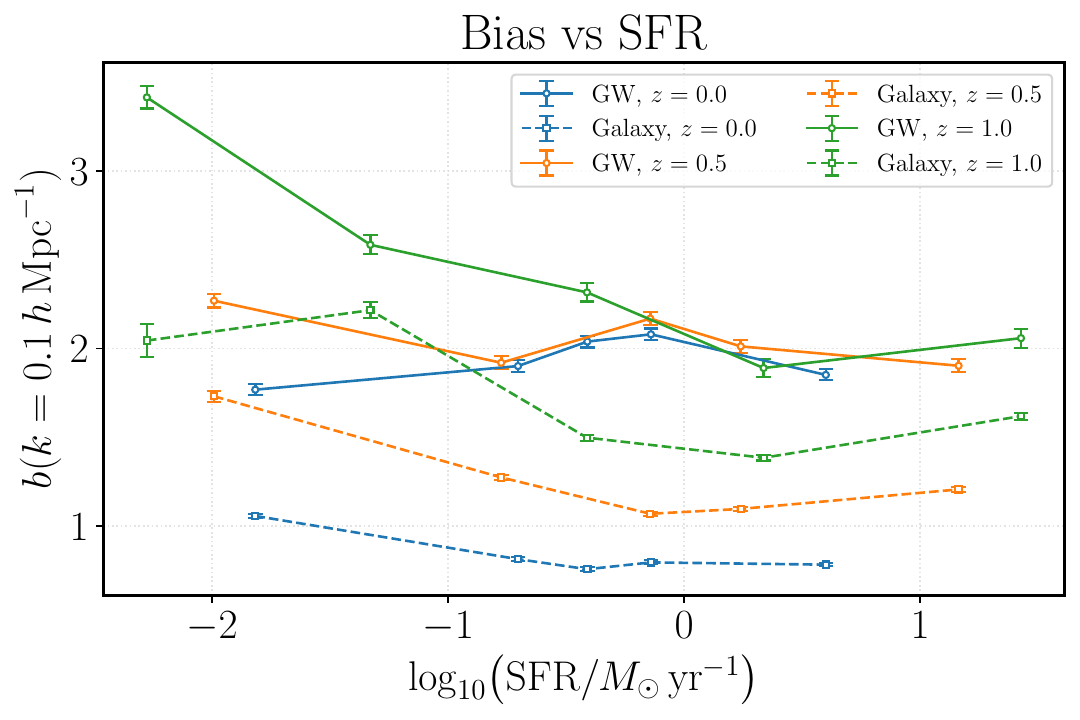}
 \caption{The best-fit bias of galaxies and sirens $b(k)$ as a function of SFR across redshifts. The bias is obtained from a linear fit to the auto-power spectra, over the range $0.04 \le k \le 1.0\,h\,\mathrm{Mpc}^{-1}$, and evaluated at $k=0.10\,h\,\mathrm{Mpc}^{-1}$ for equal-count bins in $\log_{10}(\mathrm{SFR}/M_\odot\,\mathrm{yr}^{-1})$. Error bars represent the $1\sigma$ uncertainty on the fitted bias at this scale.
    At each redshift, the siren host population is divided into five equal-count quantile bins in $\log_{10}(\mathrm{SFR})$.
    The resulting bin edges in $\log_{10}(\mathrm{SFR}/M_\odot\,\mathrm{yr}^{-1})$ for redshifts of z=\{0.0, 0.5, 1.0\} are
    $\{-2.77,-0.865,-0.543,-0.277,-0.004,1.205\}$,
    $\{-2.78,-1.203,-0.347,0.060,0.419,1.906\}$ and 
    $\{-2.75,-1.800,-0.859,0.036,0.637,2.216\}$, respectively.
    This figure corresponds to the $\alpha_{\rm CE}=5$ model with the FMR (M54) metallicity prescription.}
  \label{fig:bgw-sfr-allz}
\end{figure}

\begin{figure}[h]
    \centering
    \includegraphics[width=\linewidth]{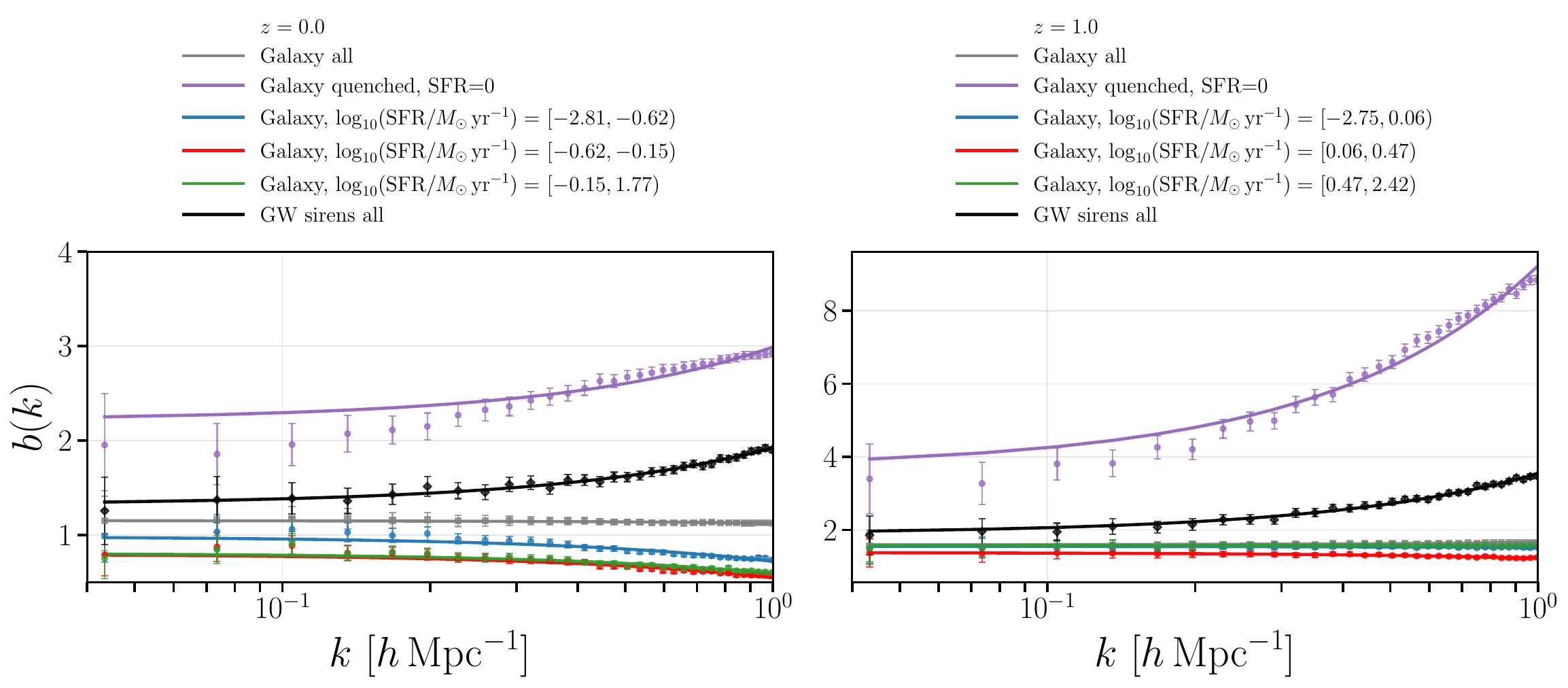}
    \caption{Bias $b(k)$ of galaxies and GW sources at $z=0$ (left) and $z=1$ (right). The grey curve shows the bias of the full galaxy sample, while the purple curve corresponds to non-star-forming galaxies (SFR$=0$). Star-forming galaxies (SFR$>0$) are further divided into three equal-count bins in $\log_{10}(\mathrm{SFR}/M_\odot\,\mathrm{yr}^{-1})$ at each redshift (blue, red, and green).
    The black curve shows the GW bias inferred from the full siren catalog. Points show the measured bias with corresponding error bars and curves show the best-fit bias model obtained over $0.04 \le k \le 1.0\,h\,\mathrm{Mpc}^{-1}$. This figure corresponds to the $\alpha_{\mathrm{CE}}=5$ model with the M54 (FMR) metallicity prescription. 
    }
    \label{fig:bias_sfr_bins}
\end{figure}

Figure~\ref{fig:bgw-sfr-allz} shows how different biases in our star forming samples vary with star formation rate for different redshifts, evaluated as before at the fixed scale $k=0.1\,h\,\mathrm{Mpc}^{-1}$ using equal-count bins in $\log_{10}(\mathrm{SFR}/M_\odot\,\mathrm{yr}^{-1})$. In contrast to the similar monotonic trend observed with stellar mass, the bias parameters here exhibit different trends for different redshifts. In particular, in low redshifts, star-forming galaxy bias is very weakly correlated with SFR, and the GW bias mirrors that but at $z=1$ GW bias exhibits a clear amplification for low SFR hosts. The SFR effect is better illustrated in Figure~\ref{fig:bias_sfr_bins}, where we compare the scale dependence of GW bias for the total GW sample versus galaxy bias in different SFR bins. Here, non-star-forming (referred to as quenched hereafter and in the plots) galaxies are separated from star-forming systems (using unequal-count bins), and we observe that the quenched galaxies exhibit substantially higher bias than their star-forming counterparts. Interestingly, the remaining non-zero SFR galaxy bins exhibit broadly similar clustering amplitudes, and the GW bias lies above these star-forming populations but is closer to them than to the quenched population. The higher bias observed for quenched systems may also be attributed to their tendency to reside in more massive halos, which is correlated with higher stellar mass (left plot in Figure \ref{fig:galaxy_distributions_M_SFR_Z}). We additionally found that although our host galaxy samples include a relatively larger fraction of high mass galaxies compared to the underlying population, they did not show a higher non-star-forming fraction. For example, at $z = 0$, the quenched fraction is actually lower in the GW host sample, and the mean SFR doesn't change much either, implying that the enhanced GW bias is not primarily driven by this subcategory of hosts. It is also worth noting again that Figures \ref{fig:ssfr_vs_stellar mass} and \ref{fig:mean mass vs sfr} show substantial scatter in SFR at fixed stellar mass, as well as a non-monotonic relation between the two quantities. Consequently, binning galaxies by SFR mixes populations spanning a broad range of stellar masses, which can dilute the imprint of the strong stellar mass dependence of clustering bias. Moreover, in our model, the BBH merger rate depends less strongly on SFR than on stellar mass, particularly at low redshift (Figure \ref{fig:mergerrate_vs_galprops}). Together, these effects may help explain why variations in SFR within the low-redshift host population produce only modest changes in the corresponding clustering bias.

\subsubsection{GW bias vs. metallicity}
\begin{figure}[h]
  \centering
  \includegraphics[width=0.8\linewidth]{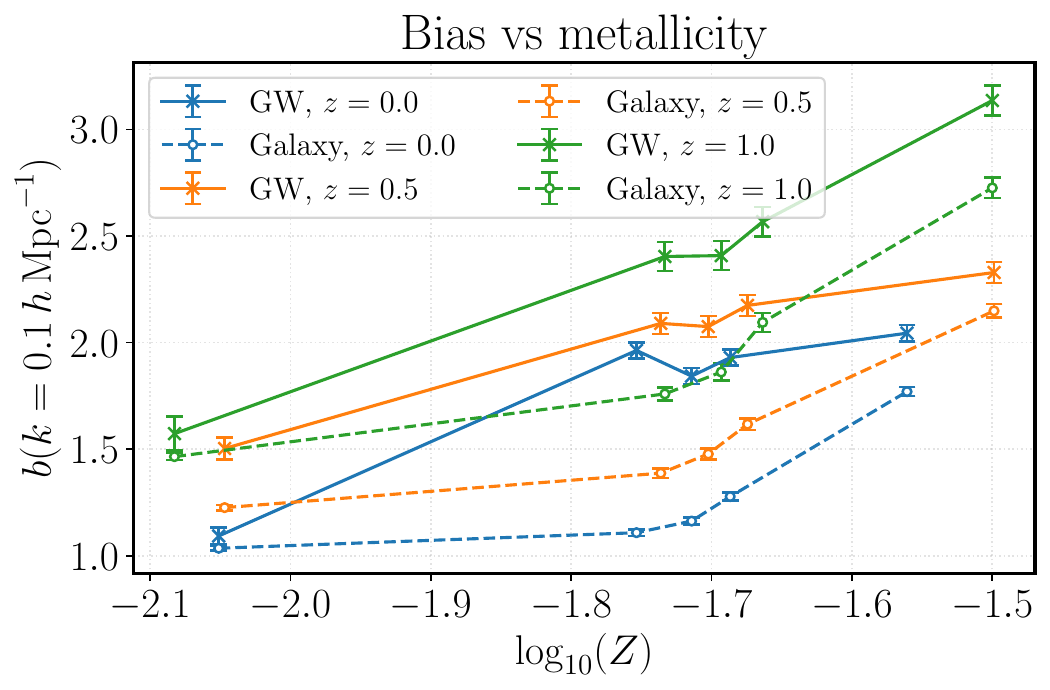}
  
     \caption{The best-fit bias of galaxies and sirens $b(k)$ as a function of a host stellar metallicity across redshifts. The bias is obtained from a linear fit to the auto-power spectra, over the range $0.04 \le k \le 1.0\,h\,\mathrm{Mpc}^{-1}$, and evaluated at $k=0.10\,h\,\mathrm{Mpc}^{-1}$ for equal-count bins in \(\log_{10}(Z)\). Error bars represent the $1\sigma$ uncertainty on the fitted bias at this scale. The \(\log_{10}(Z)\) bin edges, computed from the siren-host sample at redshifts of z=\{0.0, 0.5, 1.0\}, are
    $\{-2.323,-1.779,-1.727,-1.701,-1.672,-1.449\}$,
    $\{-2.336,-1.757,-1.715,-1.690,-1.659,-1.338\}$ and 
    $\{-2.406,-1.760,-1.707,-1.679,\\
    -1.648,-1.352\}$, respectively. This plot corresponds to the \(\alpha_{\rm CE}=5\) model using the FMR (M54) metallicity prescription.}

  \label{fig:bgw-metal-allz}
\end{figure}

Finally, in Figure~\ref{fig:bgw-metal-allz}, we present the dependence of the GW bias on host stellar metallicity across redshifts. At each redshift, the siren and galaxy samples are divided into equal-count quantile bins in $\log_{10}(Z)$, and the GW auto-power spectrum is measured for each subsample. As before, bias is obtained by fitting the tracer power spectrum relative to the dark matter power spectrum over $0.04 \le k \le 1.0\,h\,\mathrm{Mpc}^{-1}$ and evaluated at $k=0.1\,h\,\mathrm{Mpc}^{-1}$. We would like to emphasize again that the metallicity refers to the metallicity of the host galaxies at the time of the BBH merger and should not be confused with the metallicity at the epoch of black hole formation, which may have occurred substantially earlier and corresponded to a less evolved chemical state.

We can observe in this figure that the galaxy bias exhibits a visible monotonic increase with metallicity at all redshifts, similar to the dependence on stellar mass, and GW bias mirrors the same trend, only shifted to higher values. For galaxies, this behavior is consistent with the strong correlation between stellar metallicity and halo mass, and that metal-rich systems tend to reside in more massive halos that have assembled early and, therefore, exhibit elevated large scale bias.   
This amplification for BBH mergers is also not surprising since, in each band, the machine-learned merger rate model preferentially assigns higher BBH merger probabilities to both more metal-enriched and higher stellar mass galaxies (Figure \ref{fig:mergerrate_vs_galprops}). This boosts the effective clustering strength of the siren population beyond that of the general galaxy sample.

\begin{figure}[h]
    \centering
    
    \includegraphics[width=\linewidth]{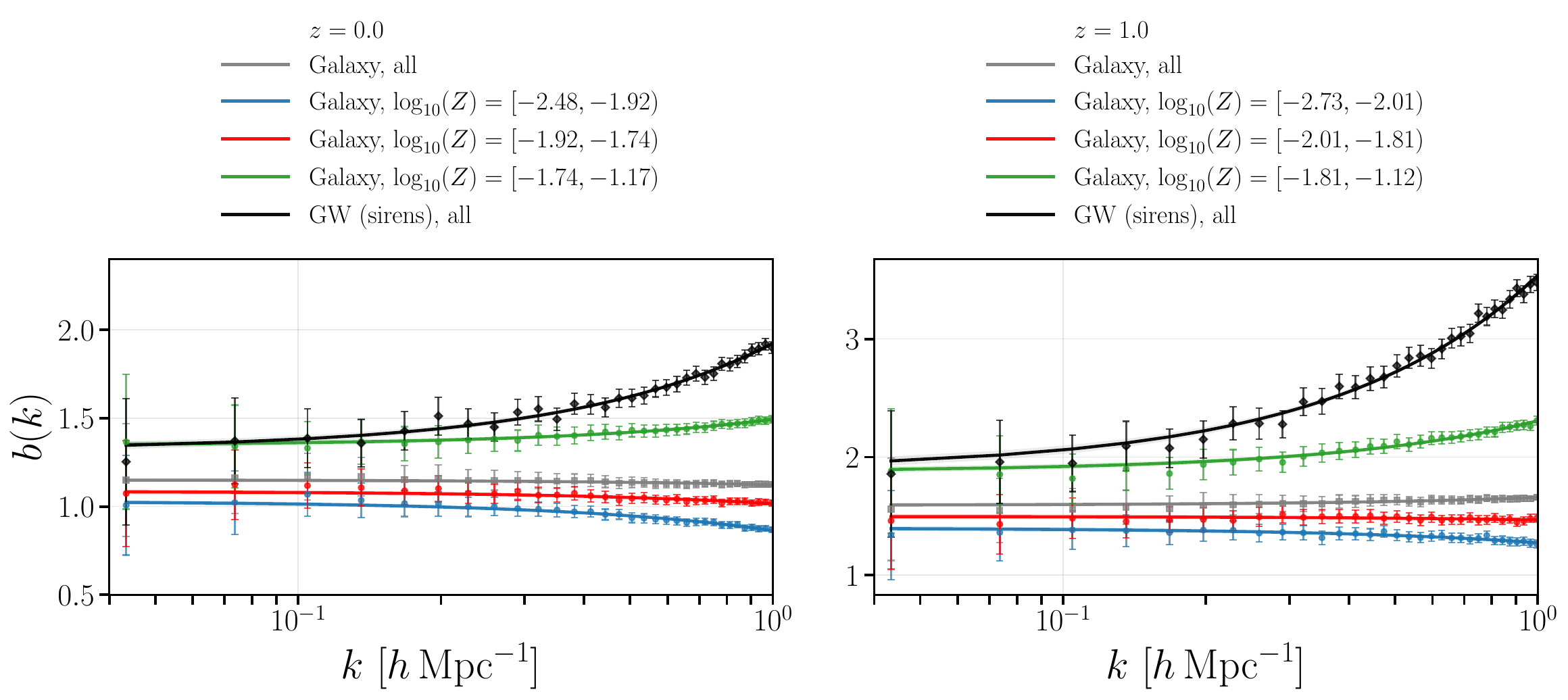}
    \caption{ Bias of galaxies and GW sources as a function of metallicity at $z=0$ and $z=1$. At each redshift, galaxies are divided into three equal-count bins in $\log_{10}(Z)$ (computed from the galaxy population only), while the GW bias is measured from the full siren catalog (black for $z=0$, gray for $z=1$). Points show the measured bias with corresponding error bars and curves show the best-fit bias model obtained over $0.04 \le k \le 1.0\,h\,\mathrm{Mpc}^{-1}$.  This figure corresponds to the $\alpha_{\mathrm{CE}}=5$ model with the M54 (FMR) metallicity prescription.
}

    \label{fig:bias_metallicity_bins}
\end{figure}

Once again to obtain additional insight, we compare the scale dependence of the bias $b(k)$ for galaxies, this time split into different metallicity bins versus GW sources for redshifts $z=0$ and $z=1$ in Figure~\ref{fig:bias_metallicity_bins}\footnote{At each redshift, galaxies are divided into equal-count metallicity bins, while the GW bias is measured from the full siren catalog.}. In both cases, higher-metallicity galaxies exhibit larger bias amplitudes, and the GW bias most closely matches the galaxy bias in the highest metallicity range (green curve). This indicates that GW hosts preferentially reside in metal-rich systems, which, as we see, are more strongly clustered. After checking the samples, we also confirmed that the mean metallicity for galaxies was within the red band, while for GW hosts, it has shifted to a value within the green band. 
  
In summary, these findings demonstrate that the clustering of GW events from stellar BBH events is highly correlated with the stellar mass and metallicity  of their host galaxies, noting that both are tightly correlated. By contrast, we did not find a strong correlation with the instantaneous star formation rate.

\section{GW bias from the phenomenological host probability model with stellar mass dependence}
\label{sec:gwbiasresult_mass_powerlaw}
\subsection{Methodology}
In this section, we adapt an analytic phenomenological model for host-galaxy probabilities to estimate the GW bias. In other words, the mock GW catalogs are generated using a simple analytical function that depends only on the stellar mass of the host galaxies, in contrast to the methodology described in Sections~\ref{sec:mergerate}--\ref{sec:mock_gw}, where a machine-learning emulator was used to estimate host-galaxy probabilities based on population-synthesis simulations and astrophysical properties. This method is analogous to that used in \cite{Dehghani_2025, Hosseini_2026}, with the key difference that in those works the host galaxies were sampled from real observational galaxy surveys (WISExSCOS, 2MPZ and Sloan Digital Sky Survey) subject to observational selection effects, whereas here they are generated from IllustrisTNG300 simulation box and therefore subject to simulation limitations. This setup allows us both to validate the features identified in those earlier analyses and to compare them with the non-analytical population-synthesis informed host-galaxy probability framework discussed previously. 

More specifically, we assign each galaxy a probability of hosting a merger according to the following broken power law:
\begin{equation}
\mathbb{P}(\text{host}\mid M_\ast) \propto
\begin{cases}
\left(\frac{M_\ast}{M_\kappa}\right)^{\alpha_l},& M_\ast < M_\kappa,\\[8pt]
\left(\frac{M_\ast}{M_\kappa}\right)^{-\alpha_h}, & M_\ast \ge M_\kappa,
\end{cases}
\label{eq:brokenPL}
\end{equation}
where $M_\ast$ is the stellar mass of the host galaxy and $M_\kappa$ is a pivot mass scale. The slopes for the broken power-law in the lower and higher range values of $M_\ast$ are adjusted by parameters $\alpha_l$ and  $\alpha_h$\footnote{We note that the parameters $\alpha_l$ and $\alpha_h$ appearing in the selection function are distinct from the common-envelope efficiency parameter $\alpha_{\mathrm{CE}}$ introduced in earlier sections.}.
 Apart from the different method for siren host selection, everything else, including density-field construction, power-spectrum estimation, Gaussian error calculations, and bias fitting is identical to Sections~\ref{sec:illustris}--\ref{sec:power_bias}.
 While this simplified model does not capture the full astrophysical complexity of binary formation and galaxy evolution, it provides a controlled analytical way to study the dependence of the GW bias on host probability-stellar mass relation. At low masses, the probability of host increases with $M_\ast$, similar to what is observed in population-synthesis analysis \cite{Artale_2019, Santoliquido_2020} (also see Figure \ref{fig:host-property-alpha-sweep}) reflecting the general positive correlation between stellar mass and star formation history, owing to the role of the SFR in both generating BBH progenitors and building up stellar mass. The subsequent fall off or flattening is then motivated by the interplay between age of the galaxy as it quenches and merger delay times (see \cite{Dehghani_2025, Hosseini_2026} for more discussion). 

For this analysis, we explored several values of $\alpha_l$ in the vicinity of unity and $\alpha_h$ near zero. These choices were motivated by best-fit parameters for the projected merger rate-stellar mass relation, obtained from the \GalR seeding framework in this study and in the previous analysis \cite{Artale_2019}.
Given different choices of $(\alpha_l,\,\alpha_h,\, M_\kappa)$ and for a series of redshift snapshots taken within the range $z \in [0, 1]$, galaxies were assigned host probabilities proportional to Eq.~\eqref{eq:brokenPL}. Using these probabilities, we generated mock GW catalogs containing $N_{\rm GW}$ events similar to Eq.~\eqref{eq:N_exp} and Eq.~\eqref{eq:N_exp_normalization} substituting $\mathbb{P}_j$ for each host in place of $\hat n_j$ \footnote{Similar to Eq.~\eqref{eq:N_exp_normalization}, a large expected number was chosen to avoid shot-noise domination. The number of galaxies in different snapshots ranged from $\sim2.1 \times 10^5$ to $\sim 2.4 \times 10^5$, and the number of siren events were taken to be $1.8\times 10^4$}.  For each realization, we then estimated the auto-power spectrum of the resulting GW catalog and determined the best-fit bias parameters by minimizing the $\chi^2$ statistic (see Eqs.~\eqref{eq:P_tt} and~\eqref{eq:chi2}) over $0.04\,\!<\!\,k\,\!<\!\,1.0\,h\,\mathrm{Mpc}^{-1}$ \footnote{As before, we also applied a stellar mass cut at $10^9 M_\odot$, so the parent catalog from which hosts were drawn included only galaxies with $M_\star > 10^9 M_\odot$.}.

\begin{figure}[h]
    \centering

    \begin{subfigure}[t]{0.48\linewidth}
        \centering
        {\small \ \ \ \ \ \ \
        $(\alpha_l,\,\alpha_h,\,M_\kappa) =
        (0.95,\,0.20,\,2.20 \times 10^{11} M_\odot)$
        \par}
        \vspace{2mm}
        \includegraphics[width=\linewidth]
        {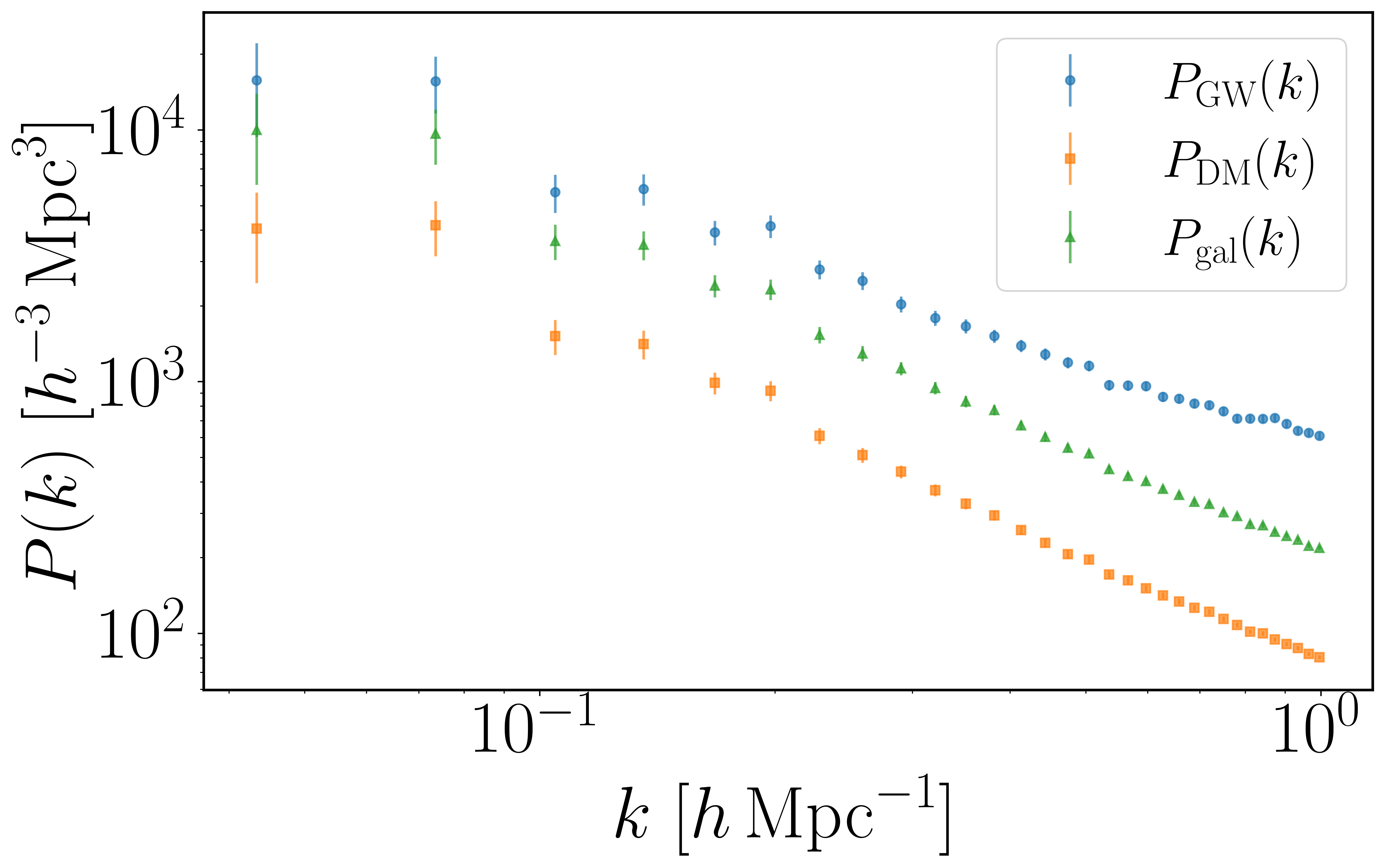}
        \label{fig:pspectra_0.95}
    \end{subfigure}
    \hfill
    \vspace{1mm}
    \begin{subfigure}[t]{0.48\linewidth}
        \centering
        {\small \ \ \ \ \ \ \
        $(\alpha_l,\,\alpha_h,\,M_\kappa) =
        (0.95,\,0.20,\,2.20 \times 10^{11} M_\odot)$
        \par}
        \vspace{2mm}
        
        \includegraphics[width=\linewidth]
        {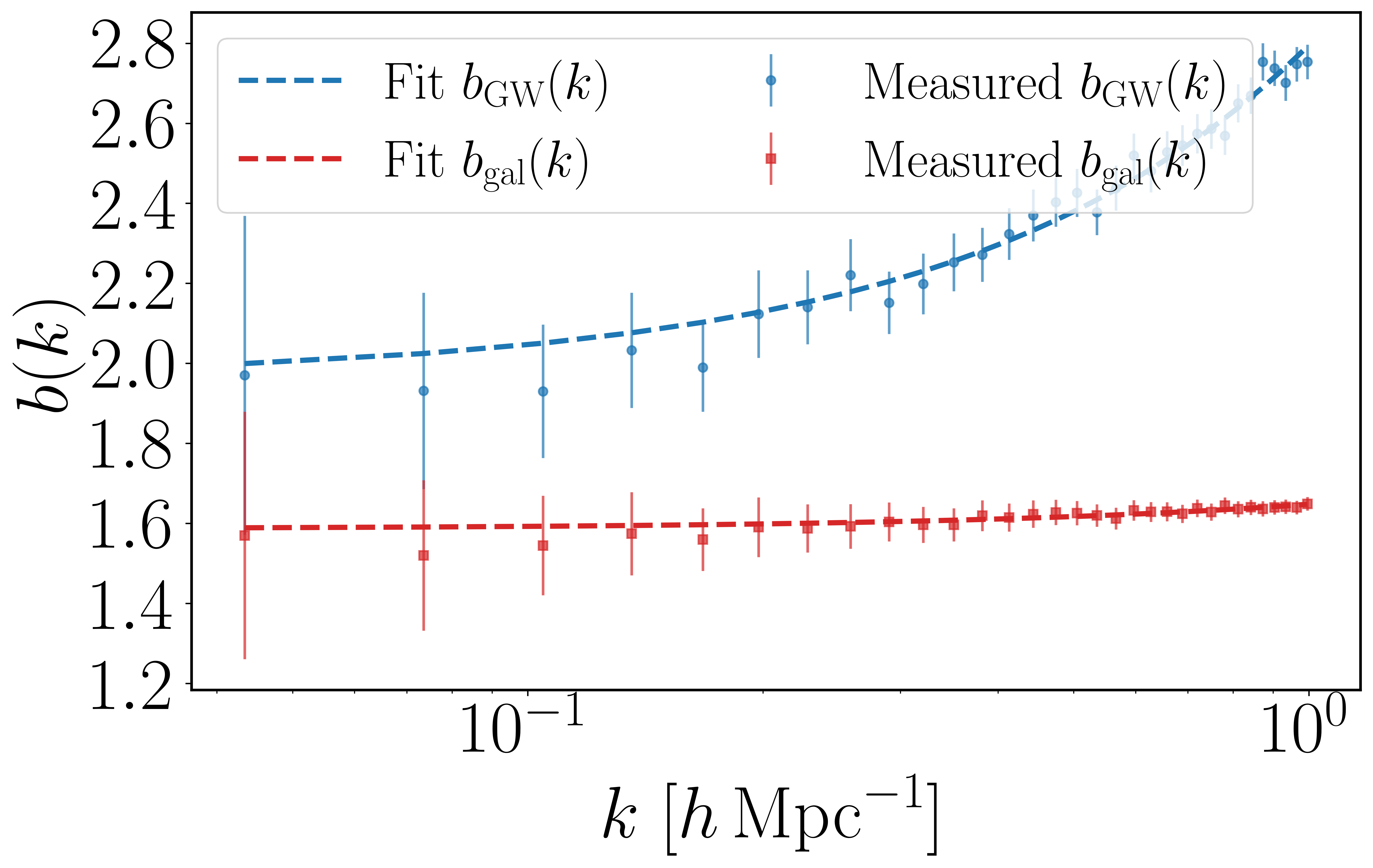}

        \label{fig:bgw_fit_0.95}
    \end{subfigure}

    \vspace{2mm}

    \begin{subfigure}[t]{0.48\linewidth}
        \centering
        {\small \ \ \ \ \ \ \
        $(\alpha_l,\,\alpha_h,\,M_\kappa) =
        (0.80,\,0.50,\,6.12 \times 10^{11} M_\odot)$
        \par}
        \vspace{2mm}
        
        \includegraphics[width=\linewidth]
        {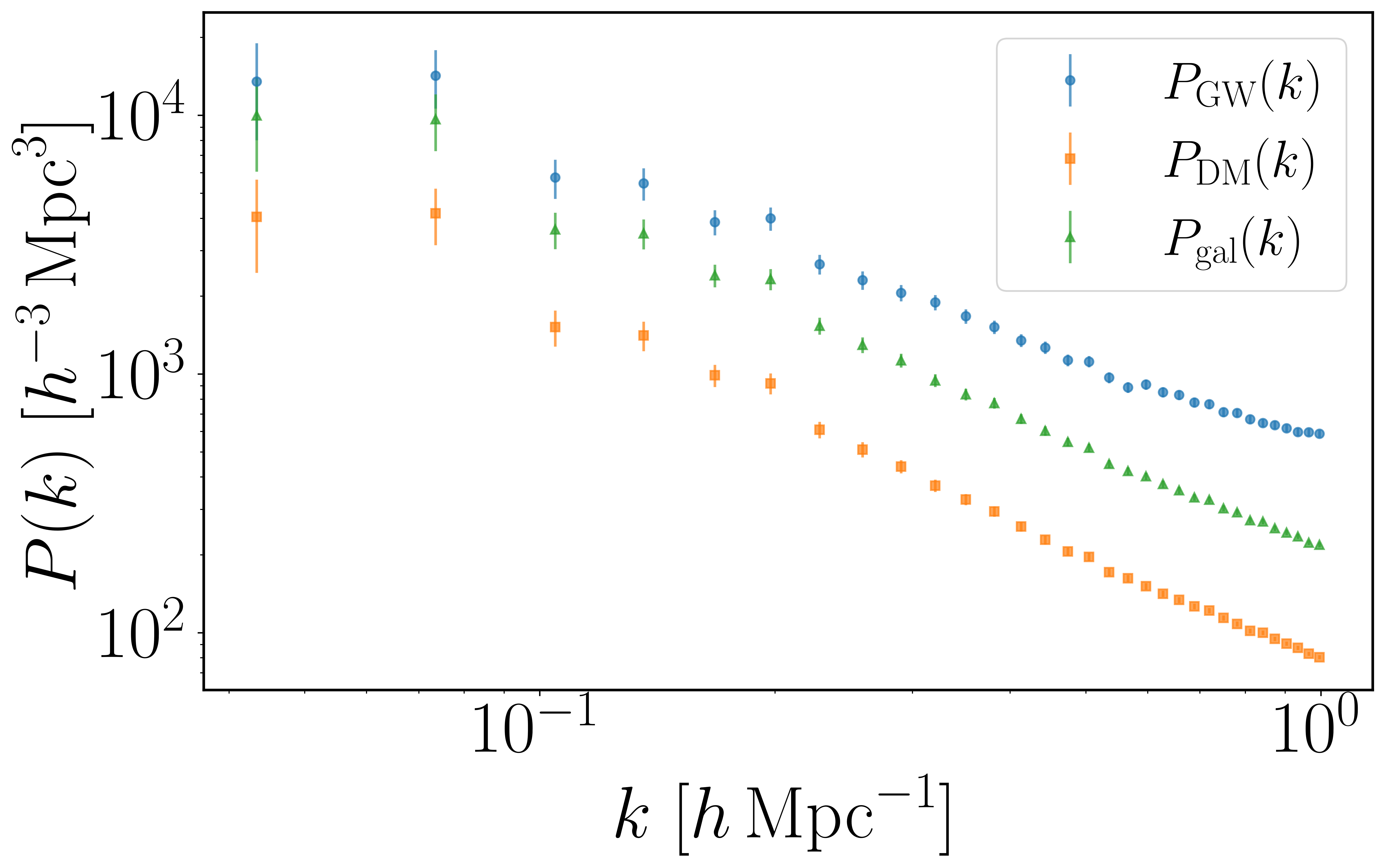}

        \label{fig:pspectra}
    \end{subfigure}
    \hfill
    \begin{subfigure}[t]{0.48\linewidth}
        \centering
        {\small \ \ \ \ \ \ \
        $(\alpha_l,\,\alpha_h,\,M_\kappa) =
        (0.80,\,0.50,\,6.12 \times 10^{11} M_\odot)$
        \par}
        \vspace{2mm}
        \includegraphics[width=\linewidth]
        {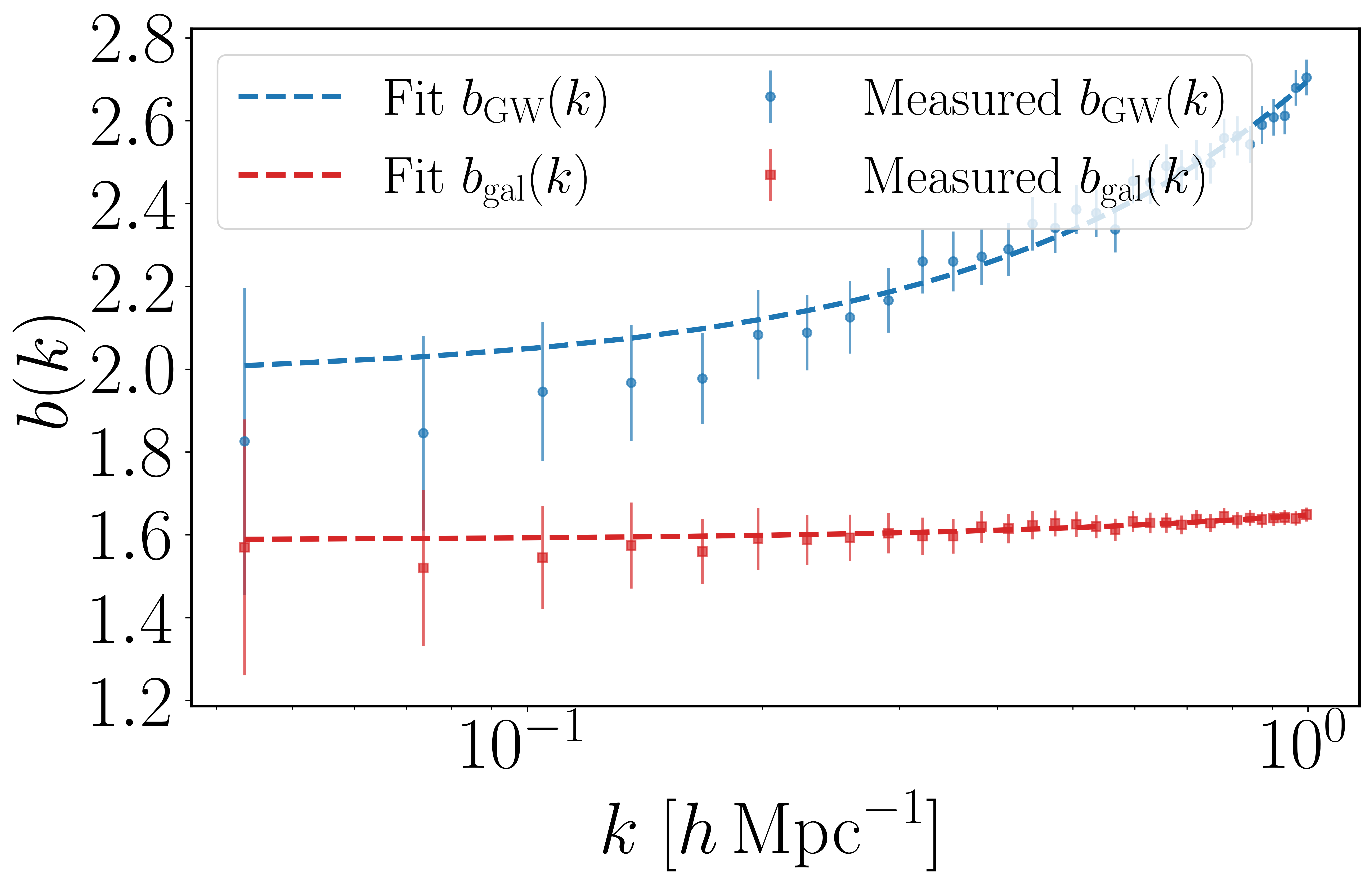}
        \label{fig:bgw_fit}
    \end{subfigure}

    \caption{
    Top panel shows the corresponding power spectra for GW sources, dark matter, and galaxies (left), and the best-fit measured GW and galaxy bias as a function of wavenumber (right) at redshift $z = 1.0$ with parameters $M_\kappa = 2.2 \times 10^{11} M_\odot$,
    $\alpha_l = 0.95$, and $\alpha_h = 0.20$. Lower panel shows the corresponding power spectra (left) and the
    best-fit inferred GW and galaxy bias (right) at $z = 1.0$ with parameters $M_\kappa = 6.12 \times 10^{11} M_\odot$, $\alpha_l = 0.80$, and $\alpha_h = 0.50$.}

    \label{fig:bgw_vs_k_fit}
\end{figure}

\subsection{Results} 

Following the procedure outlined above, we now present our findings. Since qualitative trends on how clustering depends on parameter choices $(\alpha_l,\,\alpha_h,\, M_\kappa)$ agree with previous findings in \cite{Dehghani_2025, Hosseini_2026}, we restrict ourselves to a few representative cases that highlight both shared features and differences with those results, as well as Section \ref{sec:gwbias_result_msfrz} that uses the emulator method.

We begin by displaying the clustering results for two specific examples at $z = 1.0$ corresponding to the parameter choices of 
$(\alpha_l,\,\alpha_h,\,M_\kappa) \in \{(0.95,\, 0.20,\, 2.20 \times 10^{11} M_\odot), \,(0.80,\,0.50,\,6.12 \times 10^{11} M_\odot) \}$ in Figure ~\ref{fig:bgw_vs_k_fit}. The first parameter set corresponds to the best-fitting values obtained from a broken power law model fitted to the median merger rate-stellar mass relation shown in the upper-left panel of Figure \ref{fig:mergerrate_vs_galprops}, and the second set is chosen by adopting an $\alpha_l$ close to the value reported in Table 3 (fit 1D for BBHs at $Z=1$) of \cite{Artale_2019}. Note that in that study, the model was fitted to a monotonic power law rather than a broken power law parameterization. However, as we will discuss later, in the high pivot--mass regime ($M_\kappa \gtrsim 2 \times 10^{11} \, M_\odot$), the inferred bias is not very sensitive to the precise values of $M_\kappa$ and consequently $\alpha_h$. 

The left panels in Figure \ref{fig:bgw_vs_k_fit} display the computed auto-power spectra of gravitational wave (GW) sources and galaxies, along with that of the underlying dark-matter field, while the right panels show the inferred galaxy and GW bias parameters. The GW clustering results obtained for these two specific samples are both numerically and qualitatively close ($(b_0, b_1)=(2.05,0.72)$ for upper panel and $(b_0, b_1)=(2.05,0.83)$ for lower panel). As expected, since the chosen value of pivot mass \(M_\kappa\) for both GW samples is relatively high, the seeding process preferentially selects more massive galaxies as hosts, resulting in enhanced clustering and higher power spectrum amplitudes for GW sirens compared to those of the galaxy and dark matter fields. The same trend is evident in the plots for the best bias values measured, which systematically show higher bias and scale dependence for GW sources compared to galaxies for these particular selections. These results are qualitatively consistent with the findings presented in Section \ref{sec:gwbias_result_msfrz} (see the right-hand plots of Figure \ref{fig:pk_allz} and Figure \ref{fig:scaledep_allz}). However, as we can see, the associated scale dependence is stronger in Section \ref{sec:gwbias_result_msfrz} than here, even though the mock GW sample used here is generated with a broken-power-law host-probability model whose parameters are calibrated from the projected merger rate stellar mass relation shown in the upper-left panel of Figure \ref{fig:mergerrate_vs_galprops}. This suggests that, although stellar mass is the primary driver of large scale clustering, it does not fully capture the relevant dependencies. Our machine learning model, which incorporates a joint probability based on all three galaxy properties, stellar mass, SFR, and metallicity, contributes to a stronger scale dependence. 

\begin{figure}[h]
    \centering
      \includegraphics[width=0.7\linewidth]{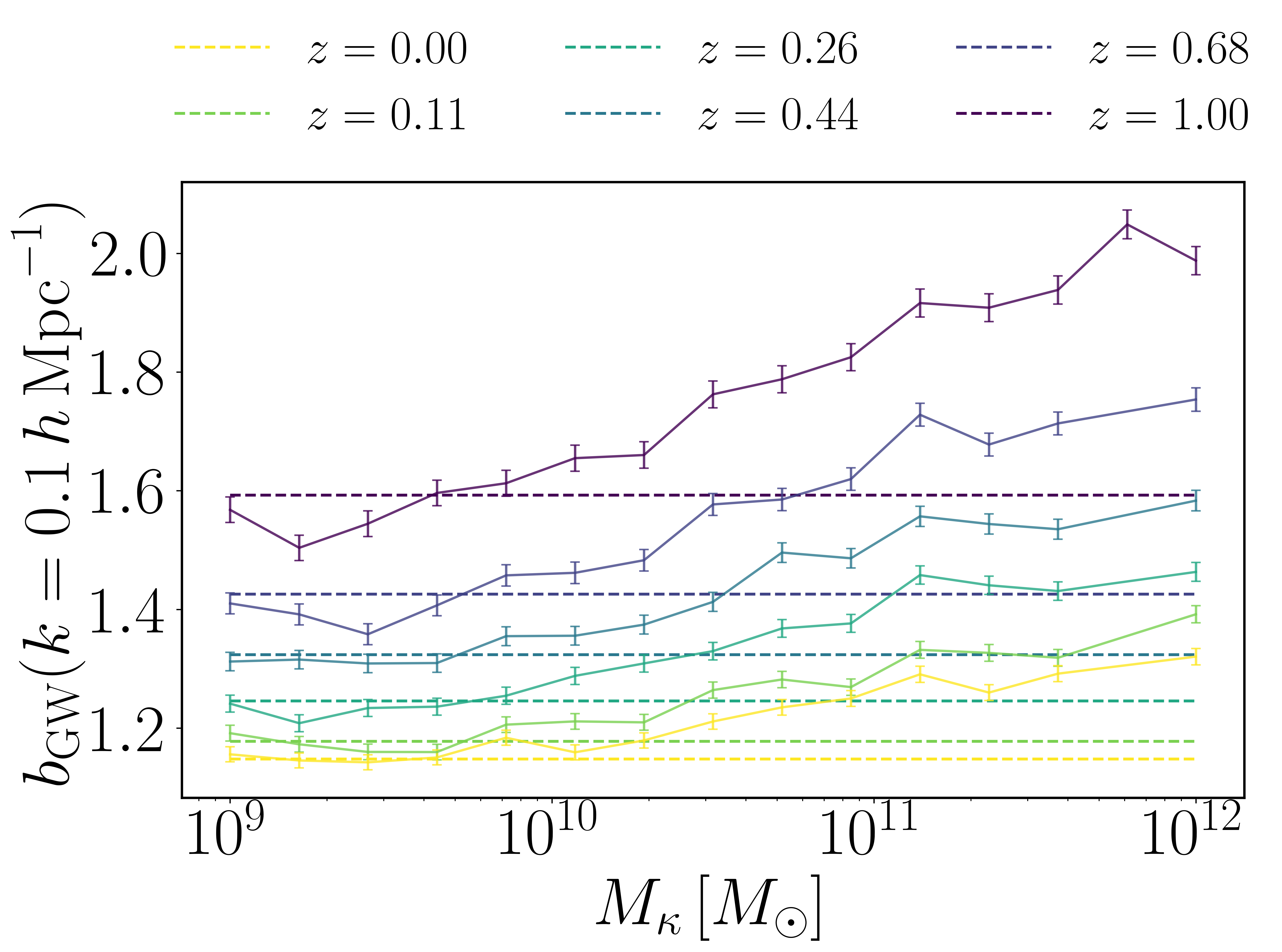}
  \caption{The points indicate the best fit GW bias as a function of pivot mass $M_\kappa$. The bias here is computed at $k = 0.1\,h\,\mathrm{Mpc}^{-1}$ and the fitting is done over $0.04 \leq k/(h\,\mathrm{Mpc}^{-1}) \leq 1$ range. The adopted parameters of the selection function are $\alpha_l = 0.80, \; \alpha_h = 0.50$. The dashed lines indicate the galaxy bias and error bars are $1\sigma$ Gaussian errors.}
  \label{fig:bgw-vs_MK}
\end{figure}

Next, Figure~\ref{fig:bgw-vs_MK} displays $b_{\rm GW}$ at $k = 0.1\,h\,\mathrm{Mpc}^{-1}$ as a function of the pivot mass $M_\kappa$ for several redshifts while keeping the slopes of the broken power law fixed at $(\alpha_l,\,\alpha_h) = (0.80,\,0.50)$. For comparison, the galaxy bias for the underlying sample from which GW events were drawn is also presented. The pivot mass $M_\kappa$ was sampled log-uniformly over the interval $[10^9,\,10^{12}]M_\odot$. As can be seen, the GW bias systematically increases with redshift. This rise is consistent with Figure \ref{fig:bias-vs-redshift}, and as discussed there, it is expected for tracer populations that are strongly correlated with stellar mass. Consequently, since GW events also trace galaxies, this results in a net rise in $b_{\rm GW}$ toward higher redshifts. However, the notable feature of this figure is that $b_{\rm GW}$ increases monotonically with pivot mass $M_\kappa$. This behavior is a direct consequence of the host-galaxy selection function: shifting the pivot mass to higher values preferentially selects more massive galaxies as GW hosts. Since more massive galaxies reside in more strongly biased dark matter halos, this selection naturally leads to an enhanced clustering amplitude and, consequently, a higher GW bias. Another notable aspect of this plot is that the GW bias converges to an approximately constant value for $M_\kappa \gtrsim 2\times 10^{11} M_\odot$. This feature arises because the number of galaxies in the galaxy samples drops significantly in this high mass range (Figure \ref{fig:galaxy_distributions_M_SFR_Z}), causing the GW bias to be very weakly sensitive to variations in the pivot mass. 

\begin{figure}[h]

    \begin{subfigure}[b]{0.5\textwidth}
    \centering
    \includegraphics[width=\textwidth]{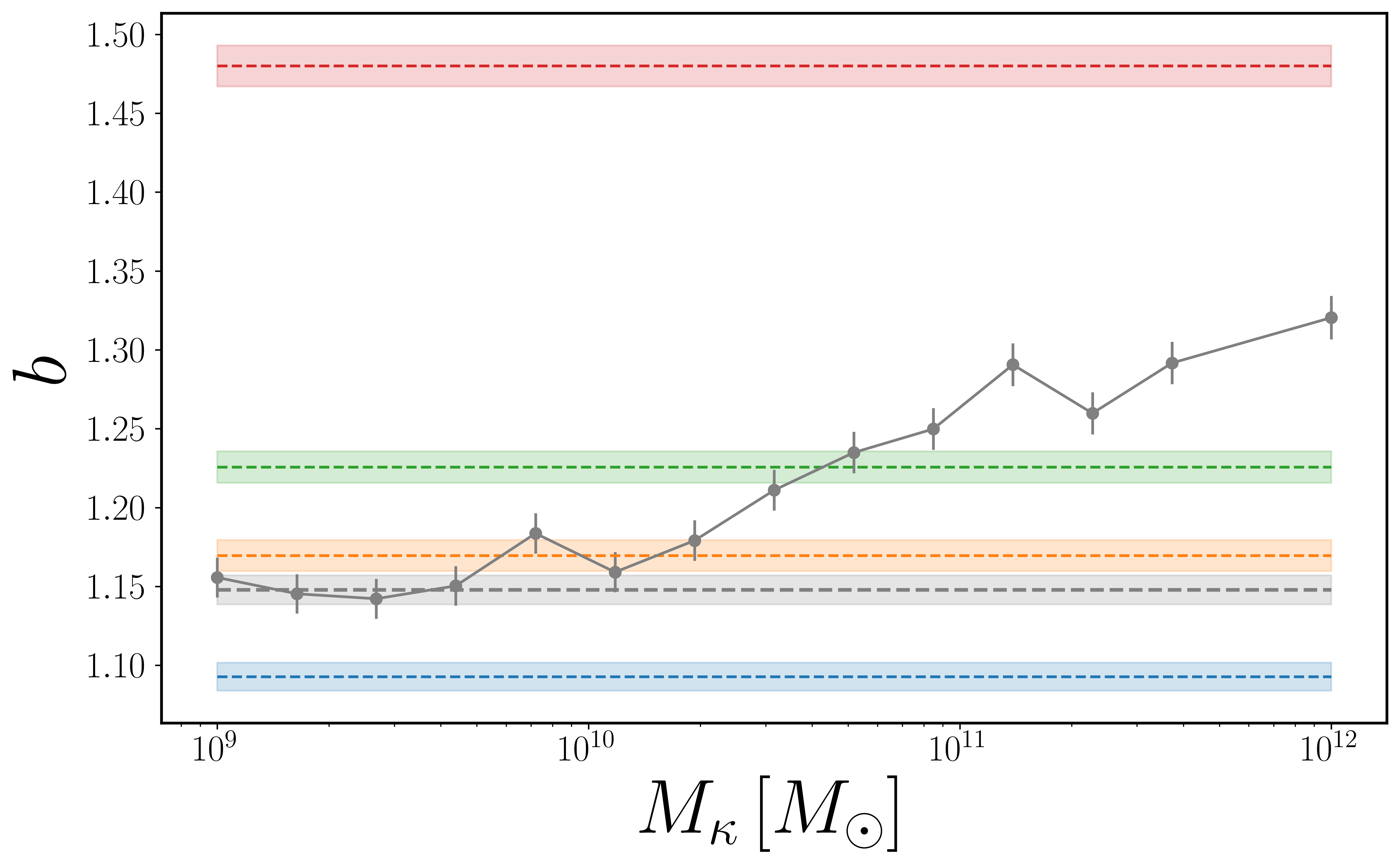}
    
    \caption{$z = 0$}
        
    \end{subfigure}
    \begin{subfigure}[b]{0.5\textwidth}
    \centering
        \includegraphics[width=\textwidth]{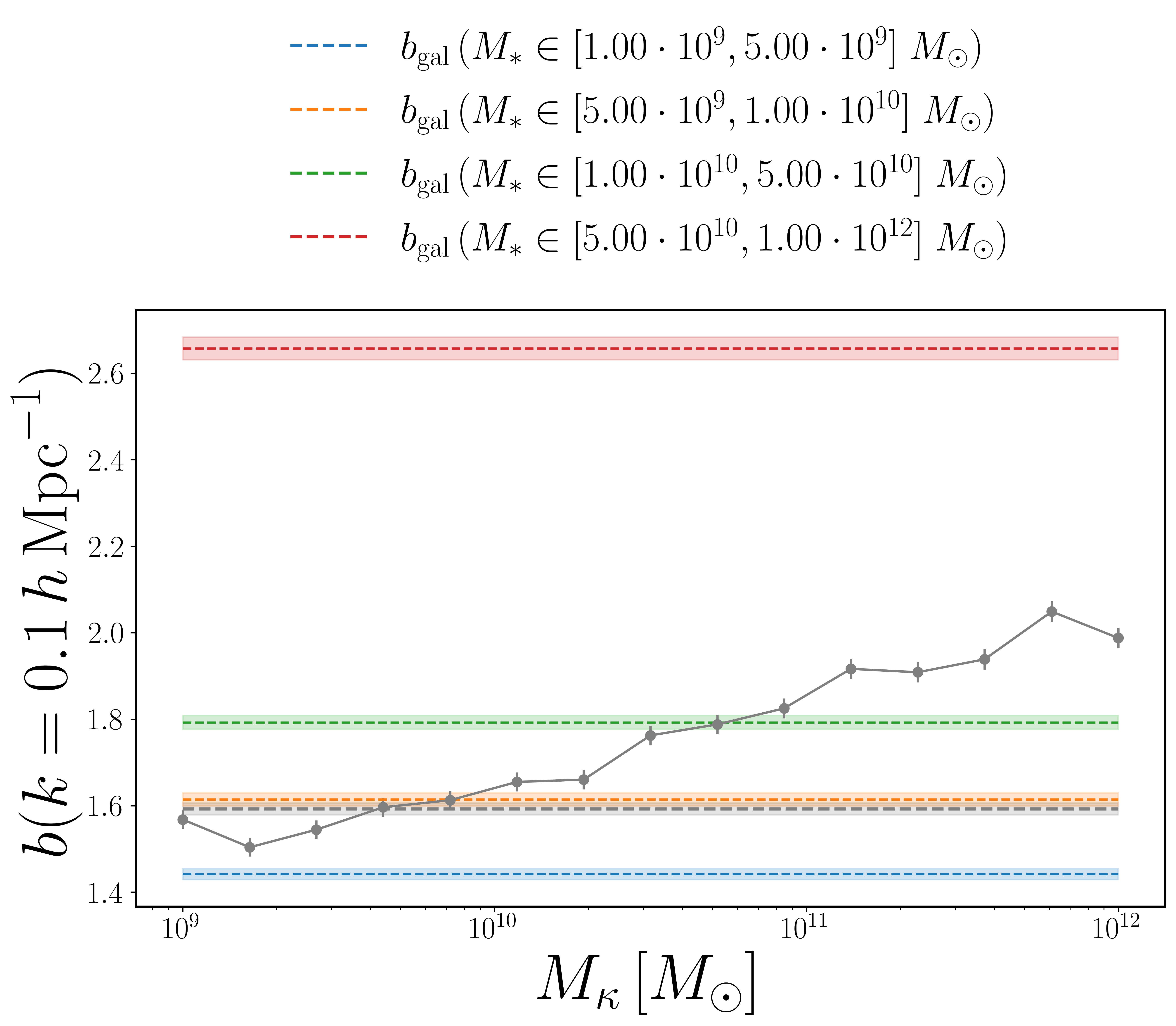}
    \caption{$z = 1$}
    \end{subfigure}
    \caption{Dashed lines show the galaxy bias for different equal--count mass bins. The corresponding bands represent the ranges of $1 \sigma$ errors. Black points and the solid line going through them represent the GW bias as a function of pivot mass $M_\kappa$. Here, the parameters of the selection function are set as $\alpha_l = 0.80, \; \alpha_h = 0.50$. }
    \label{fig:bias_gal_binned_and_gw_vs_pivot}
\end{figure}

\begin{figure}[h]
    \centering
    \includegraphics[width=0.8\textwidth]{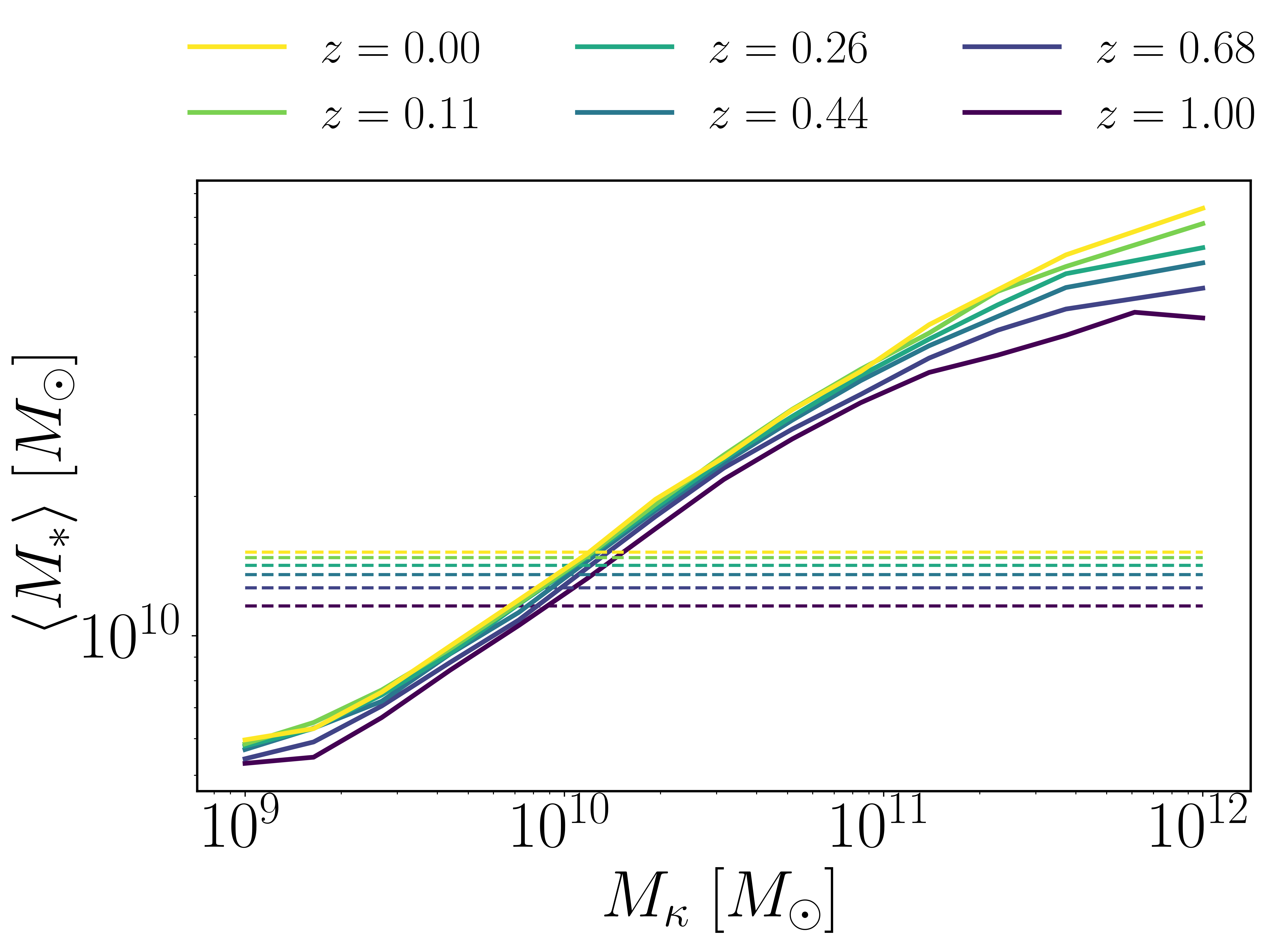}
    \caption{Solid lines represent the mean GW host-galaxy stellar mass, $\langle M_\ast\rangle_{\rm GW}$, as a function of the pivot mass $M_\kappa$ for different redshifts. The dashed lines indicate the mean stellar mass of all galaxies, $\langle M_\ast\rangle_{\rm g}$, at the corresponding redshift. Here, the adopted parameters of the selection function are $\alpha_l = 0.80, \; \alpha_h = 0.50$. }
    \label{fig:avgmass}
\end{figure}

The details of where $b_{\rm GW}$ crosses the galaxy bias, as well as its limiting behavior at low and high pivot masses, are determined by the nontrivial interplay between the slopes of the selection function and the underlying stellar mass distribution of the galaxy sample. To better understand the impact of the stellar mass distribution of the galaxy sample on the GW bias, it is useful to know how the galaxy bias itself varies with stellar mass. Figure~\ref{fig:bias_gal_binned_and_gw_vs_pivot} shows both the dependence of $b_{\rm GW}$ on the pivot mass $M_\kappa$ for the same set of slopes as the previous case ($\alpha_l = 0.80$ and $\alpha_h = 0.50$), together with the galaxy bias in several stellar mass ranges, for redshift bins $z=0$ and $z=1$. The plots show that at low pivot masses, $b_{\rm GW}$ does not reach the galaxy bias associated with the lowest stellar mass bin (blue), and at high pivot masses, it also does not rise up to the galaxy bias of the highest--mass galaxy range (red). Furthermore, the bias of the full galaxy sample (dashed black) lies just below but close to that of the second stellar mass bin (orange), $5 \times 10^9\,M_\odot < M_\star < 10^{10}\,M_\odot$, and the crossover between $b_{\rm GW}$ and the galaxy bias occurs when the pivot mass lies within the same interval. This suggests that the majority of galaxies in the sample reside in the mass range $[5\times10^9,\,10^{10}]\,M_\odot$. Consequently, when the pivot mass lies in this interval, the dominant GW hosts occupy the same stellar mass range, and the corresponding biases coincide. We also see that for $M_\kappa > 10^{11}\,M_\odot$, $b_{\rm GW}$ approaches a value above the third stellar mass bin (green) corresponding to $10^{10}\,M_\odot < M_\star < 5\times10^{10}\,M_\odot$). This implies that at larger pivot masses, the mean stellar mass of the GW host galaxies saturates above $5 \times 10^{10}\,M_\odot$ but well below $10^{12} \,M_\odot$, leading to the asymptotic behavior (right tails of the curves) observed in Figure~\ref{fig:bias_gal_binned_and_gw_vs_pivot}.
Similarly, at low pivot masses, the mean stellar mass of the host galaxies of GW must be saturating somewhere above $5\times10^{9}\,M_\odot$ but close to or slightly below the average stellar mass of the sample. Thus, the GW bias saturates above the blue range and is closer to the dashed black (and orange) line. This interpretation is further supported by Figure~\ref{fig:avgmass}, which shows the dependence of the average stellar mass of the GW hosts, $\langle M_\ast\rangle_{\rm GW}$, as the pivot mass $M_\kappa$ (solid lines) varies together with the mean stellar masses of the full galaxy sample, $\langle M_\ast\rangle_{\rm g}$ (dashed lines). We find that when the pivot mass moves below the mean galaxy stellar mass, the mean GW host stellar mass always remains above $5\times10^{9} M_\odot$ within the mass range of the orange band in Figure~\ref{fig:bias_gal_binned_and_gw_vs_pivot}. It also shows that $\langle M_\ast\rangle_{\rm GW}$ coincides with the mean stellar mass of the full galaxy sample when the pivot mass falls within $[5\times10^9,\,10^{10}]\,M_\odot$. At larger pivot masses, $\langle M_\ast\rangle_{\rm GW}$ converges to a value in the range 
$[5\times10^{10},\,10^{11}]\,M_\odot$, consistent with $b_{\rm GW}$ exceeding 
$b_{\rm gal}$ for galaxies with stellar masses in $[10^{10},\,5\times10^{10}]\,M_\odot$ (green band in Figure~\ref{fig:bias_gal_binned_and_gw_vs_pivot}). At the same time, the fact 
that $\langle M_\ast\rangle_{\rm GW}$ remains well below $10^{12}\,M_\odot$ 
could account for the asymptotic behavior in which $b_{\rm GW}$ stays below the red 
curve in Figure~\ref{fig:bias_gal_binned_and_gw_vs_pivot}, in agreement with 
previous findings~\cite{Dehghani_2025, Hosseini_2026}.

\begin{figure}[h]
    \begin{subfigure}[b]{0.5\textwidth}
    \centering
        \includegraphics[width=\textwidth]{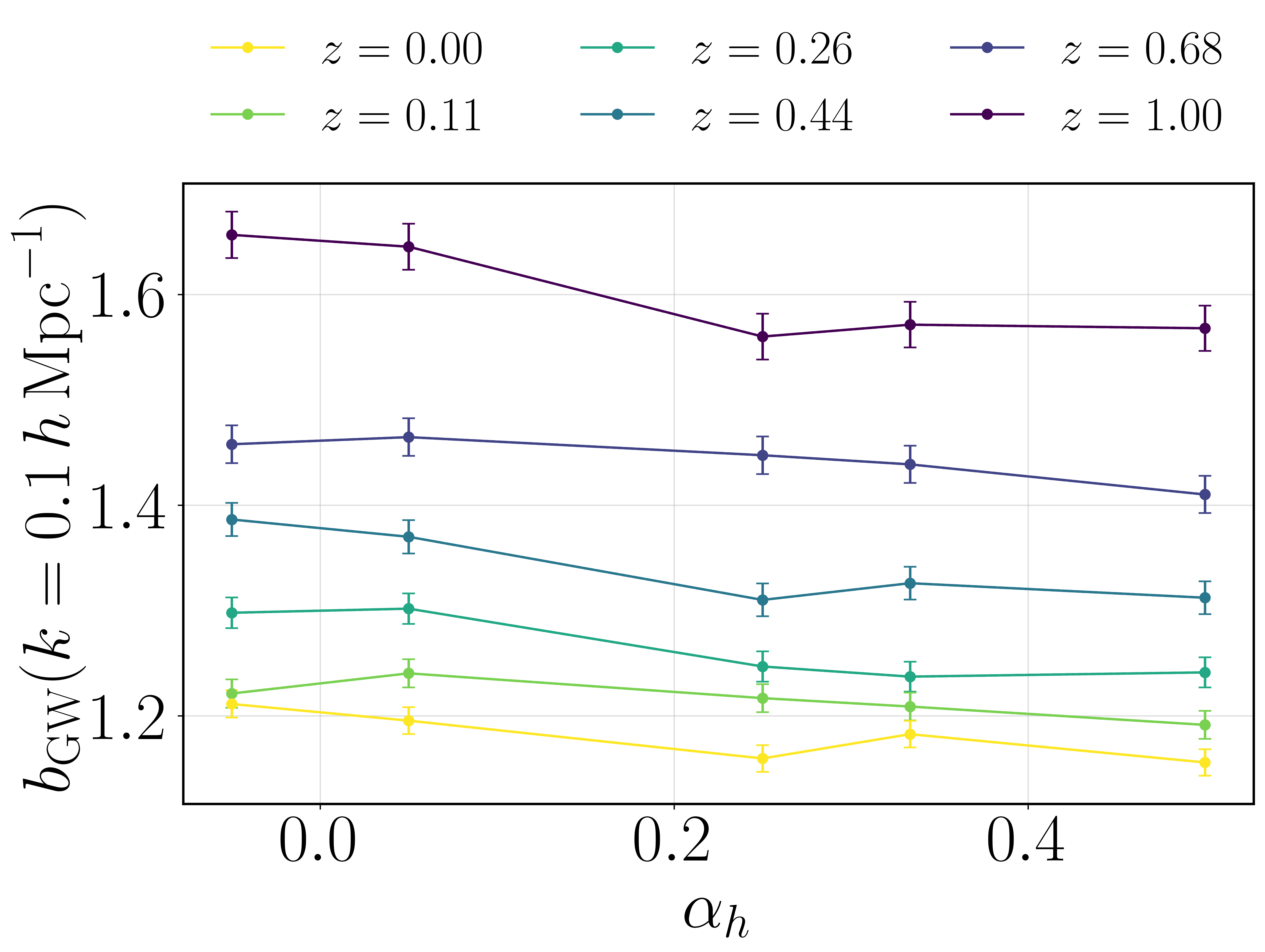}
    \caption{$M_{\kappa} = 10^9 M_\odot$}
    \end{subfigure}
    \begin{subfigure}[b]{0.5\textwidth}
    \centering
        \includegraphics[width=\textwidth]{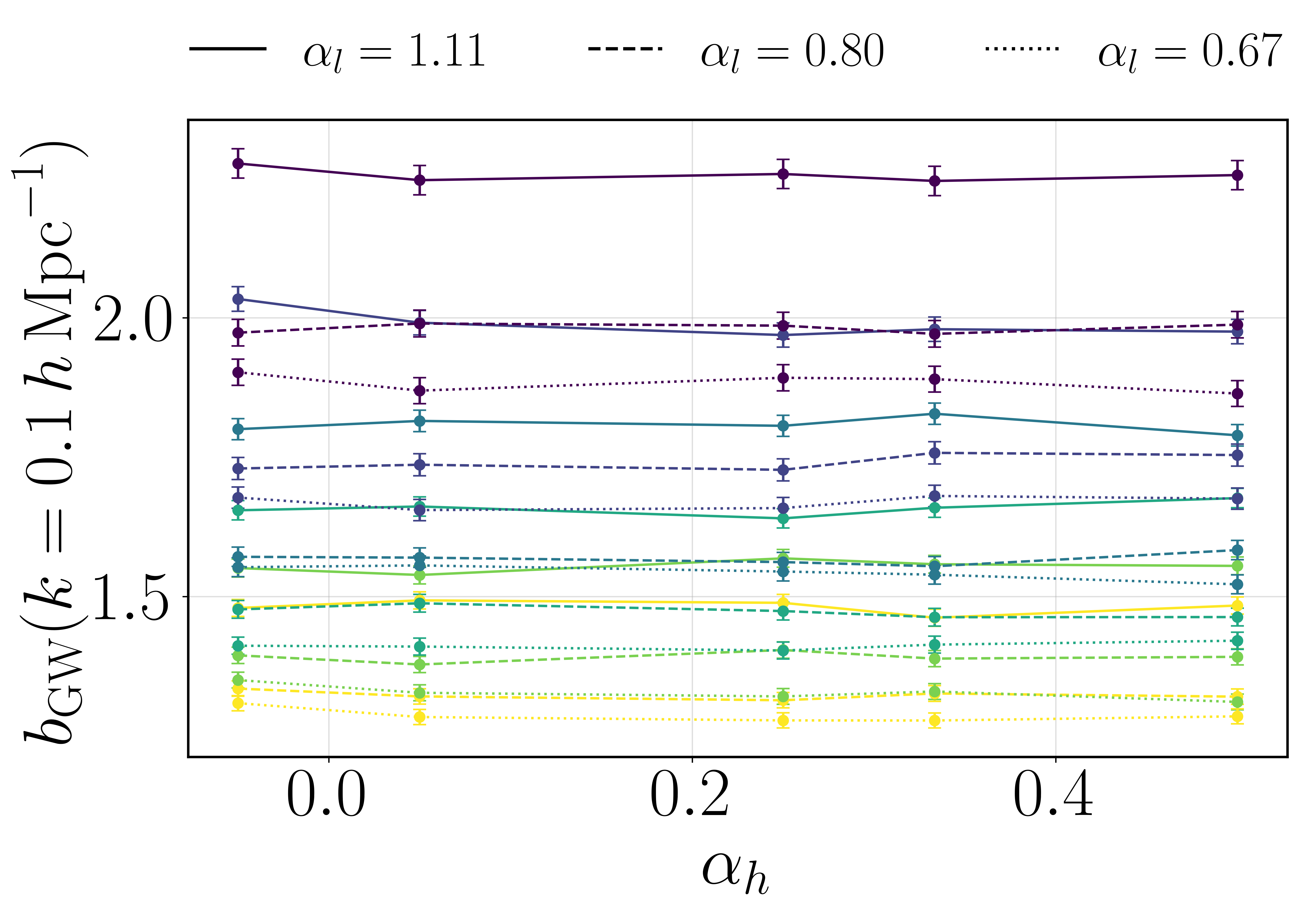}
    \caption{$M_{\kappa} = 10^{12} M_\odot$}
    \end{subfigure}
    \caption{GW bias as a function of the slope parameters at fixed values of the pivot mass, $M_\kappa = 10^9\,M_\odot$ (left panel) and $10^{12}\,M_\odot$ (right panel), for several redshifts. The slope on the low-mass side of the power law is chosen as $\alpha_l \in \{0.67,\, 0.80,\, 1.11\}$, corresponding to progressively steeper, approximately linear increases. The slope of the high-mass segment $\alpha_h$, defined with reversed sign, is varied from slightly below zero, $\alpha_h = -0.05$ (nearly flat but weakly rising), up to $\alpha_h = 0.50$ (corresponding to a more pronounced decline).}
\label{fig:bgw-dh}
\end{figure}

Next, in Figure~\ref{fig:bgw-dh} we examine how the GW bias changes when the pivot mass $M_\kappa$ is fixed at its two limiting values while the slopes $\alpha_l$ and $\alpha_h$ vary. The left panel of Figure~\ref{fig:bgw-dh} shows the dependence of the GW bias on $\alpha_h$ for $M_\kappa = 10^9\,M_\odot$, while the right panel shows the corresponding dependence for $M_\kappa = 10^{12}\,M_\odot$. In both panels, different colors correspond to different redshifts, and different line styles denote different values of $\alpha_l$. Since these pivot masses lie at the lower and upper boundaries of the galaxy stellar mass range, \footnote{Aside from a small population of extremely massive galaxies, nearly the entire allowed stellar mass range falls within $10^{9}\,M_\odot < M_\ast < 10^{12}\,M_\odot$.} each panel effectively probes the behavior of only one branch of the broken power law selection function. This is clearly reflected in the plots as well. In the left panel, where $M_\kappa = 10^9\,M_\odot$, the low-mass slope $\alpha_l$ has essentially no impact on host selection, and consequently, the curves corresponding to different line styles completely overlap. In contrast, in the right panel, where $M_\kappa = 10^{12}\,M_\odot$ is located, the high-mass slope $\alpha_h$ no longer affects the selection function, and the GW bias therefore remains nearly unchanged as $\alpha_h$ varies. Aside from these points, another expected behavior when varying the slopes of the GW host-galaxy selection function is that amplifying high-mass weighting increases the bias, while low-mass weighting reduces it. For example, we observe that at all redshifts, the GW bias is higher in the right panel ($M_\kappa = 10^{12}\,M_\odot$) than in the left panel ($M_\kappa = 10^{9}\,M_\odot$). We also observe that for low $M_\kappa$ (i.e., in the left panel, where the knee lies below the median galaxy stellar mass $M_\star$), increasing $\alpha_h$, corresponding to a suppression of more massive host galaxies, causes the bias to decrease monotonically at all redshifts. On the other hand, in the right panel for high $M_\kappa$ (i.e., when the knee lies above the median galaxy stellar mass $M_\ast$) and $\alpha_l$ governs the host selection, enhancing the contribution of high-mass galaxies (larger $\alpha_l$) increases $b_{\rm GW}$. Consequently, the curve corresponding to $\alpha_l = 1.11$ lies highest, while that with $\alpha_l = 0.67$ lies lowest at all redshifts. 

Another interesting feature of the right panel of Figure~\ref{fig:bgw-dh} is that it once again allows for a direct comparison with the non-analytical 3D population- synthesis based model discussed in the previous section, where additional galaxy properties, such as star formation rate and metallicity, also affect the host probability. 

As can be seen here, at $z=1$, when the stellar mass weighting is relatively high by taking a large pivot mass $M_\kappa = 10^{12}\,M_\odot$ together with $\alpha_l = 1.11$, the resulting GW bias rises to $b_{\rm GW}\gtrsim 2.2$, showing an enhancement of about $40\%$ relative to $b_g$. This is higher than the value $b_{\rm GW} \sim 2.07$ (an enhancement of about $30\%$) obtained in 
Figure~\ref{fig:bias-vs-redshift}. However, as we see by lowering 
$\alpha_l\sim 0.80$, $b_{\rm GW}$ also approaches the value that the emulator produces. Naively, if our emulator framework were driven solely by the stellar mass-merger rate relation when selecting host galaxies, then the best-fit broken power law to the merger rate relation in Figure~\ref{fig:mergerrate_vs_galprops} would correspond to $(M_\kappa, \alpha_l) = (6.12 \times 10^{11}\,M_\odot,\,0.80)$, which should theoretically imply an even weaker weighting toward high-mass galaxies since it has a lower $M_\kappa$; but as we mentioned above, beyond some $M_\kappa$ threshold the GW bias becomes insensitive to that variation, as the number of very massive galaxies becomes increasingly small. So just setting a high value of $M_\kappa$ and adjusting for $\alpha_l$ reproduces the bias amplitude. A similar GW bias result is also visible at $z=0$ for $\alpha_l\sim 0.80$. In this case, the analytical stellar-mass-only model shown in Figure~\ref{fig:bgw-dh} yields $b_{\rm GW}\sim 1.3$--$1.4$ ($b_{\rm GW}/b_g\sim 20\%$), consistent with Figures~\ref{fig:bias-vs-redshift} and~\ref{fig:bias_sfr_bins}.

These findings provide further evidence that stellar mass is a strong factor in determining the amplitude of the bias, although the scale dependence of clustering, as mentioned before, may not be fully determined by stellar mass alone. Additional galaxy properties such as SFR and metallicity further influence the host selection through more complex joint probability distributions than simple power laws, leading to further enhancement or suppression of the GW bias, yet at leading order a simple analytical model can already go a long way. Ultimately, frameworks such as the one introduced in Section~\ref{sec:gwbias_result_msfrz}, which explicitly incorporate these multi-dimensional dependencies, offer a more robust and informative tool for future investigations in this field.

\subsection{Impact of choice of $k_{\textrm{max}}$}

\begin{figure}[h]
  \centering
    \includegraphics[width=0.8\textwidth]{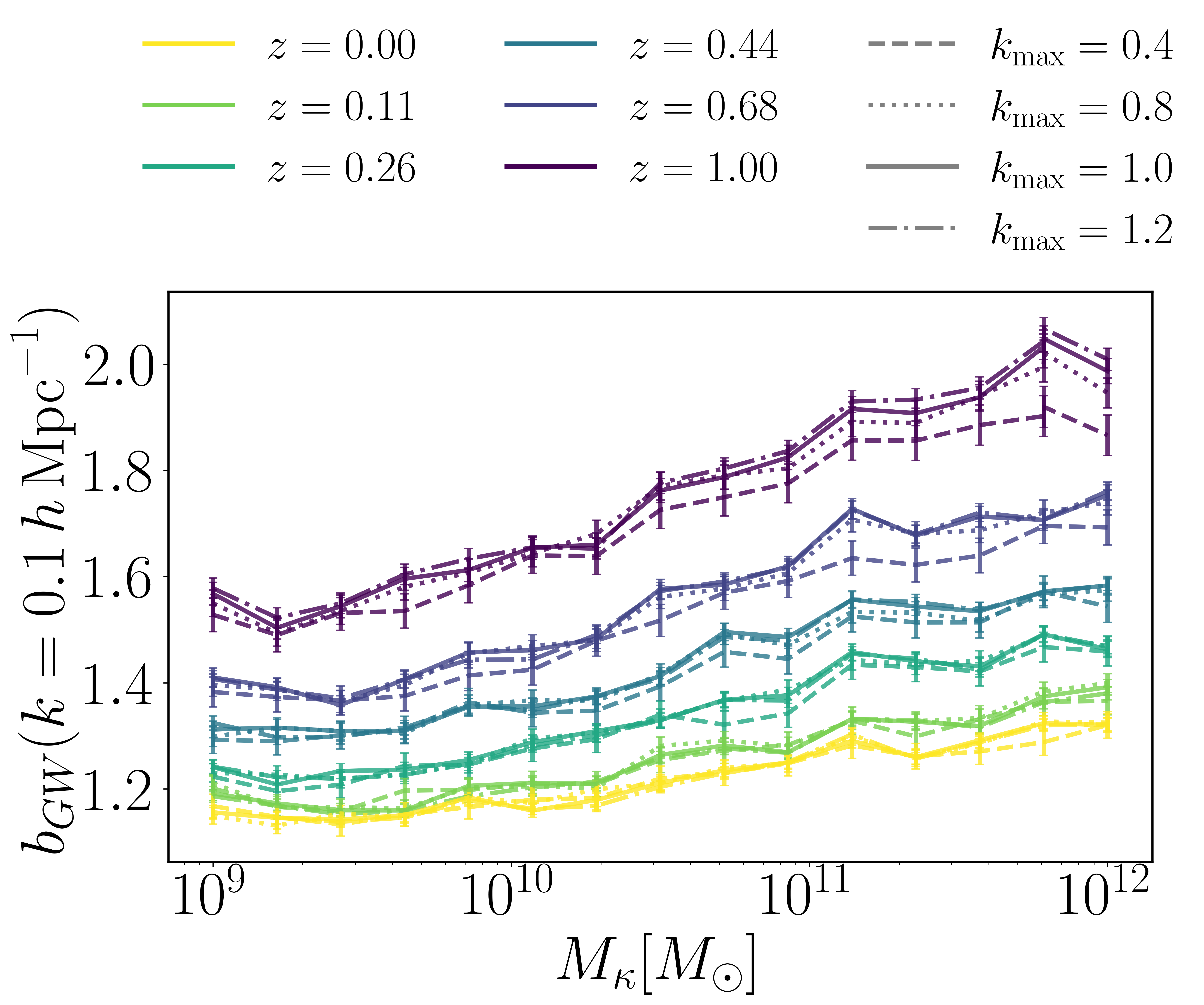}
  \caption{GW bias, $b_{\rm GW}$ is shown as a function of the pivot mass $M_\kappa$ for several choices of the maximum wavenumber $k_{\rm max}$ used in the fit, with $0.04\,h\,\mathrm{Mpc}^{-1} < k < k_{\rm max}$, and for different redshifts. The selection function parameters are fixed to $\alpha_l = 0.80$ and $\alpha_h = 0.50$.}
  \label{fig:bgw-vs_kmax}
\end{figure}

As we mentioned earlier in the plots and results discussed here, the GW bias reported was calculated using Equation \ref{eq:P_tt}, where the best-fit bias parameters $b_0$ and $b_1$ were determined in the range $0.04 < k < 1.0 \, h\,\mathrm{Mpc}^{-1}$. Therefore, we also examined how the choice of the upper bound of the fitting range, $k_{\rm max}$, may affect the amplitude $b_{\rm GW}$, comparing to Fig.\ 15 in \cite{Dehghani_2025}. Figure~\ref{fig:bgw-vs_kmax} shows the dependence of the GW bias on the pivot mass $M_\kappa$ for several values of $k_{\rm max}$. For each combination of $M_\kappa$ and $k_{\rm max}$, the bias is calculated for $\alpha_l = 0.80$ and $\alpha_h = 0.50$. We found that, at all choices of redshifts and pivot masses, $b_{\rm GW}$ shows a slight increase as $k_{\rm max}$ is raised. While the increase is not significant, it is more noticeable for higher values of $M_\kappa$. This is expected because extending the fit to higher $k$ includes increasingly nonlinear scales, where scale-dependent bias becomes more important. The increase is more pronounced for larger $M_\kappa$, as higher pivot masses give more weight to massive galaxies that reside in highly biased haloes. Consequently, in this regime, the contribution of these strongly clustered hosts on small scales is higher, leading to a larger inferred GW bias when higher-$k$ modes are included in the fit.
This behavior is also consistent with Figure~15 of \cite{Dehghani_2025}: increasing $k_{\rm max}$ does not significantly modify the qualitative dependence of $b_{\rm GW}$ on $M_\kappa$ but it can lead to small changes in its amplitude. Here, we can also observe the corrections at higher redshifts that were not visible in Figure~15 of \cite{Dehghani_2025}, since that analysis was performed using an observed survey limited to $z \leq 0.4$. In particular, we see that at higher redshifts the differences in $b_{\rm GW}$ between different values of $k_{\rm max}$ become more pronounced. This behavior may reflect the fact that at higher redshift, the host galaxies are more strongly biased tracers of the matter field. Consequently, nonlinear and scale-dependent bias effects become more significant, making the inferred $b_{\rm GW}$ more sensitive to the choice of $k_{\rm max}$. 

\section{Conclusion}
\label{sec:conclusion}
In this work, we developed a simulation-based framework to model the clustering of gravitational wave (GW) sources and to estimate the GW bias and its dependence on host-galaxy astrophysical properties, enabling a self-consistent connection between stellar binary black hole (BBH) merger distributions, galaxy evolution, and large scale structure. This study complements our previous analysis based on seeding GW events into observed galaxy surveys, where we generated mock BBH GW catalogs using phenomenological host-selection functions, initially depending only on stellar mass \cite{Dehghani_2025}, and later extending it to include star formation rate and metallicity \cite{Hosseini_2026} as well. While observational surveys have the advantage of naturally incorporating the astrophysical processes that link large scale structure to galaxy properties, providing important insights into how these connections can affect tracer clustering, they are inevitably limited by the incompleteness and observational selection effects of the surveys. This motivated the present study, where we applied two complementary GW seeding approaches to the cosmological hydrodynamical simulation IllustrisTNG300 \cite{2015A&C....13...12N,Pillepich_2017,Springel_2017,Nelson_2017,Naiman_2018,Marinacci_2018} to construct catalogs for GW events. We then analyzed the GW bias for these catalogs within a physically self-consistent setting using our dedicated bias calculation framework.

In the first approach (section \ref{sec:gwbias_result_msfrz}), we trained a machine-learning emulator on \\ \GalR data \cite{Santoliquido_2022}, which are obtained by seeding compact binary systems from the MOBSE population-synthesis code \cite{10.1093/mnras/sty1999} into mock galaxy samples based on merger trees from the EAGLE cosmological simulation. We then applied the trained emulator to the IllustrisTNG300 simulation to predict galaxy-specific BBH merger rates as functions of stellar mass, star formation rate, metallicity, and redshift. These predicted rates were then used to  populate the full three-dimensional simulation volume at multiple redshifts, generating mock GW catalogs and providing the basis for the derivation of a physically motivated description of GW source clustering that incorporates galaxy evolution and compact-binary formation physics and is not affected by observational selection effects.

In the second approach (section \ref{sec:gwbiasresult_mass_powerlaw}), we applied a broken power law phenomenological stellar mass dependent host probability functions introduced in \cite{Dehghani_2025} to assign GW events to individual galaxies in IllustrisTNG300 simulation and construct corresponding GW mock catalogs. This allowed us to not only compare our results with those based on observed galaxy surveys, but also to directly contrast a purely mass-based prescription with a fully astrophysics-informed, machine-learning driven merger rate framework described in the first approach, all within the same simulated large scale structure. This way, we could also check for the additional influence of other host-galaxy properties correlated with stellar mass on the inferred GW bias.

Our results are consistent with the hypothesis that stellar mass is the dominant driver of GW clustering, but they also indicate that it does not fully determine the GW bias.
By comparing the 3-dimensional machine-learned merger rate framework and the phenomenological stellar mass dependent host-selection framework (with parameters adjusted to be close to those of population-synthesis models and within the same simulated large scale structure), we find that both approaches predict qualitatively and quantitatively similar trends in GW clustering while exhibiting some differences in the scale dependence of the inferred bias. In particular, although both frameworks preferentially select GW hosts in more massive galaxies, the 3D population-synthesis based machine-learned merger rate model generally predicts a larger GW bias on small scales than can be reproduced by stellar mass weighting alone (see, for example, the right plot in Figure \ref{fig:bias_mass_bins} versus the right plots in Figure \ref{fig:bgw_vs_k_fit} for $z=1$). 
This indicates that additional host-galaxy properties and their correlations with the environment contribute to the clustering of GW sources beyond what is captured by a purely mass-based prescription. Nevertheless, both prescriptions predict that the GW bias is more enhanced than the galaxy bias, which reflects the fact that GW events in these models were preferentially seeded in higher stellar mass galaxies, which reside in more highly biased halos. In other words, stellar mass serves as an effective proxy for host halo mass, and more massive halos are intrinsically more biased tracers of the density field. This behavior is consistent with expectations from hierarchical structure formation and agrees closely with our previous studies using observational photometric surveys as the base for host galaxies in \cite{Dehghani_2025} and subsequent spectroscopic analysis \cite{Hosseini_2026}. There is also a steepening of the mass-bias relation with redshift (see Figure \ref{fig:bgw-vs-mstar-allz}), which arises because massive systems become increasingly rare and strongly biased at earlier cosmic times, leading to higher GW bias for the highest-mass hosts. The persistence of enhancement and this redshift evolution across both phenomenological and population-synthesis driven merger rate models highlights the promise of GW bias as a tracer of the evolving large scale structure. 

We also investigated the explicit dependence of GW and galaxy bias on the star formation rate in our TNG300 mock catalogs. We found that non-star-forming galaxies exhibit substantially higher bias than star-forming systems, as well as the GW hosts (Figure~\ref{fig:bias_sfr_bins}). Furthermore we found that within the star forming galaxies, as well as GW hosts (in the 3D machine-learned selection model), there are relatively weaker clustering variations as SFR varies, especially at low redshifts (Figure \ref{fig:bgw-sfr-allz} and \ref{fig:bias_sfr_bins}). 

We performed a similar analysis with respect to metallicity and found that stellar metallicity exhibits a noticeable correlation with both galaxy and GW bias in the 3D model. $b_{\rm g}$ and $b_{\rm GW}$ both increase monotonically with metallicity at all redshifts (Figure \ref{fig:bgw-metal-allz} and \ref{fig:bias_metallicity_bins}), with the trend becoming steeper and shifting to higher bias toward $z=1$. Metal-rich galaxies reach significantly higher bias compared to lower-metallicity systems. This behavior reflects the strong correlation between metallicity, stellar mass, and assembly history in IllustrisTNG300: metal-rich galaxies are typically early-forming systems residing in more massive halos and highly clustered environments. However, we also observe in the same metallicity bands $b_{\rm GW}>b_{\rm g}$ and a difference in scale dependence, supporting the idea that the merger rate is not just linearly and directly conditioned on metallicity but rather has a non-trivial dependence on the three properties (as seen in Figure \ref{fig:mergerrate_vs_galprops}) that effectively select GW sources in progressively more biased halos. This metallicity dependence was not clearly detected in our earlier SDSS-based analysis \cite{Hosseini_2026}, likely due to the limitations of the uncertainties in observational metallicity estimates. By contrast, the present simulation-based framework provides a more self-consistent check on merger rate predictions, allowing the metallicity effect to emerge naturally. Our results, therefore, highlight metallicity as a physically meaningful secondary tracer of clustering. 

Taken together, these findings establish a consistent picture of bias enhancement for GW sources and more generally for any tracer in which host stellar mass provides the dominant contribution, while also showing that additional galaxy properties, particularly metallicity, contribute secondary effects not fully 
captured by stellar-mass-only models. From a cosmological perspective, these results emphasize that future GW large scale structure analyses must carefully model not only host stellar mass but also additional galaxy properties that 
correlate with the environment and merger probability. As next-generation detectors 
such as ET and CE deliver vast GW catalogs, incorporating physically motivated merger rate models into GW bias frameworks will be essential for extracting reliable cosmological and astrophysical information from GW clustering. The 
pipeline developed here provides a foundation for such efforts, enabling forward modeling of GW populations directly from galaxy formation physics and offering a pathway toward precision GW cosmology.

\section*{Acknowledgments}
We express our gratitude to Illustris TNG300 for providing data. We are especially grateful to Filippo Santoliquido for generously providing the \GalR data. We also thank M. Celeste Artale, Suvodip Mukherjee, Amir Dehghani and J. Leo Kim for insightful discussions. The computational aspects of this research were executed on the computing clusters at the Perimeter Institute (PI) for Theoretical Physics with support from the Canada Foundation for Innovation. In this work, several Python packages have been utilized, namely: \verb|Pypower|\footnote{\url{https://github.com/cosmodesi/pypower}}, \verb|Pylians|\footnote{\url{https://pylians3.readthedocs.io/en/master/}}\cite{Pylians}, \verb|Astropy| \footnote{\url{http://www.astropy.org}}\cite{astropy:2013, astropy:2018, astropy:2022}, \verb|NumPy| \footnote{\url{http://www.numpy.org}}, \verb|SciPy|\footnote{\url{http://www.SciPy.org}}, \verb|Healpy|\footnote{\url{https://healpy.readthedocs.io/}} and \verb|Matplotlib|\footnote{\url{http://www.Matplotlib.org}}.
The research carried out by DSH and GG is supported by the Discovery Grant from the Natural Sciences and Engineering Research Council of Canada (NSERC). DSH is supported by the University of Waterloo. DSH's and GG's research is also supported by Perimeter Institute for Theoretical Physics (PI). Research at PI is supported by the Government of Canada through the Department of Innovation, Science and Economic Development Canada and led by the Province of Ontario through the Ministry of Research, Innovation and Science. TK is supported by the Massachusetts Institute of Technology, and NC is supported by the Indian Institute of Science. TK's and NC's contributions to this project were carried out while being hosted and supported by the Perimeter Institute for Theoretical Physics through the PSI Start program. AK was supported as a CITA National Fellow by the Natural Sciences and Engineering Research Council of Canada (NSERC), funding reference \#DIS-2022-568580.

 \section*{Data Availability}
The data underlying this article will be available from the authors upon request.

\appendix
\section{Machine Learning Model Comparison (NN vs.\ GBM)}
\label{sec:AppendixA}
To assess the robustness of our population-synthesis calibrated reconstruction of the merger rate $\log_{10}(n_{\rm GW})$ from the host-galaxy properties
$(\log M_\star,\ \log{\rm SFR},\ \log Z)$, we trained and evaluated the following two machine learning models on the \GalR dataset and chose the better-performing one for our main analysis.

\begin{itemize}
    \item \textbf{Gradient Boosting Regressor (GBM)}, an ensemble of decision trees
    trained with a learning rate $\eta$ and $n_{\rm est}$ boosting stages, using polynomial features of \\$(\log M_\star,\ \log{\rm SFR},\ \log Z)$ 
    as input.
    \item \textbf{Fully connected Neural Network (NN)},  a feed--forward network
 with three hidden layers of sizes $(64,\,64,\,32)$, ReLU activations, and
 dropout $p=0.1$, acting on the same base feature set
    $(\log M_\star,\ \log{\rm SFR},\ \log Z)$.
\end{itemize}

Both models were trained using the same input features ($\log M_\star,\ \log {\rm SFR},\ \log Z$) and the same
90\%-10\% training-test split at each redshift
($z \in \{0.0, 0.1, 0.2, 0.5, 1.0\}$).
This controlled comparison allowed us to isolate differences in model behavior rather than differences in data selection or preprocessing.

We quantitatively and qualitatively evaluated and compared the performance of the two Machine-Learning models in predicting the number of mergers against the true number of mergers from the \GalR dataset.

\subsection{Quantitative Performance Metrics}
First, to ensure that our machine learning reconstruction of the merger rate is not limited by suboptimal model choices, we conducted a systematic hyperparameter search for both the GBM regressor and the NN at each redshift. The best performing configurations are summarized in Table \ref{tab:gbm_hyperparams_M54A5} (GBM) and \ref{tab:nn_hyperparams_M54A5} (NN), where the optimal settings were selected using a held-out 10\% test set. 
We then used the best performing configurations for the main training process, and Table~\ref{tab:NNvsGBMmetrics} summarizes the performance metrics for both 
GBM and NN models. We report the mean squared error (MSE), mean absolute error (MAE), and the coefficient of determination $R^2$ computed on the test set. As can be seen, the GBM achieves higher $R^2_{\rm test}$ and lower $\mathrm{MSE}_{\rm test}$ at all redshifts, demonstrating that it provides a more accurate and robust mapping between host-galaxy properties and merger rates. For this reason, the GBM is adopted as the fiducial model throughout the paper.

\begin{table}[h]
\centering
\small
\begin{tabular}{c c c c c c c c c}
\hline
$z$ &
$n_{\rm est}$ & $\eta$ & $d_{\max}$ &
subsample & $n_{\rm leaf}$ & $n_{\rm split}$ &
$R^2_{\rm test}$ & MSE$_{\rm test}$ \\
\hline
0.0 & 900 & 0.05 & 5 & 0.9 & 10 & 10 & 0.692 & 0.149 \\
0.1 & 900 & 0.05 & 5 & 0.9 & 10 & 10 & 0.675 & 0.118 \\
0.2 & 900 & 0.05 & 5 & 0.9 & 10 & 10 & 0.687 & 0.148 \\
0.5 & 600 & 0.05 & 5 & 0.9 & 10 & 50 & 0.673 & 0.167 \\
1.0 & 600 & 0.05 & 5 & 0.9 & 10 & 10 & 0.708 & 0.144 \\
\hline
\end{tabular}
\caption{Best performing GBM hyperparameters for the FMR
host-galaxy model (M54, $\alpha_5$), evaluated at each redshift. Here, $n_{\rm est}$ is the number of boosting trees; 
$\eta$ is the learning rate; 
$d_{\max}$ is the maximum tree depth; 
the subsample fraction controls stochastic regularization; 
$n_{\rm leaf}$ and $n_{\rm split}$ set the minimum number of samples in leaf nodes and split nodes, respectively.  
As described in the text, $R^2_{\rm test}$ and $\mathrm{MSE}_{\rm test}$ are performance metrics evaluated on the held-out test set.
}
\label{tab:gbm_hyperparams_M54A5}
\end{table}

\begin{table}[h]
\centering
\small
\begin{tabular}{c c c c c c}
\hline
$z$ &
$\eta_{\rm NN}$ & seed & $N_{\rm epoch}$ &
$R^2_{\rm test}$ & MSE$_{\rm test}$ \\
\hline
0.0 & 0.001 & 2 & 238 & 0.657 & 0.166 \\
0.1 & 0.001 & 2 & 292 & 0.642 & 0.133 \\
0.2 & 0.001 & 2 & 112 & 0.635 & 0.175 \\
0.5 & 0.001 & 1 & 114 & 0.642 & 0.180 \\
1.0 & 0.001 & 1 & 139 & 0.690 & 0.151 \\
\hline
\end{tabular}
\caption{Best performing NN hyperparameters for the FMR
host-galaxy model (M54, $\alpha_5$), selected from a grid over seeds and learning rates. Here, $\eta_{\rm NN}$ is the learning rate, the \texttt{seed} controls weight
initialization, and $N_{\rm epoch}$ is the number of epochs before early
stopping. Similar to the previous table, $R_{\rm test}^2$ and ${\rm MSE}_{\rm test}$ report performance on the
held-out test set.
}
\label{tab:nn_hyperparams_M54A5}
\end{table}

\begin{table}[h]
\centering
\begin{tabular}{cccccccc}
\hline
$z$ & MSE$_{\rm NN}$ & MAE$_{\rm NN}$ & $R^2_{\rm NN}$ 
   & MSE$_{\rm GBM}$ & MAE$_{\rm GBM}$ & $R^2_{\rm GBM}$ & $N_{\rm gal}$ \\
\hline
0.0 & 0.168 & 0.305 & 0.658 & 0.137 & 0.273 & 0.721 & 87033 \\
0.1 & 0.130 & 0.269 & 0.645 & 0.108 & 0.244 & 0.704 & 102727 \\
0.2 & 0.177 & 0.311 & 0.631 & 0.138 & 0.275 & 0.710 & 84670 \\
0.5 & 0.178 & 0.311 & 0.647 & 0.155 & 0.289 & 0.693 & 67542 \\
1.0 & 0.152 & 0.294 & 0.687 & 0.133 & 0.274 & 0.726 & 66877 \\
\hline

\end{tabular}
\caption{Comparison of Neural Network and Gradient Boosting performance 
for the M54 (FMR)--$\alpha_5$ model.}
\label{tab:NNvsGBMmetrics}
\end{table}

\subsection{Comparison of Machine Learning Reconstructions}
\begin{figure}[h]
  \centering
  \includegraphics[width=\textwidth]{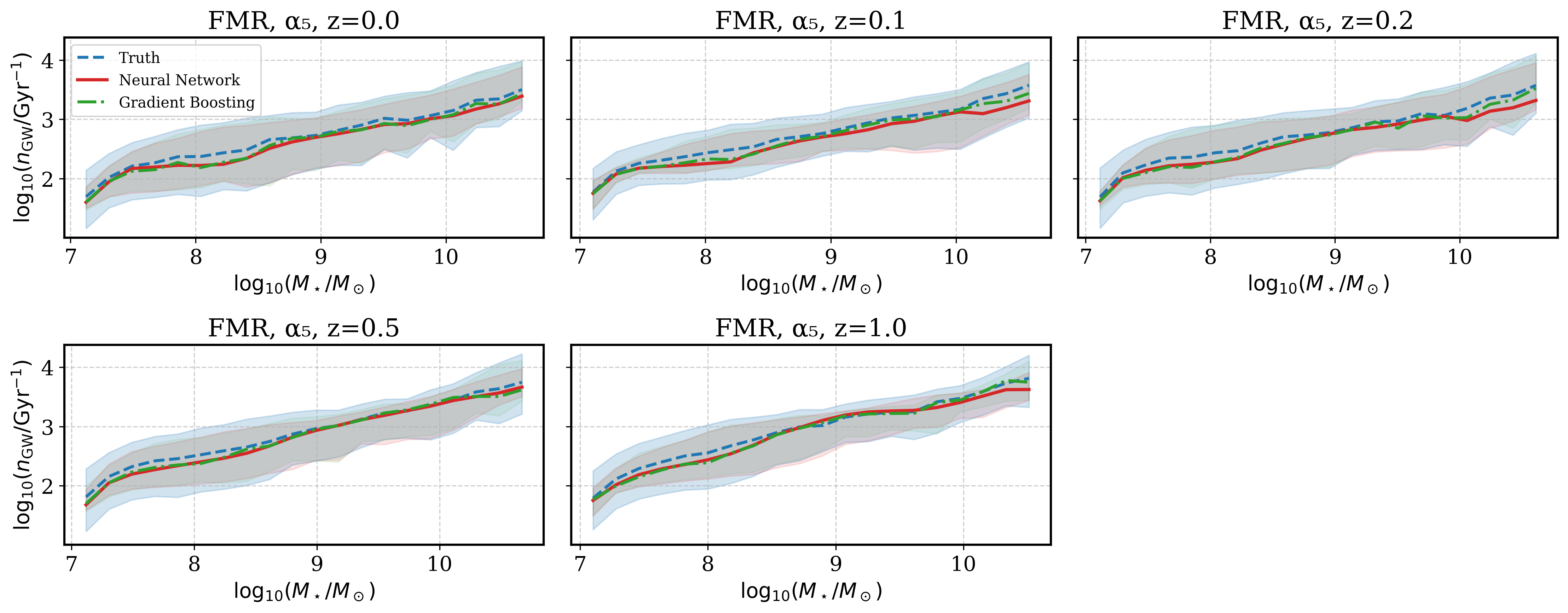}
  \caption{
    Comparison between \GalR data
    (``Truth'', dashed blue), the neural network (solid red), and the
    gradient boosting regressor (dash-dotted green) as a function of stellar mass.
    Each panel corresponds to a different redshift, from $z=0$ to $z=1$,
    for the FMR host-galaxy model (M54) with $\alpha_5$.
    The curves show the median $\log_{10}(n_{\mathrm{GW}}/\mathrm{Gyr}^{-1})$ for 
    bins in $\log_{10}(M_\star/M_\odot)$, with shaded regions indicating the
    16th-84th percentile range. Note that in these plots the other two physical properties 
(SFR and metallicity) are not binned and are free; the statistics represent the full galaxy population projected onto the mass axis.}
  \label{fig:M54A5_Mstar_truth_NN_GBM}
\end{figure}

\begin{figure}[h]
  \centering
  \includegraphics[width=\textwidth]{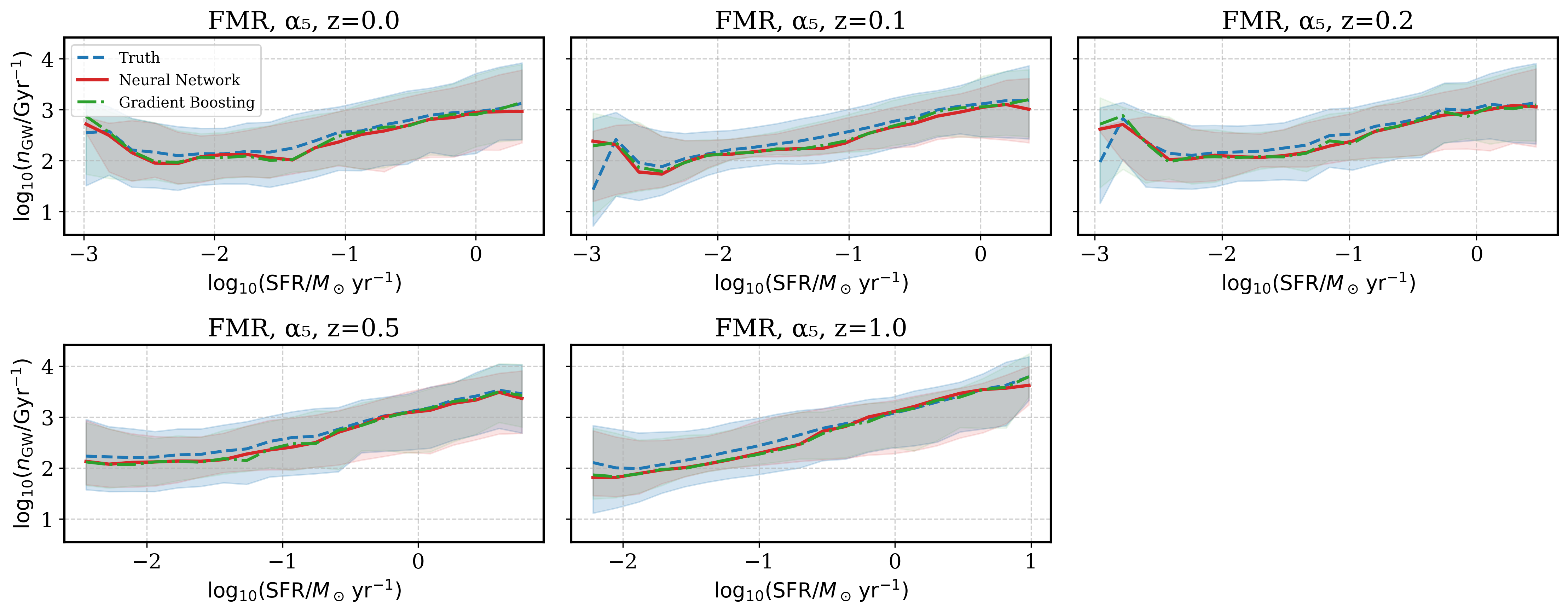}
  \caption{
    Comparison between \GalR data
    (``Truth'', dashed blue), the neural network (solid red), and the
    gradient boosting regressor predictions (dash-dotted green) as a function of SFR.
    Each panel corresponds to a different redshift, from $z=0$ to $z=1$,
    for the FMR host-galaxy model (M54) with $\alpha_5$.
    The curves show the median $\log_{10}(n_{\mathrm{GW}}/\mathrm{Gyr}^{-1})$ in
    bins of $\log_{10}(\mathrm{SFR}/M_\odot \mathrm{yr}^{-1})$, with shaded regions indicating the 16th-84th percentile range. Note that in these plots the other two physical properties 
(stellar mass and metallicity) are not held fixed; the statistics represent the full galaxy population projected onto the SFR axis.
  }
  \label{fig:M54A5_SFR_truth_NN_GBM}
\end{figure}

\begin{figure}[h]
  \centering
  \includegraphics[width=\textwidth]{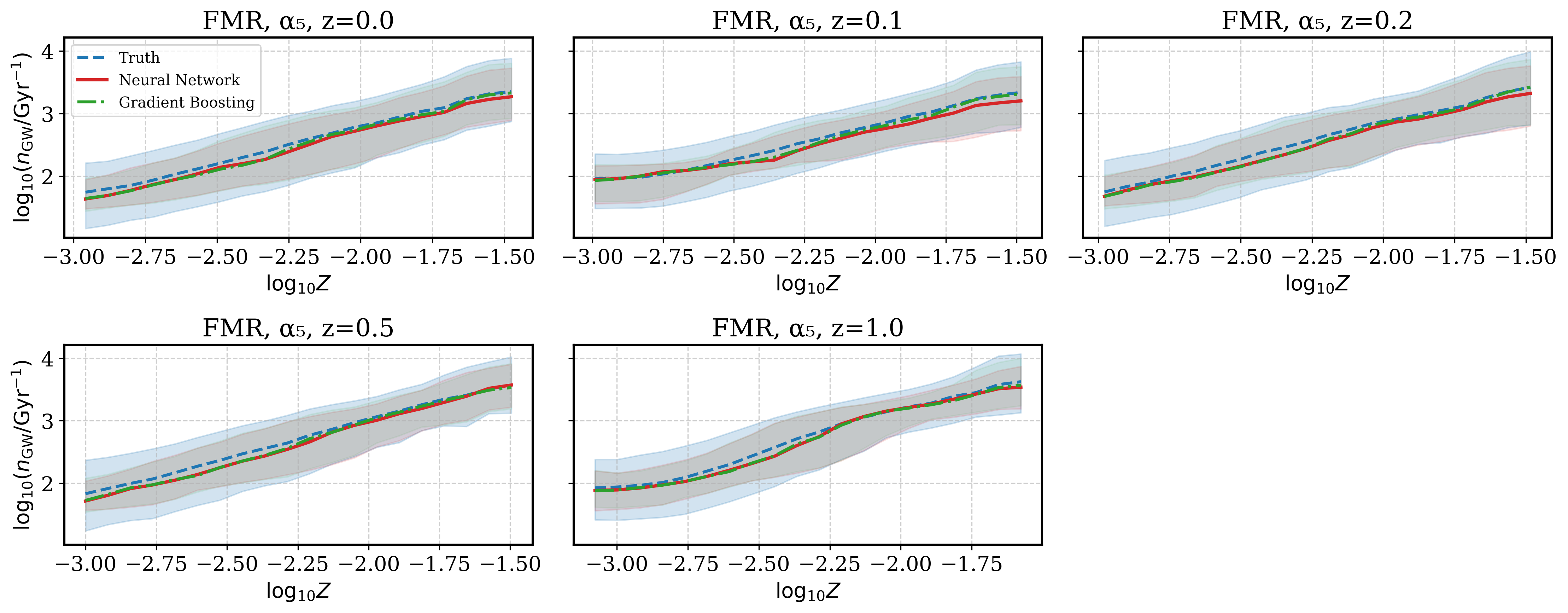}
  \caption{
    Comparison between \GalR data
    (``Truth'', dashed blue), the neural network (solid red), and the
    gradient boosting regressor predictions (dash-dotted green) as a function of metallicity.
    Each panel corresponds to a different redshift, from $z=0$ to $z=1$,
    for the FMR host-galaxy model (M54) with $\alpha_5$.
    The curves show the median $\log_{10}(n_{\mathrm{GW}}/\mathrm{Gyr}^{-1})$ in
    bins of $\log_{10}(Z)$, with shaded regions indicating the 16th-84th percentile range. Note that in these plots the other two physical properties 
(stellar mass and SFR) are not held fixed; the statistics represent the full galaxy population projected onto the metallicity axis.
  }
  \label{fig:M54A5_Z_truth_NN_GBM}
\end{figure}

Figures~\ref{fig:M54A5_Mstar_truth_NN_GBM}, \ref{fig:M54A5_SFR_truth_NN_GBM}, and \ref{fig:M54A5_Z_truth_NN_GBM} show for a given galaxy property (mass/SFR/metallicity), a 
three–way comparison between the ``Truth'' merger rates from \GalR (dashed blue), the neural network predictions (solid red), and the gradient boosting predictions (dash-dotted green) for redshifts $z \in \{0,0.1,0.2,0.5,1\}$, the FMR model (M54, $\alpha_5$). In each panel, we bin galaxies in the horizontal variable ($M_\star$/SFR/$Z$) and plot the median
$\log_{10}(n_{\rm GW}/{\rm Gyr}^{-1})$ together with the 16th-84th percentile
range.  This provides a distribution-level comparison between the two ML models and the underlying ``Truth'' behavior across all redshifts and all three physical axes: stellar mass (Figure~\ref{fig:M54A5_Mstar_truth_NN_GBM}), star formation rate 
(Figure~\ref{fig:M54A5_SFR_truth_NN_GBM}), and metallicity
(Figure~\ref{fig:M54A5_Z_truth_NN_GBM}). As can be seen, the gradient boosting regressor consistently tracks the true median trends very closely. The NN model performs well too, but it shows slightly more deviations at higher mass and higher SFR tails, where it tends to underestimate the true signal. Because the GBM achieves (i) lower test–set MSE and MAE, (ii) higher $R^2$, and (iii) a visibly more accurate reproduction of the galaxy-population trends, we adopt the GBM model throughout the main analysis.

\begin{figure}[!htbp]
    \centering
    \includegraphics[width=\textwidth]{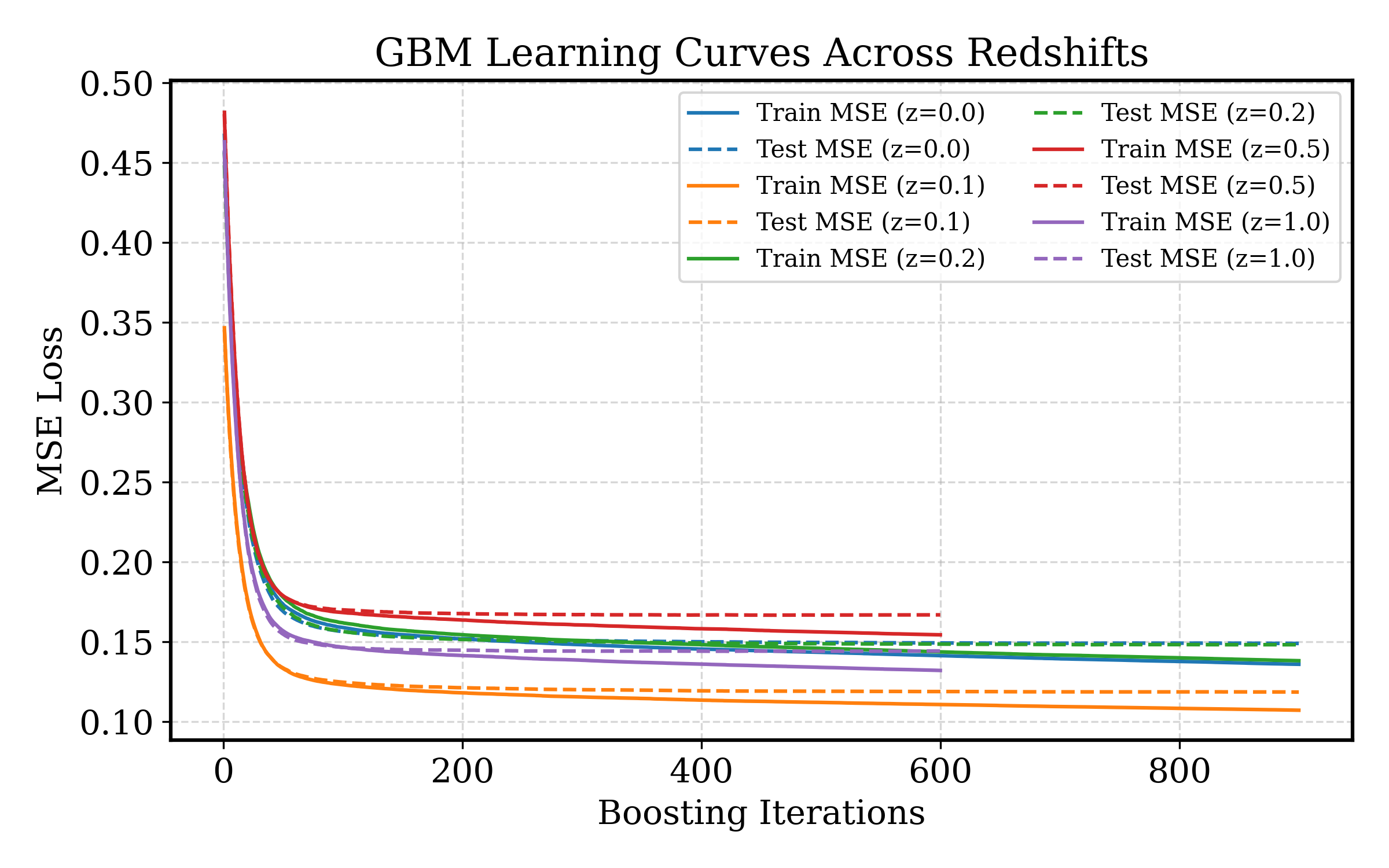}
     \caption{GBM learning curve showing training and test loss as a function of boosting iterations.}
    \label{fig:gbm-learning_residual}
\end{figure}

Having identified GBM as the primary emulator for our analysis, we present its learning curve in Figure ~\ref{fig:gbm-learning_residual}.

\section{Merger rate distribution derived for IllustrisTNG300 using alternative population-synthesis models} \label{sec:mergerrate_other models}

\begin{figure}[htbp]
    \centering
    \includegraphics[width=\textwidth]{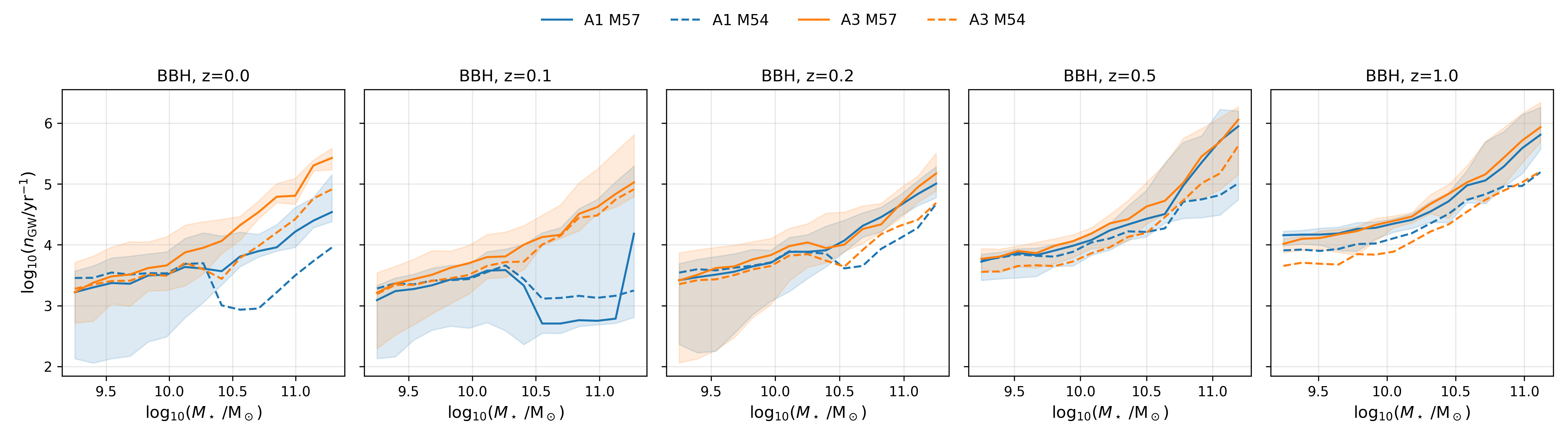}
    \includegraphics[width=\textwidth]{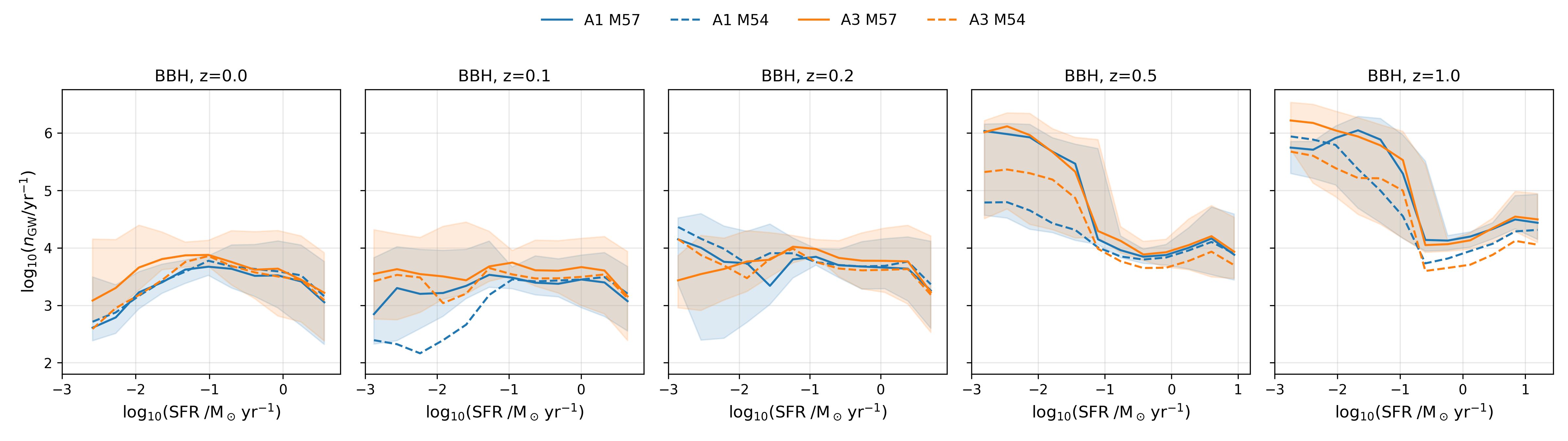}
    \includegraphics[width=\textwidth]{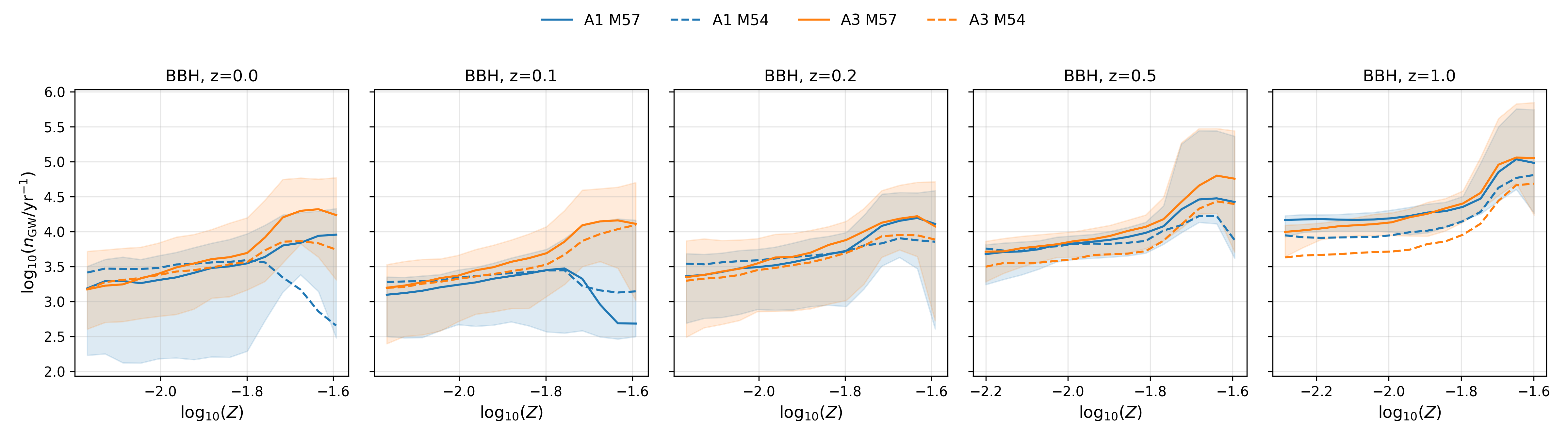}
    \caption{
        \textbf{BBH merger rate versus host-galaxy properties for different astrophysical models.}
        Rows correspond to stellar mass (top), SFR (middle), and metallicity (bottom).
        Columns correspond to redshifts $z=\{0.0,\,0.1,\,0.2,\,0.5,\,1.0\}$.
        Solid curves use the mass-metallicity relation (MZR; Model M57), while dashed curves use the fundamental metallicity relation (FMR; Model M54).
        Colors show stellar-evolution models $\alpha_{CE}=1$($A_1$) and $\alpha_{CE}=3$($A_3$).
        Shaded bands indicate the 16--84\,percentile range for models.
    }
    \label{fig:three_panel_merger_rates}
\end{figure}

Plots in Figure~\ref{fig:three_panel_merger_rates} show the variation of the median of $\log_{10}(n_{\rm GW})$ (after implementing our ML algorithm in IllustrisTNG300 galaxy sample) based on different seeding models used in \GalR. The distribution of the merger rate is presented as a function of metallicity, star formation rate, and stellar mass for two common envelope efficiency parameter choices, $\alpha_1$ and $\alpha_3$ (different from the one considered in the main text, which was $\alpha_5$), and for both metallicity prescriptions: the mass-metallicity relation (MZR$\to$ M57, shown with solid curves) and the fundamental metallicity relation (FMR $\to$ M54, shown with dashed curves). Shaded bands denote the 16--84th percentile range for each model. Each row spans five redshift snapshots, illustrating how the sensitivity of merger rate to galaxy properties evolves from $z=1$ to today. As illustrated, varying the population-synthesis parameters and the metallicity prescription can significantly affect the merger rate predictions at low redshift.

\bibliographystyle{JHEP}
\bibliography{paper}

\end{document}